\documentclass[12pt]{article}
\usepackage{graphicx, setspace}
\usepackage{mdframed}
\usepackage{xcolor}

\usepackage{amsmath, amssymb}
\usepackage{titlesec}
\usepackage{soul}
\usepackage[dvipsnames]{xcolor}
\usepackage{graphicx}
\usepackage{needspace} %this ensures that my section titles remain as one block and are not broken into 2 bits even at the end of a page
\usepackage{listings}
\usepackage[colorlinks=true,linkcolor=blue]{hyperref}%
\usepackage{longtable} %multipage tabular
\usepackage{lettrine}
\usepackage{ebgaramond}
\usepackage{subcaption}
\usepackage{pdfpages}
\usepackage{wrapfig}

\usepackage{lmodern}

\usepackage{xurl}

\usepackage[titles]{tocloft}

\usepackage[total={5.5in, 8.5in},top=3cm,bottom=4cm,left=3cm,right=3cm]{geometry}

\usepackage{array}
\usepackage{fancybox}

\usepackage[round,sectionbib]{natbib}
\usepackage{chapterbib}

\makeatletter
\renewcommand\maketitle
  {\begin{center}
   {
   \Large\scshape{\@title}
   }
   \end{center}
  }
\makeatother

\titleformat{\section}[display]
    {\clearpage\flushright}
    {\fontsize{96}{50}\selectfont\thesection.}
    {-5pt}
    {\Huge}
    [\vspace{-1.5ex} \hspace{1.3ex} \rule{0.8\textwidth}{0.2pt} ]
\titlespacing*{\section}{0pt}{0pt}{\baselineskip}

\titleformat{\subsection}[block]
    {\Large\scshape}
    {}
    {0pt}
    {\needspace{5\baselineskip}\rule{0.8\textwidth}{0.2pt} \\ \vspace{0.5ex} \thesubsection. \quad }
    [\vspace{-1.5ex} \rule{0.8\textwidth}{0.2pt} ] 
\titlespacing{\subsection}{-1cm}{\baselineskip}{\baselineskip}

\renewcommand{\thesubsubsection}{\arabic{subsubsection}}
\titleformat{\subsubsection}{\scshape}{}{0.2ex}
{\thesubsubsection. \quad}
[ \vspace{-2ex} \rule{\textwidth}{0.15pt} ]

\DeclareCaptionLabelFormat{custom}
{%
      #1 \textbf{(#2)}
}
\DeclareCaptionLabelSeparator{custom}{--}
\DeclareCaptionFormat{custom}
{%
    #1#2\small #3
}
\newenvironment{sectionauthor}
{\begin{flushright}\it}
{\end{flushright}}

\hypersetup{
  colorlinks,
  citecolor=Blue,
  linkcolor=Blue,
  urlcolor=Blue}

\definecolor{codegreen}{rgb}{0,0.6,0}
\definecolor{codegray}{rgb}{0.5,0.5,0.5}
\definecolor{codepurple}{rgb}{0.58,0,0.82}
\definecolor{codekeyword}{rgb}{0.5, 0.5, 0.2}
\definecolor{Orchid}{rgb}{0.686,0.447,0.690}

\definecolor{CLIprompt}{rgb}{0.4,0.6,0}
\definecolor{xspeckeyword}{rgb}{0.6, 0.4, 0.3}

\lstdefinestyle{terminal}{    
    commentstyle=\it\color{MidnightBlue},
    keywordstyle=\color{MidnightBlue},
    numberstyle=\tiny\color{codegray},
    stringstyle=\bfseries\color{codepurple},
    basicstyle=\ttfamily\footnotesize,
    breakatwhitespace=false,         
    breaklines=true,                 
    captionpos=b,                    
    keepspaces=true,                 
    numbers=none,                                      
    showspaces=false,                
    showstringspaces=false,
    showtabs=false,                  
    tabsize=2,
    otherkeywords = {user@here:\$, XSPEC12>,PLT>, PyXspec>},
    keywordstyle={\color{CLIprompt}\bfseries},
    xleftmargin=.25in,
    xrightmargin=.25in
}

\lstdefinestyle{file}{    
    commentstyle=\color{codegreen},
    keywordstyle=\color{codekeyword},
    numberstyle=\tiny\color{codegray},
    stringstyle=\bfseries\color{codepurple},
    basicstyle=\ttfamily\footnotesize,
    breakatwhitespace=false,         
    breaklines=true,                 
    captionpos=b,                    
    keepspaces=true,                 
    numbers=left,                    
    numbersep=10pt,                  
    showspaces=false,                
    showstringspaces=false,
    showtabs=false,                  
    tabsize=2,
    xrightmargin=.3in
}
 
\lstdefinelanguage{xspec}{
    basicstyle=\ttfamily\small,
    columns=fullflexible,
    morecomment=[s][\color{Orchid}\bfseries]{[}{]},
    morecomment=[l]{\#},
    morecomment=[l]{;},
    commentstyle=\color{gray}\ttfamily,
    morekeywords = [2]{lmod, cpd, setplot, dummyrsp, dummy, fakeit, model, Model},
    keywordstyle = [2]{\color{xspeckeyword}\bfseries},
    otherkeywords = {XSPEC12>,PLT>},
    keywordstyle = {\color{CLIprompt}\bfseries}
}

\makeatletter \def\lst@gkeywords@sty{\color{CLIprompt}\bfseries} % here to fix what seems to be a bug with lstlisting where I give a straightforward definition to whatever internal function it's missing

\newcommand{\rin}[1][]{%
  \ifx\relax#1\relax
    \ensuremath{r_{\rm in}}%
  \else
    \ensuremath{r_{{\rm in}, #1}}%
  \fi
}

\newcommand{\rout}[1][]{%
  \ifx\relax#1\relax
    \ensuremath{r_{\rm out}}%
  \else
    \ensuremath{r_{{\rm out}, #1}}%
  \fi
}

\newcommand{\rg}[1][]{%
  \ifx\relax#1\relax
    \ensuremath{r_{g}}%
  \else
    \ensuremath{r_{g, #1}}%
  \fi
}

\newcommand{\risco}[1][]{%
  \ifx\relax#1\relax
    \ensuremath{r_{\rm ISCO}}%
  \else
    \ensuremath{r_{#1{\rm (ISCO)}}}%
  \fi
}

\newcommand{\rl}[1][]{%
  \ifx\relax#1\relax
    \ensuremath{r_{\rm L}}%
  \else
    \ensuremath{r_{#1{\rm (L)}}}%
  \fi
}

\newcommand{\rms}[1][]{%
  \ifx\relax#1\relax
    \ensuremath{r_{\rm MS}}%
  \else
    \ensuremath{r_{#1{\rm (MS)}}}%
  \fi
}

\begin{document}

\includepdf{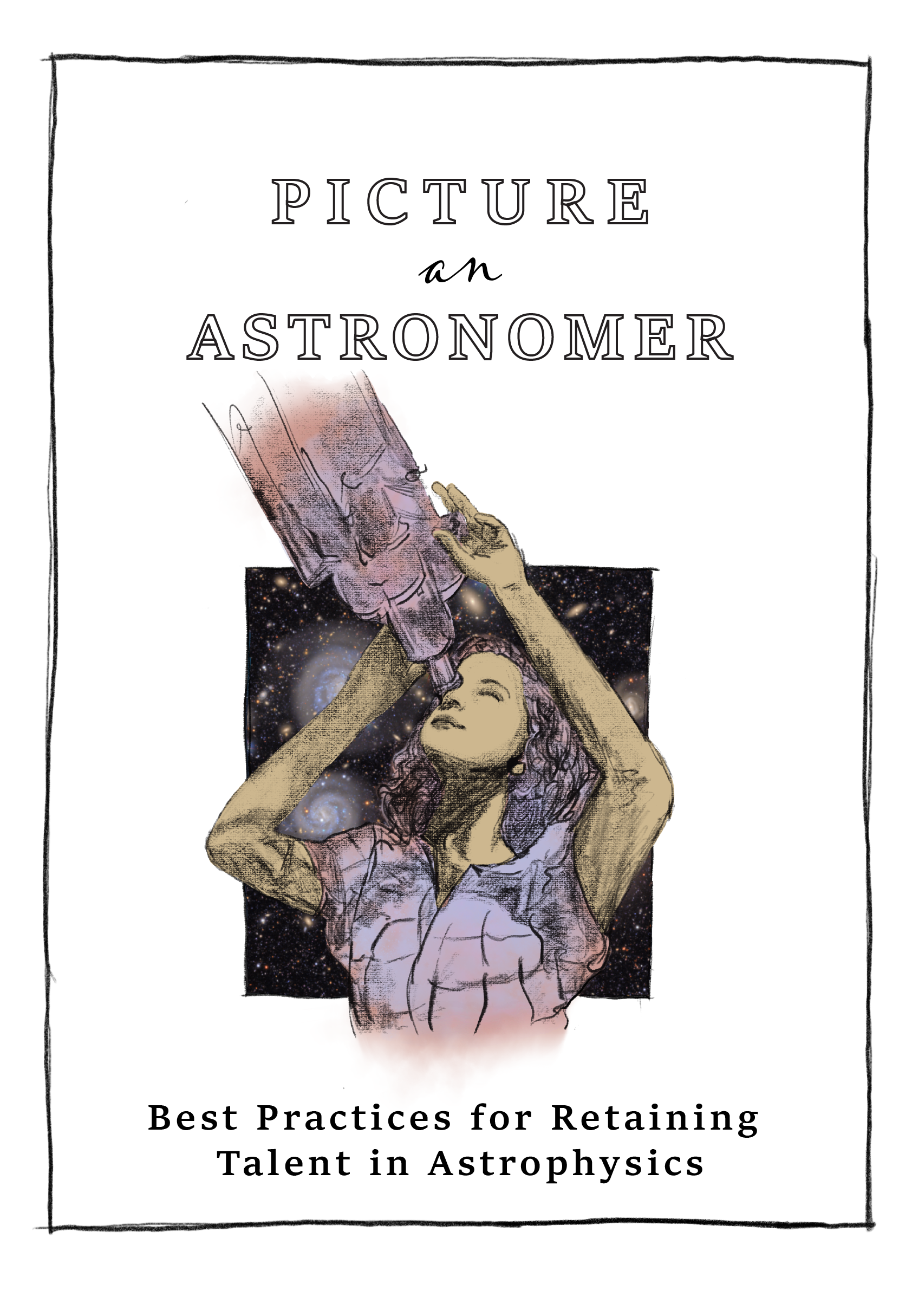}

\pagebreak

\thispagestyle{empty}
% \strut \vspace{35pt} %\vspace{40pt}
% \strut \vspace{20pt} %\vspace{40pt}

\Huge{\textsc{Picture an Astronomer:}}\\ 
\large{\textsc{Best Practices for Retaining Talent in Astrophysics}}\\
\\

% \vspace{10pt}
\vspace{7pt}
\normalsize{\textit{Edited by:} \href{https://orcid.org/0000-0002-5283-933X}{Ava Polzin}$^{1,2}$ and \href{https://orcid.org/0000-0001-7160-3632}{Katherine E. Whitaker}$^{3,4}$}\\
\vspace{10pt}

\normalsize{\textit{Foreword by:} \href{https://orcid.org/0000-0002-0745-9792}{C. Megan Urry}$^{5,6}$}\\
\vspace{10pt}

% \begin{center}
\normalsize{\href{https://orcid.org/0009-0000-7177-9697}{Henna Abunemeh}$^{7}$, \href{https://orcid.org/0009-0003-1240-2330}{Sanyukta Agarwal}$^{8}$, \href{https://orcid.org/0009-0008-1965-9012}{Aadya Agrawal}$^{7}$, 
\href{https://orcid.org/0009-0003-2076-6118}{Nathaniel Alden}$^{2,9}$, \\Ann-Marsha Alexis$^{10}$, \href{https://orcid.org/0000-0001-9348-9738}{Sydney Andersen}$^{11}$, \href{https://orcid.org/0000-0002-0517-9842}{Melanie Archipley}$^{1,2}$, \href{https://orcid.org/0000-0002-8320-2198}{Yasmeen Asali}$^{12}$, \\\href{https://orcid.org/0000-0002-4449-9152}{Katie Auchettl}$^{13,14}$, \href{https://orcid.org/0000-0002-5108-6823}{Bradford Benson}$^{1,2}$, \href{https://orcid.org/0000-0002-7707-1996}{Binod Bhattarai}$^{15}$, \href{https://orcid.org/0000-0002-4826-8642}{Sarah Biddle}$^{16}$,
\\\href{https://orcid.org/0000-0003-2404-2427}{Madison Brady}$^{17}$, \href{https://orcid.org/0000-0001-5228-6598}{Katelyn Breivik}$^{10,18}$, Disha Chakraborty$^{8}$, \href{https://orcid.org/0009-0006-4036-4919}{Mikel Charles}$^{19,20}$, \\\href{https://orcid.org/0000-0001-8813-4182}{Hsiao-Wen Chen}$^{1,2}$, Josephine Chishala$^{21,22}$, \href{https://orcid.org/0000-0002-7155-679X}{Anirudh Chiti}$^{23}$, \\Panagiota Eleftheria Christopoulou$^{24}$, \href{https://orcid.org/0000-0002-5995-9692}{Mi Dai}$^{25,26,27}$, \href{https://orcid.org/0000-0001-6793-2572}{Flaminia Fortuni}$^{28}$, \\\href{https://orcid.org/0000-0002-1819-0215}{Shanika Galaudage}$^{29,30}$, \href{https://orcid.org/0000-0002-2411-2766}{Daniel Glazer}$^1$, Anika Goel$^{31}$, \href{https://orcid.org/0000-0002-5726-5216}{Andrea Gokus}$^{32}$, \\\href{https://orcid.org/0000-0002-5612-3427}{Jenny E. Greene}$^{33}$, \href{https://orcid.org/0009-0007-8812-1630}{Ryn Grutkoski}$^{1,2}$, Yiqing Guo$^{34,35}$, Joseph Guzman$^{36}$, \\\href{https://orcid.org/0000-0002-0965-7864}{Ren\'{e}e Hlo\v{z}ek}$^{37,38}$,
\href{https://orcid.org/0000-0002-1496-6514}{Lindsay R. House}$^{39,40}$,
\href{https://orcid.org/0009-0006-5765-1607}{Lillian N. Joseph}$^{41}$, Molly Beth Jourdan$^{42}$, \href{https://orcid.org/0000-0002-1384-9949}{Tanvi Karwal}$^{1,2,43}$, \href{https://orcid.org/0009-0003-0415-404X}{Zuzanna Kocjan}$^{44}$, \href{https://orcid.org/0000-0001-8682-1421}{Emily Koivu}$^{19,20}$, Varun Kore$^{8}$, \\\href{https://orcid.org/0000-0003-4307-634X}{Andrey Kravtsov}$^{1,2,43}$, \href{https://orcid.org/0009-0000-4830-1484}{Keerthi Kunnumkai}$^{10}$, \href{https://orcid.org/0009-0007-8974-9016}{Shalini Kurinchi-Vendhan}$^{45}$, \\\href{https://orcid.org/0000-0002-2450-1366}{Johannes U. Lange}$^{46}$, \href{https://orcid.org/0000-0003-3217-5967}{Sarah R. Loebman}$^{15}$, \href{https://orcid.org/0009-0009-9730-0483}{Kira Lund}$^{47}$, \href{https://orcid.org/0009-0004-2625-5527}{Julie Malewicz}$^{48}$,
\\\href{https://orcid.org/0009-0006-1832-6713}{Olivia McAuley}$^{49}$,
\href{https://orcid.org/0000-0002-6187-4866}{Rebecca McClain}$^{50,20}$, \href{https://orcid.org/0000-0003-4248-6128}{Stephen McKay}$^{51}$, \href{https://orcid.org/0009-0008-5622-6857}{Emily McPike}$^{52,53}$, \\\href{https://orcid.org/0000-0003-3585-3356}{Cassidy Metzger}$^{54}$, \href{https://orcid.org/0000-0002-8530-9765}{Lamiya A. Mowla}$^{55,56}$, \href{https://orcid.org/0009-0009-2307-2350}{Katherine Myers}$^{57}$, \href{https://orcid.org/0000-0002-7524-374X}{Erica Nelson}$^{58}$, \\\href{https://orcid.org/0000-0003-0280-6617}{Aline Novais}$^{59}$, \href{https://orcid.org/0009-0007-8900-7178}{Camilla Nyhagen}$^{59}$, \href{https://orcid.org/0000-0001-5636-3108}{Micah Oeur}$^{15}$, \href{https://orcid.org/0009-0002-6065-3292}{Lou Baya Ould Rouis}$^{60,61}$, \\\href{https://orcid.org/0009-0003-6803-2420}{Paarmita Pandey}$^{50,19}$, \href{https://orcid.org/0009-0009-2685-4067}{Raagini Patki}$^{62}$, \href{https://orcid.org/0000-0002-8131-7020}{Sonu Tabitha Paulson}$^{63}$, \href{https://orcid.org/0009-0000-5561-9116}{Haile M. L. Perkins}$^7$, \\Ashi Poorey$^{7}$, Izabella Pozo$^{64}$, \href{https://orcid.org/0000-0002-4964-2691}{Heather L. Preston}$^{1}$, \href{https://orcid.org/0000-0002-5104-5263}{Pazit Rabinowitz}$^{31}$, \\\href{https://orcid.org/0000-0003-3953-1776}{Alexandra S. Rahlin}$^{1,2}$, \href{https://orcid.org/0000-0003-0343-0121}{Janiris Rodriguez-Bueno}$^{7}$, \href{https://orcid.org/0000-0001-6535-1766}{Francisco Rodr\'{i}guez Montero}$^{1,2}$,
\href{https://orcid.org/0000-0001-8023-4912}{Huei Sears}$^{65}$, \href{https://orcid.org/0000-0002-0415-3077}{\'{A}lvaro Segovia Otero}$^{66}$, \href{https://orcid.org/0000-0001-9710-9827}{Uliana Solovieva}$^{67}$, \href{https://orcid.org/0000-0002-6748-6821}{Rachel Somerville}$^{68}$, \\\href{https://orcid.org/0000-0003-3430-3889}{Jessica Speedie}$^{69}$, \href{https://orcid.org/0000-0003-2539-8206}{Tjitske Starkenburg}$^{34,35,40}$,
\href{https://orcid.org/0009-0005-6146-7151}{Laura Stiles-Clarke}$^{70}$, \href{https://orcid.org/0000-0003-0478-0473}{Chin Yi Tan}$^{2,9}$, \\\href{https://orcid.org/0000-0003-4209-1599}{Yu-Hsuan Teng}$^{44}$, \href{https://orcid.org/0000-0001-6746-9936}{Tanya Urrutia}$^{71}$, \href{https://orcid.org/0000-0001-8638-2780}{Padmavathi Venkatraman}$^{7}$, \href{https://orcid.org/0000-0003-1535-4277}{Margaret E. Verrico}$^{7,72}$, \\\href{https://orcid.org/0000-0002-4186-6164}{Amanda Wasserman}$^{7,40}$, \href{https://orcid.org/0000-0003-2369-2911}{Claire E. Williams}$^{73,74}$,
\href{https://orcid.org/0000-0002-7759-0585}{Tony Wong}$^{7}$, Shirin Gul Zaidi$^{52,53,75}$, and \href{https://orcid.org/0000-0001-8298-5859}{Chantene Zichterman}$^{76}$}
% \end{center}

% \vfill
% \begin{center}
%     \textit{All art by Julie Malewicz.}
% \end{center}

% \vspace{20pt}

% \scriptsize{
% \textit{Cover:} Illustration of Vera Rubin, based on the 1948 picture of her at the Vassar College Observatory from the \href{https://archivesspace.carnegiescience.edu/repositories/3/resources/26/digitized}{Carnegie Science Vera C. Rubin Photograph Collection}. The background is one of the first light images from the NSF-DOE Vera C. Rubin Observatory released June 2025. It is the first major observatory to be named after a woman.
% }

\pagebreak
\thispagestyle{empty}

\newgeometry{top=3cm,bottom=4cm,left=2.5cm,right=2.5cm}
\scriptsize{\setstretch{1.0}
$^1$ \hspace{1pt} Department of Astronomy \& Astrophysics, University of Chicago, Chicago, IL, USA\\
$^2$ \hspace{1pt} Kavli Institute for Cosmological Physics, University of Chicago, Chicago, IL, USA\\
$^3$ \hspace{1pt} Department of Astronomy, University of Massachusetts, Amherst, MA, USA\\
$^4$ \hspace{1pt} Cosmic Dawn Center (DAWN), Copenhagen, DK\\
$^5$ \hspace{1pt} Department of Physics, Yale University, New Haven, CT, USA\\
$^{6}$ \hspace{1pt} Yale Center for Astronomy \& Astrophysics, Yale University, New Haven, CT, USA\\
$^7$ \hspace{1pt} Department of Astronomy, University of Illinois Urbana-Champaign, Urbana, IL, USA\\
$^{8}$ \hspace{1pt} Department of Physics \& Astronomy, University of Kansas, Lawrence, KS, USA\\
$^{9}$ \hspace{1pt} Department of Physics, University of Chicago, Chicago, IL, USA\\
$^{10}$ Department of Physics, Carnegie Mellon University, Pittsburgh, PA, USA\\
$^{11}$ Department of Astronomy, University of Washington, Seattle, WA, USA\\
$^{12}$ Department of Astronomy, Yale University, New Haven, CT, USA\\
$^{13}$ School of Physics, University of Melbourne, Parkville, VIC AUS\\
$^{14}$ Department of Astronomy and Astrophysics, University of California, Santa Cruz, CA, USA\\
$^{15}$ Department of Physics, University of California, Merced, Merced, CA, USA\\
$^{16}$ Center for Astrophysics $|$ Harvard \& Smithsonian, Cambridge, MA, USA\\
$^{17}$ Department of Physics and Astronomy, Michigan State University, East Lansing, MI, USA\\
$^{18}$ McWilliams Center for Cosmology and Astrophysics, Carnegie Mellon University, Pittsburgh, PA, USA\\
$^{19}$ Department of Physics, The Ohio State University, Columbus, OH, USA\\
$^{20}$ Center for Cosmology and AstroParticle Physics, The Ohio State University, OH, USA\\
$^{21}$ Department of Physics and Astronomy, Botswana International University of Science and Technology, Palapye, BW\\
$^{22}$ Department of Physics, School of Mathematics and Natural Sciences, Copperbelt University, Kitwe, ZM\\
$^{23}$ Kavli Institute for Particle Astrophysics \& Cosmology, Stanford University, Stanford, CA, USA\\
$^{24}$ Department of Physics, University of Patras, Patras, GR\\
$^{25}$ LSST Interdisciplinary Network for Collaboration and Computing Frameworks, Tucson AZ, USA\\
$^{26}$ Pittsburgh Particle Physics, Astrophysics, and Cosmology Center, University of Pittsburgh, Pittsburgh, PA, USA\\
$^{27}$ Department of Physics \& Astronomy , University of Pittsburgh, Pittsburgh, PA, USA\\
$^{28}$ INAF - Osservatorio Astronomico di Roma, Monte Porzio Catone (Roma), IT\\
$^{29}$ Laboratoire Lagrange, Universit\'e C\^ote d'Azur, Observatoire de la C\^ote d'Azur, CNRS,  Nice, FR\\
$^{30}$ Laboratoire Artemis, Universit\'e C\^ote d'Azur, Observatoire de la C\^ote d'Azur, CNRS,  Nice, FR\\
$^{31}$ Department of Astronomy, Indiana University Bloomington, Bloomington, IN, USA\\
$^{32}$ Department of Physics, Washington University in St. Louis, St. Louis, MO, USA\\
$^{33}$ Department of Astrophysical Sciences, Princeton University, Princeton, NJ, USA\\
$^{34}$ Department of Physics \& Astronomy, Northwestern University, Evanston, IL, USA\\
$^{35}$ Center for Interdisciplinary Exploration and Research in Astrophysics, Northwestern University, Evanston, IL, USA\\
$^{36}$ Chicago Astronomer, chicagoastronomer.com, Chicago, IL, USA\\
$^{37}$ Dunlap Institute for Astronomy and Astrophysics, University of Toronto, Toronto, ON, CA\\
$^{38}$ David A. Dunlap Department of Astronomy and  Astrophysics, University of Toronto, Toronto, ON, CA\\
$^{39}$ Data Science Institute, University of Chicago, Chicago, IL, USA\\
$^{40}$ NSF-Simons AI Institute for the Sky (SkAI), Chicago, IL, USA\\
$^{41}$ College of Science and Health, Benedictine University, Lisle, IL, USA\\
$^{42}$ Kenwood Academy, Chicago Public Schools, Chicago, IL, USA\\
$^{43}$ Enrico Fermi Institute, University of Chicago, Chicago, IL, USA \\
$^{44}$ Department of Astronomy, University of Maryland, College Park, MD, USA\\
$^{45}$ Max Planck Institute for Astronomy, Heidelberg University, Heidelberg, Baden-W\"{u}rttemberg, DE\\
$^{46}$ Department of Physics, American University, Washington, DC, USA\\
$^{47}$ Department of Physics, University of Bath, Claverton Down, Bath, UK\\
$^{48}$ Department of Physics, Georgia Institute of Technology, Atlanta, GA, USA\\
$^{49}$ Department of Physics, Bryn Mawr College, Bryn Mawr, PA, USA\\
$^{50}$ Department of Astronomy, The Ohio State University, Columbus, OH, USA\\
$^{51}$ Department of Physics, University of Wisconsin-Madison, Madison, WI, USA\\
$^{52}$ Graduate Center, City University of New York, New York, NY, USA\\
$^{53}$ Department of Astrophysics, American Museum of Natural History, New York, NY, USA\\
$^{54}$ Department of Physics and Astronomy, Dartmouth College, Hanover, NH, USA\\
$^{55}$ Whitin Observatory, Department of Physics and Astronomy, Wellesley College, Wellesley, MA, USA\\
$^{56}$ Center for Astronomy, Space Science, and Astrophysics, Independent University Bangladesh, Dhaka, BD\\
$^{57}$ Department of Physics and Astronomy, York University, Toronto, ON, CA\\
$^{58}$ Department for Astrophysical and Planetary Science, University of Colorado, Boulder, CO, USA\\
$^{59}$ Division of Astrophysics, Department of Physics, Lund University, Lund, SE\\
$^{60}$ Department of Astronomy, Boston University, Boston, MA, USA\\
$^{61}$ Institute for Astrophysical Research, Boston University Boston, MA, USA\\
$^{62}$ Department of Astronomy, Cornell University, Ithaca, NY, USA\\
$^{63}$ Instituto de Astronom\'{i}a, Universidad Nacional Aut\'{o}nomo de M\'{e}xico, Ciudad de M\'{e}xico, MX\\
$^{64}$ Department of Physics and Astronomy, State University of New York at Stony Brook, Stony Brook, NY, USA\\
$^{65}$ Department of Physics and Astronomy, Rutgers, the State University of New Jersey, Piscataway, NJ, USA\\
$^{66}$ Department of Astronomy, Tsinghua University, Beijing, PRC\\
$^{67}$ Department of Psychology, University of Chicago, Chicago, IL, USA\\
$^{68}$ Center for Computational Astrophysics, Flatiron Institute, New York, NY, USA\\
$^{69}$ Department of Earth, Atmospheric, and Planetary Sciences, Massachusetts Institute of Technology, Cambridge, MA, USA\\
$^{70}$ Department of Astronomy and Physics, Saint Mary's University, Halifax, NS, CA\\
$^{71}$ Leibniz-Institut f\"{u}r Astrophysik Potsdam, Potsdam, Brandenburg, DE\\
$^{72}$ Center for AstroPhysical Surveys, National Center for Supercomputing Applications, Urbana, IL, USA\\
$^{73}$ Department of Physics and Astronomy, University of California Los Angeles, Los Angeles, CA, USA\\
$^{74}$ Mani L. Bhaumik Institute for Theoretical Physics, University of California Los Angeles, Los Angeles, CA, USA\\
$^{75}$ Niels Bohr Institute, University of Copenhagen, Copenhagen, DK\\
$^{76}$ Department of Organismal Biology \& Anatomy, University of Chicago, Chicago, IL, USA%\\
% $^{61}$ Department of Public Administration, University of Illinois Springfield, Springfield, IL, USA %left off of last entry in authorship form (for CZ)

}
\thispagestyle{empty}

\pagebreak
\restoregeometry
\normalsize

\thispagestyle{empty}

\strut \vspace{150pt}

\scriptsize{
\textit{Cover:} Illustration of Vera Rubin, based on the 1948 picture of her at the Vassar College Observatory from the \href{https://archivesspace.carnegiescience.edu/repositories/3/resources/26/digitized}{Carnegie Science Vera C. Rubin Photograph Collection}. The background is one of the first light images from the NSF-DOE Vera C. Rubin Observatory released June 2025. It is the first major observatory to be named after a woman.
}

\vspace{80pt} 

\normalsize
\begin{center}
    \textit{All art by Julie Malewicz.}
\end{center}

\pagebreak
\setcounter{page}{1}

\tableofcontents
\pagebreak
\section{Foreword \label{sec:foreword}}

\vspace{-40pt}
\begin{center}
    \includegraphics[width=0.4\linewidth]{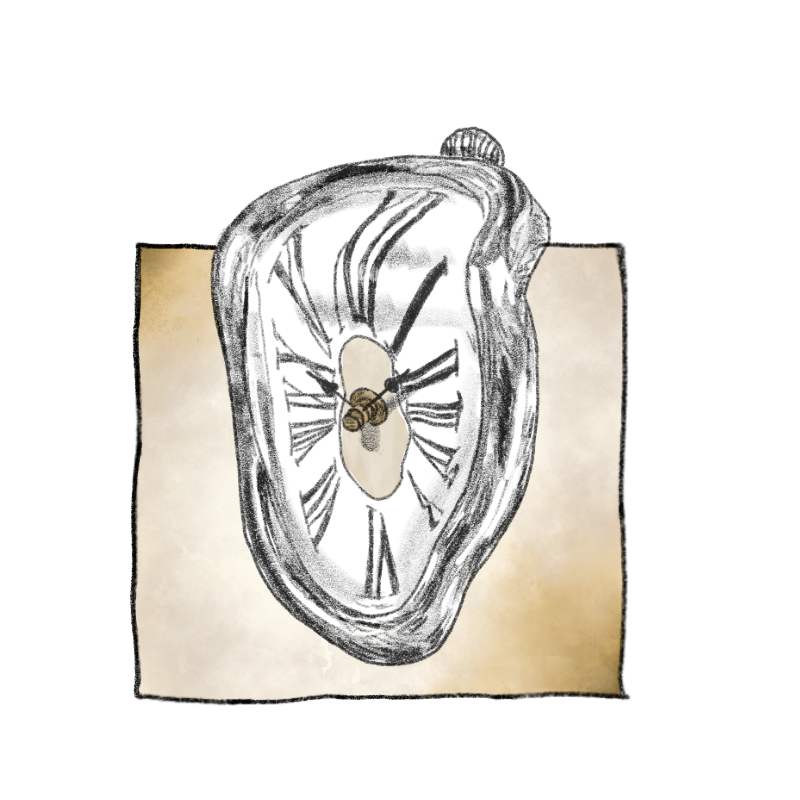}
\end{center}
\vspace{-40pt}

\begin{sectionauthor}
\normalsize{
\href{https://orcid.org/0000-0002-0745-9792}{C. Megan Urry}$^{1,2}$}

\vspace{15pt}

\scriptsize{
$^1$ Department of Physics, Yale University, New Haven, CT, USA\\
$^{2}$ Yale Center for Astronomy \& Astrophysics, Yale University, New Haven, CT, USA\\

}
\normalsize{}
\end{sectionauthor}
\vspace{20pt}

When I first entered the world of astronomy four decades ago, the dearth of women was very noticeable, though there were still more than in the physics classrooms I inhabited. For example, I was one of only four women among the 100+ graduate students in the Johns Hopkins Physics Department. 

Still, I didn’t think too much about it. For some reason, it didn’t occur to me there could be particular barriers for women entering the sciences. As a beneficiary of the second wave feminism of the 1960s and 70s, and as the child of endlessly encouraging parents, my expectations were sky high. So were my standards. I wasn’t sure I’d be able to succeed in my career plans but I didn’t think about anyone or anything external stopping me. If I’d thought harder about why there were so few women, I’d probably have hypothesized that women just weren’t very interested in physics and astronomy.

I was too na\"{i}ve to understand how wrong that was. 

With hindsight, I remember classmates who were interested in physics but dropped away for one reason or another---perhaps they thought they weren’t good enough (even though they were doing as well as the boys in the class), or they were attracted by other interests that seemed more compelling or attainable (see Eileen Pollack’s wonderful book, ``The Only Woman in the Room: Why Physics Is Still a Boys’ Club''). I didn’t think, then, about women being driven away by an environment that was unwelcoming at best, and hostile at worst. 

I was usually one of the stronger students in my physics classes. Still, people suggested I would get ahead mainly because of \textit{advantages} given to women. The same language persists today---we are routinely told that women, people of color, LGBTQ+, and other outsiders have risen to leadership positions only because they are ``DEI hires.''

Meanwhile, I felt like a fish out of water, or at least a red fish in a sea full of blue fish. That’s why simply hearing the name of a woman astronomer was enormously reassuring. At conferences, I would notice the few other women, each like a prize, an encouragement, an unspoken sign that what I was trying to do wasn’t crazy. I rarely talked to these women (especially those senior to me), but their existence meant the world to me. Anne Kinney, in her hilarious after-dinner speech\footnote{\url{https://www.stsci.edu/files/live/sites/www/files/home/events/event-assets/1992/_documents/1992-workshop-women-at-work-last-words-kinney.pdf}} at the 1992 Women in Astronomy meeting in Baltimore, described her discovery that Margaret Burbidge had been elected President of the AAS, and how hugely meaningful that was for her, despite knowing little to nothing about her. ``The name Margaret Burbidge was the light at the end of a very long, dark tunnel for me,'' she said, ``If I had never heard that name, I am not sure that I would be in the field today.'' Instead, Anne went on to a brilliant career working on the Hubble Space Telescope and at NASA and the NSF.

At some point, I started to think about all the women---and other outsiders---who weren’t able to stick it out, who lacked the support or resources that I had. I was, like Anne, so thirsty for role models. Weren’t we all? Far from having an easier path, we were running a tough race while handicapped with extra-heavy weights.

One particular conversation when I was a postdoc in the 1980s clarified both the ubiquity and inaccuracy of the upside-down notion that women had it easier than men. It started with the usual statement from a male colleague that, thanks to affirmative action, I would have no difficulty advancing in the field (a claim wildly contrary to the lack of encouragement I experienced to that point). I challenged him to substantiate that view. He launched into a story about a woman hired as faculty at a top university despite her complete lack of qualifications, and despite overwhelming competition from an outstanding young man for whom the job had actually been intended. But an interfering Dean had insisted that this woman be added to the short list and then insisted that she be hired. I might have believed this story---after all, such stories were commonplace---but when I asked who the woman was and what her research area was, the storyteller didn’t know any details. Wait, I said, you don’t know who she is or what she does but you are sure she was unqualified? ``Everyone knows this is true,'' he responded.

As scientists, we know this isn’t how evidence and scientific analysis is supposed to work. Before coming to conclusions, we seek facts and are skeptical of broad claims---you don’t just accept some story because it aligns with your beliefs. Later, I happened to meet someone who had been on the actual search committee for that position. When I recounted the story to him, he---a person who was there, who participated in the deliberations and the decision to hire this woman---told me the story was flat out wrong. In fact, the woman had been on the short list from the get-go and was hired because she was the strongest candidate by far. According to this first-hand account, she blew the rest of them out of the water, including the young man who was a supposed shoo-in.

This jolted me into a new awareness of the realities of my profession. I began to see that women\footnote{Forty years ago, I understood gender as a binary concept; it wasn’t until later that I appreciated its fluidity and range. I am grateful to the friends and family and colleagues who enlightened me.} were judged differently than men---indeed, much more harshly---which the social science literature confirms. We were less likely to be seen as academic stars, more likely to be criticized or overlooked. At the Space Telescope Science Institute (STScI), created in 1981, the tenure-track staff included only one woman (Neta Bahcall) among the first 60 people hired. This was despite the fact that women received 10-20\% of the PhDs in astronomy in the 1980s. I was the third woman hired onto the tenure track, after Anne Kinney. She and I started asking, ``Why so few?'' This prompted the Director, Riccardo Giacconi, to start asking his own questions, which led to STScI hiring Melissa McGrath, Stefi Baum, Laura Danly, Anuradha Koratkar, and many others, adding to amazing women on the operations staff, like Olivia Lupie, Vicki Balzano, and Pat Parker.

Still, numbers were not enough. All the women at STScI were underpaid relative to the men. No matter how hard we worked, how many technical contributions we made, how many papers we wrote, how much grant money we brought in, we were not afforded the opportunities given to our male counterparts, we were not nominated for internal or external recognition. The discrimination was fairly blatant.

In response, the women banded together, at all levels---tenure-track, scientist track, research assistants, technical staff. We started having monthly lunches. At first, much time was spent venting, but eventually, we started digging into what was going on. We read about statistics, unconscious bias, stereotype threat. With the strong support of STScI management, we organized the first-ever conference for women in astronomy, in 1992. The response---from women, at least---was overwhelming. We quickly filled to capacity (limited by the size of the STScI auditorium) and had to turn people away. We even had representatives from England, Russia, Italy, and Australia. Some male allies helped organize the meeting, attended it, and contributed a great deal, but at the same time, other male colleagues were less supportive. One famous professor at Harvard reportedly laughed when he received our invitation to the meeting---why, he scoffed, would he want to attend such a thing?

Of the 200 attendees, 150 were women\footnote{Conference photo at \url{https://www.stsci.edu/files/live/sites/www/files/home/events/event-assets/1992/_images/1992-workshop-women-at-work-group-photo.jpg}; identification of participants at \url{https://www.stsci.edu/files/live/sites/www/files/home/events/event-assets/1992/_documents/1992-workshop-women-at-work-photo-list.pdf}, using this numbered outline \url{https://www.stsci.edu/files/live/sites/www/files/home/events/event-assets/1992/_documents/1992-workshop-women-at-work-photo-outline.pdf}.}. None of us had ever been in the presence of so many women astronomers before. It induced in us a mental paradigm shift, analogous to how the famous 1968 Apollo 8 photo reframed humanity’s view of the Earth. We started to think differently about the lack of women on faculties, on speaker rosters, on lists of prize recipients. There wasn’t a dearth of talented women---they were simply under-recognized, not first to come to mind, ignored, passed over.

After the meeting, we organized the collective inputs from break-out sessions into the Baltimore Charter for Women in Astronomy\footnote{\url{https://www.stsci.edu/stsci/meetings/WiA/BaltoCharter.html}. The first draft was written by the wonderful feminist and author, the late Sheila Tobias, whose language made the document soar. She also suggested the very appropriate epigraph, ``Women hold up half the sky.''}. This two-page manifesto stated the problem, recommended some simple improvements such as greater transparency and affirmative approaches, and called on everyone in the profession to be part of the solution. Despite the modesty of the Charter language, only a few institutions agreed to endorse it. I’m not sure why the Charter seemed to induce resistance, including among some women; perhaps there was simply little urgency on the part of the majority of astronomers. Thanks largely to the advocacy of Debbie Elmegreen, the American Astronomical Society (AAS) did agree to endorse the ``goals of the Baltimore Charter.''

Nonetheless, we had turned a corner. The focus was no longer, What is the problem? It was: Time to make change.

We documented the statistics, explored relevant issues in the Committee on the Status of Women in Astronomy’s newsletter STATUS\footnote{\url{https://aas.org/comms/cswa/STATUS}} and the AASWOMEN listserv\footnote{\url{https://aas.org/comms/cswa/AASWOMEN}} (supplemented later by the wonderful Women in Astronomy blog\footnote{\url{https://womeninastronomy.blogspot.com/}}), invited experts to CSWA\footnote{The Committee on the Status of Women in Astronomy is one of the standing advocacy committees of the American Astronomical Society; home page \url{https://aas.org/comms/cswa}.} sessions at AAS meetings, put forward names for women speakers at conferences, and nominated women for prizes. (Huge thanks are due to CSWA Chairs and committee members, including those I worked with most closely, Pat Knezek and Joan Schmelz, for their dedicated efforts.) And things did begin to change faster. Networking among women and collective activism were the key.

Despite the 1992 meeting’s initial focus on women, we understood early on that the frame was far broader than gender. It was obvious that outsider status---which could follow from ``gender, gender identity or expression, race, color, national or ethnic origin, religion or religious belief, age, marital status, sexual orientation, disabilities, veteran status, or any other reason not related to scientific merit''\footnote{From the American Astronomical Society statement on ``Shirt-gate,'' \url{https://aas.org/posts/news/2014/11/aas-issues-statement-shirtgateshirtstorm}.}---conveyed an extra burden. The 2003 follow-on meeting in Pasadena explicitly addressed issues with regard to other minority groups\footnote{The Pasadena Recommendations can be found at \url{https://aas.org/comms/cswa/news/pasadenarecs}.}, as did the third meeting in 2009, in College Park, Maryland. By the time of the Inclusive Astronomy meeting in Nashville in 2015, the commonalities and differences among these groups were regularly discussed, and broadening participation to all talent had become the theme. The recommendations from these meetings got commensurately longer and more detailed. \textit{Picture an Astronomer} takes an important step beyond those earlier efforts.

From the 1990s to the 2000s to the present day, the number of women earning PhDs in astronomy increased steadily, from less than 20\% to more than 40\%\footnote{Statistics from the American Institute of Physics, \url{https://www.aip.org/statistics/percent-of-bachelors-degrees-and-doctorates-in-astronomy-earned-by-women-classes-1992-through-2022}, and from the American Physical Society, \url{https://www.aps.org/learning-center/statistics/diversity}.}. Interestingly, the percentage of women remains roughly double the percentage of women in physics, even though the two fields require the same knowledge base and skill set. The percentages also vary enormously from one country to the next. Even across western Europe, where standards of living and levels of education are quite similar, the percentage of women in astronomy varies widely\footnote{Regina Jorgenson, who attended the 1992 Women in Astronomy meeting, received a Watson Fellowship to study differences in participation of women in astronomy in Europe. As described in the June 2000 issue of STATUS (\url{https://aas.org/comms/cswa/STATUS}), historically Catholic countries (Italy, France, Spain) have far greater percentages of women astronomers than Protestant countries (Germany, Netherlands, United Kingdom).}. These facts indicate the participation of women is a cultural issue rather than a reflection of ability or interest.

So while ``women in astronomy'' is too narrow a frame---which \textit{Picture an Astronomer} recognizes beautifully---it does provide a handy metric for tracking change.

Today, our field is far more sophisticated in its understanding of these issues. Most colleagues are familiar with the concepts of unconscious bias, of excellence following from inclusion rather than being in tension with it, and of the need for active intervention to re-balance the playing field. Indeed, I would claim that Astronomy has led the way among the sciences. And though Chemistry and Biology graduate a higher percentage of women PhDs, Astronomy doesn’t have the fall-off-the-cliff profile of those fields after the postdoc years. We should be proud of what we as a community have accomplished---while still aspiring to do better.

As the demographics in our field have changed, the culture has changed. To give just a few examples, in the past decade the AAS held Town Halls on sexual harassment in astronomy\footnote{``Town Hall: Harassment in the Astronomical Sciences,'' 2016 January AAS meeting \#227, Session 116, 12:45-1:45, Osceola C, in Kissimmee, Florida.} and racism in astronomy\footnote{``Town Hall: Racism = Prejudice + Power: A Discussion of Racism in the Field of Astronomy,'' 2017 January AAS meeting \#229, Session 136, 4:30-5:30, Texas A, in Grapevine, Texas.}. These plenary discussions could never have happened at AAS meetings decades earlier, when mention of gender or discrimination was dismissed as a social issue not central to the business of astronomy, instead relegated to the “choir” who attended CSWA or CSMA\footnote{Committee on the Status of Minorities in Astronomy, one of the standing advocacy committees of the American Astronomical Society; home page: \url{https://aas.org/comms/csma}.} sessions. I was particularly struck by Caitlin Casey’s Pierce Prize lecture at the January 2019 AAS meeting in Seattle, Washington, where she not only highlighted her brilliant work on dust-obscured galaxies but also listed at the bottom of each slide some of the obstacles she encountered as a student and postdoc\footnote{\url{https://caitlinmcasey.github.io/plenary_jan2019.html}}. Every single one of those was something I too had personally experienced, a quarter century earlier. But most importantly, her talk changed the conversation: at subsequent meetings, many talks---by men as well as women---have referenced career obstacles and hardships, in a way that builds community and inspires those coming up behind. This would not have happened had our field not changed in fundamental ways.

Another benefit of changing demographics is that many more colleagues are engaged in efforts to expand participation in the field. That in turn means an ever-diminishing burden on individual activists. Just look at the rich content in the weekly AASWOMEN emails, contributed by dozens and dozens of readers---these days it’s hard to read it all! Yes, we still hear the same stories. Young women still experience the same things Caitlin Casey, I and too many others did. But numbers make a difference---they change attitudes---even if change is slow and difficult. You have to keep pushing. With every year, astronomy becomes more diverse, more exciting, more able to connect with the public, and more engaged in improving access.

At the January 2000 AAS meeting, I hosted a CSWA session centered around the recent MIT report on inequities among male and female faculty\footnote{The Chair of the MIT Senate that released A Study on on the Status of Women Faculty in Science at MIT (\url{https://web.mit.edu/fnl/women/women.html}), in 1999, was Prof. Lotte Bailyn, faculty member in the MIT Sloan School of Management (and mother of Yale astronomer Charles Bailyn), who was an invited speaker in the January 2000 CSWA session.}. One of the speakers, Prof. Claude Canizares (my former postdoctoral advisor), raised the question ``When will we know we have succeeded?'' I didn’t have an answer then, but he did, and it’s the right one: \textit{when we reach parity}. When a group attains representation in the inner sanctum in the same proportion as their presence in the talent pool, then you know the playing field is level\footnote{For example, if girls and boys are equally prepared in high school physics and math, why are only 20\% of undergraduate physics majors women? If women earn half the PhDs in the life sciences, why are they fewer than 20\% of newly hired assistant professors? If Black children are 15\% of high school students, why are they only a few percent of STEM majors or PhDs? To me, this differential attrition indicates a biased system.}. I personally think it’s basic justice that everyone have equal opportunity. Though some worry about lowering standards---infuriating, I know, and contrary to evidence---the larger the pool of talent, the higher standards will be. 

Parity of experience between the dominant population and outsiders is still elusive. There is still work to be done. \textit{Picture an Astronomer} reminds us that insufficiently recognized talent is still knocking at the door and that we have some distance to go before we can declare victory.

How will we get to parity? I look to history for some lessons. First, we have to play the long game. Persistence is essential. It can be tempting, when one has right on one’s side, to think that simply making a good argument should be enough. It should be, but it rarely is. One has to make the argument over and over, to reset norms, often incrementally, and to be vigilant against the inevitable backlash. It’s like driving some electric cars: when you take your foot off the accelerator, it feels like hitting the brakes in a conventional car---you can't rely on momentum to carry you forward. I remember how Vera Rubin---world-famous astronomer and role model---spent many years nominating dozens of younger women for honors and recognition. Even though few of those nominations succeeded at first, eventually some did, then more and more did. It didn’t stop Vera from trying again and again. When you sit on selection committees, you see how lopsided the discussions can be. If there is preferential selection, it mostly benefits the majority side---that is, the (usually white male) selectors elevate their (usually white male) proteg\'{e}s above outsiders.

Another important lesson of history is whether we should \textit{preach} or \textit{teach}. By ``preaching'' I mean saying what you believe, in the language you personally resonate with, while ``teaching'' is saying what you want your students to learn, in language they will hear. For changing minds---especially those that aren’t inclined to change---teaching works better than preaching, even if preaching may be more satisfying in the moment. Think about your audience, try to motivate them by addressing what they care about. For example, in the context of equity, there are many different reasons to favor increased participation in science. Whether or not you think excellence is the ``right'' motivation, it might for some audiences be the argument that moves the needle.

History teaches us that the time scales for social change are long---often 50 years or longer for cultural change within institutions like academia or government---basically, a human life span. These days it’s hard to imagine a time when women couldn’t vote, yet opinion pieces 150 years ago were full of ``good'' arguments against giving women the vote\footnote{The issue of women’s suffrage has reappeared in present-day politics, with rising interest among some conservatives in walking back women’s right to vote.}. It took more than 70 years from the Seneca Falls convention in 1848 to the passage of the 19th amendment to the constitution in 1920 (thanks to the single vote of a Tennessee legislator who changed his ``no'' to a ``yes'' after being implored to do so by his mother). The Equal Rights Amendment, which explicitly prohibits discrimination on the basis of sex, was introduced in 1923, re-introduced in 1971 and approved by supermajorities in both houses of Congress, but ratified by only 35 of the required 38 states. It probably wouldn’t pass Congress today.

Likewise, the battle to abolish slavery took more than 200 years. In the U.S., Rhode Island restricted slavery in 1652, followed more than a century later by Vermont (1777) and Pennsylvania (1780); the 13th amendment abolishing slavery passed in 1865. England outlawed slavery in 1833, France in 1848, Brazil in 1888, Mauritania in 1981. Change is always harder and slower than it should be.

Think of this struggle as a lifelong push. What you do will help, and no one thing you do will solve the problem right away. As the abolitionist minister Theodore Parker first said in 1853, and Martin Luther King rephrased, ``The arc of the moral universe is long but it bends toward justice.''

Let me be very clear: I am not saying that it is okay that change takes so long. Rather, I am saying we need to be not merely patient but relentless, to prevent inertia or backlash from defeating or further delaying that long arc toward justice.

It helps to have a clear vision for the future. Twenty years from now, approximately half the PhDs in astronomy and physics (and all other fields) should go to women, and the same percentage of new faculty hired should be women. Similarly, people of color, LGBTQ+ colleagues, military veterans, and religious minorities should not be disadvantaged by identity. Parity in career progression would be evidence that widespread discrimination is in the past. Twenty years from now, I believe, all astronomers will recognize the influence of expectations shaped by unequal treatment in the past and will be working to ensure equity.

Achieving this future won’t be easy. And it could be like special relativity: the closer you get to the speed of light, the more energy you need to accelerate.

Consider equity in society generally. It’s often seen as a zero-sum game: if one group gains, another group loses. The aim should be for people to do what they love and are good at. That is a world of increased productivity and expanding knowledge in which everyone prospers.

Back to the present moment: focus on a part of the problem you feel equipped to tackle. Many organizations work to get girls into science, engineering, computing, and mathematics. Join one, give an outreach talk, donate, write op-eds. Or you can influence inequity at one of the chokepoints. I used to run the Hubble Space Telescope proposal review, and thus heard hundreds of discussions of the relative merits of submitted proposals. The truth is the vast majority would have yielded amazing science but limits on observing time dictated that only a small fraction could be accepted. Neill Reid’s seminal study showed that HST proposals led by women were far less likely to be accepted than those led by men\footnote{Reid 2014, ``Gender-based Systematics in HST Proposal Selection,'' PASP, 126, 923 (\url{https://doi.org/10.1086/678964})}.

How are those tough decisions made? Successful proposals inevitably need a champion among the decision-makers---someone who will fight hard for a particular proposal. A classic paper by Madeleine Heilman based on equivalent resum\'{e}s showed that, when women are a small fraction of the pool, they rarely rise to the top of a ranked list, whereas when they reach the 30-40\% level, they are ranked at the top that corresponding fraction of time\footnote{Heilmann 1980, ``The impact of situational factors on personnel decisions concerning women: Varying the sex composition of the applicant pool,'' Organizational Behavior and Human Performance, Volume 26, Issue 3, 
December 1980, Pages 386-395 (\url{https://www.sciencedirect.com/science/article/abs/pii/0030507380900744}).}. Find a champion, be a champion. There are countless ways to make a difference. Choose yours.

\textit{Picture an Astronomer} goes beyond what has come before. The authors---dedicated and committed young scientists---have done what scientists should do: they examined the evidence, summarized the trends, and now they point the way forward. Their wisdom and attention to detail will help us all. Read this book.

\vspace{20pt}
\strut \hfill \textit{December 27, 2025; New Haven, CT, USA}
\pagebreak
\section{Introduction} \label{sec:intro}

\vspace{-30pt}
\begin{center}
    \includegraphics[width=0.37\linewidth]{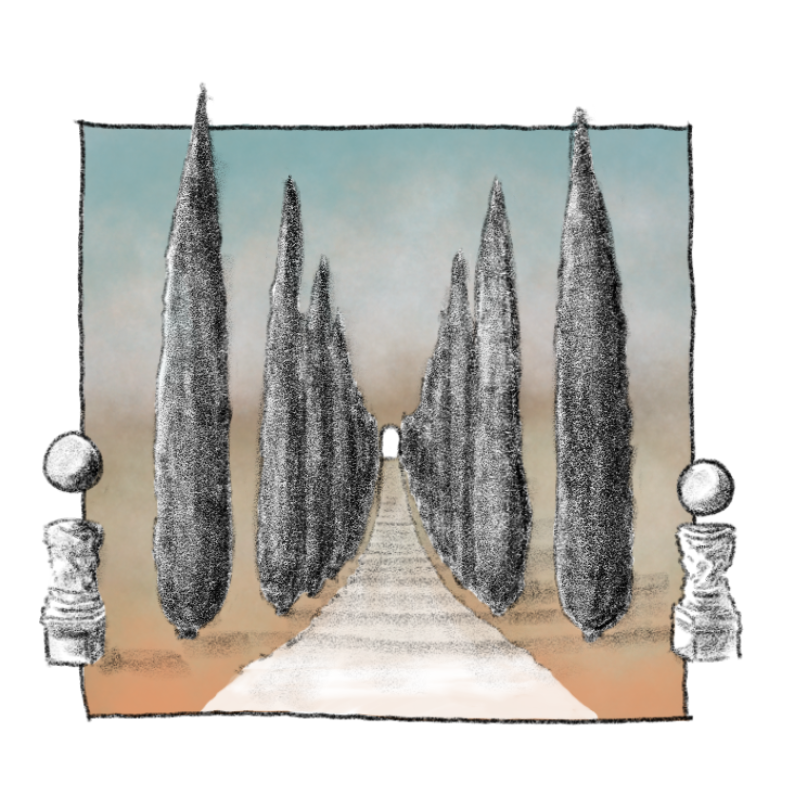}
\end{center}
\vspace{-30pt}

Women are consistently underrepresented in astrophysics yet are simultaneously subject to disproportionate attrition at every career stage. This disparity between demonstrated efficacy in job performance and ultimate career outcome was the primary motivation for the Picture an Astronomer\footnote{\url{https://pictureanastronomer.github.io}} series, which included both targeted public outreach to increase representation of women in astrophysics and high-level, solution-oriented discussions among professional astronomers. In March 2025, more than 200 astronomers came together in a hybrid-format symposium focused on the state of the field for female scientists, combining scientific exchange with discussions of policies and practices to strengthen retention of talent in the field. This white paper is the result of those discussions.

The Picture an Astronomer symposium was made possible by generous funding from the University of Chicago Women's Board, the Kavli Institute for Cosmological Physics, and the University of Chicago Department of Astronomy \& Astrophysics. The views expressed in this white paper, however, are those of the authors alone and do not necessarily reflect those of the sponsoring organizations.  Each section stems from small-group discussions during and after the symposium and is attributable only to its listed authors.
\pagebreak
\section{The Insidiousness of Bias and Importance of Perception} \label{sec:bias}

\vspace{-40pt}
\begin{center}
    \includegraphics[width=0.4\linewidth]{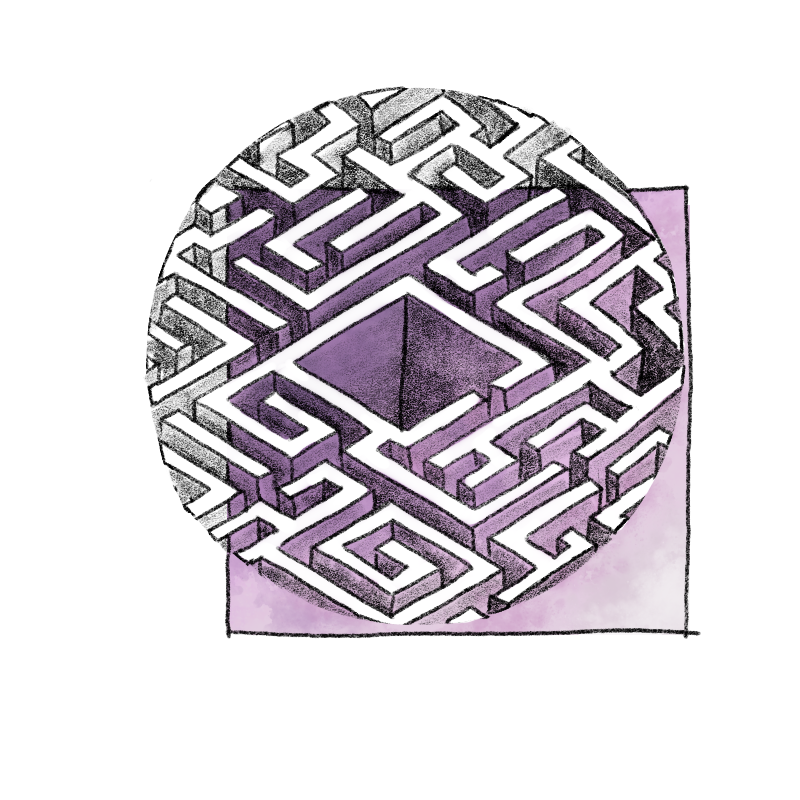}
\end{center}
\vspace{-40pt}

\begin{sectionauthor}
\normalsize{
\href{https://orcid.org/0009-0003-2076-6118}{Nathaniel Alden}$^{1,2}$, \href{https://orcid.org/0000-0002-1496-6514}{Lindsay R. House}$^{3,4}$, \href{https://orcid.org/0009-0006-1832-6713}{Olivia McAuley}$^{5}$, \href{https://orcid.org/0000-0002-5283-933X}{Ava Polzin}$^{1,6}$, \\\href{https://orcid.org/0000-0001-8813-4182}{Hsiao-Wen Chen}$^{1,6}$, and \href{https://orcid.org/0000-0002-2411-2766}{Daniel Glazer}$^6$}

\vspace{15pt}

\scriptsize{
$^1$ Kavli Institute for Cosmological Physics, University of Chicago, Chicago, IL, USA\\
$^{2}$ Department of Physics, University of Chicago, Chicago, IL, USA\\
$^{3}$ Data Science Institute, University of Chicago, Chicago, IL, USA\\
$^{4}$ NSF-Simons AI Institute for the Sky (SkAI), Chicago, IL, USA\\
$^{5}$ Department of Physics, Bryn Mawr College, Bryn Mawr, PA, USA\\
$^6$ Department of Astronomy \& Astrophysics, University of Chicago, Chicago, IL, USA\\

}
\normalsize{}
\end{sectionauthor}
\vspace{20pt}

\subsection{Introduction}
One great challenge of discussing the retention of any underrepresented group in science--we focus on women and their experiences here--is that much of what informs interactions, hiring decisions, who is seen as capable and who is simply \textit{seen}, ... is driven by societal and cultural beliefs. These biases, both implicit and explicit, are pervasive, notoriously hard to combat, and often unconscious\footnote{Implicit association tests offer an avenue to understand one's own biases: \url{https://implicit.harvard.edu/implicit/takeatest.html}.}. The often-quoted study by \citet{Goldin.Rouse.2000} highlights orchestra auditions as a controlled setting analogous to double-blind academic review, where the impact of bias can be directly assessed. The article is most famous for a footnote stating that a truly ``blind'' audition sometimes required women to remove high heels or involved carpeting to muffle shoe sounds--subtle cues that could otherwise reveal gender and undermine the intended anonymity. While this effect was not experimentally verified, the aside has become a powerful illustration of implicit bias and has helped frame curative interventions, recognizing that many biases that shape perceptions and decisions are, in fact, unconscious.
 
While awareness of these problems has helped to reduce the prevalence of overt sexism in academia \citep{overt_sexism},
the underlying biases themselves continue to manifest in subtle, but damaging ways. 
It is readily observed that women and minorities form a disproportionately small percentage of the academic workforce; in 2018, for example, there were roughly half as many graduate-level women in the physical sciences as graduate level men \citep{NSF_WMD_2021}.
However, the reasons for this are not so clear--while some propose it is due to bias towards women and minorities \citep{perception_of_womeninscience},
others propose that it is instead a byproduct of societal norms and expectations, which guide people's choice of a career path from a young age \citep{Ceci_life_factors}.

Though some cultural strides have been made, such as the increased prevalence of children's drawings of scientists that feature women over the last decades \citep{Miller.etal.2018}, societal norms and expectations remain entrenched. For instance, despite equivalent mathematical capacity \citep{Kersey2019}, girls report lower confidence in math \citep{Raabe.Block.2024} and less interest in mathematical careers than equally skilled boys \citep{Breda2023}. Their performance is further shaped by gendered social perceptions of ability \citep{Eble2022}. These results are especially notable given the perhaps surprising context that gendered stereotypes are weaker for math than for computer science, engineering, and especially physics \citep{Miller.etal.2024}, which is the most stereotyped. 

Persistent disparities point to the powerful role of early-emerging cultural beliefs about intellectual ability \citep{bian2025early}. In particular, the stereotype that boys and men are more likely to possess innate intellectual talent, or \textit{brilliance}, emerges cross-culturally as early as age six \citep{bian2017gender,shu2022gender,kim2024gender,okanda2022gender,zhao2022acquisition} and operates both explicitly and implicitly \citep{Napp.Breda.2022,storage2020adults}, steering girls away from activities framed for ``really, really smart'' children. Another entrenched belief is that success in certain fields, such as astrophysics, requires innate brilliance rather than effort \citep{Leslie.etal.2015,meyer2015women,hannak2023field,muradoglu2024culture}. As early as elementary school, students begin to believe that doing well in disciplines like math require being brilliant \citep{jenifer2024you}, contributing to gender achievement gaps \citep{cimpian2016have}. Ultimately, these biased beliefs demotivate high school girls from pursuing ``brilliance'' fields \citep{ito2018factors} and undermine women’s aspirations by lowering their self-efficacy and sense of belonging \citep{bian2018messages}.

Societal beliefs and stereotypes don't just impact self-perception and subsequent aspiration; they also govern professional behavior and evaluation. Women are, for instance, less likely to engage in self-promotion that can be critical for visibility and career advancement \citep{Peng2025} and use less promotional language on average in scientific papers \citep{Lerchenmuellerl6573}. At astronomical conferences, women are significantly less likely than men to ask questions \citep{Davenport.etal.2014,Pritchard.etal.2014,Schmidt.etal.2016}, though these results are not corrected for seniority, and the effect may reflect that women are, on average, more junior due to changing demographics and disproportionate attrition at each career stage.
In the classroom, where the cultural expectation that women are accommodating and nurturing conflicts with their role as an authority and expert, they are rated more harshly.  This bias emerges even when teaching is identical, as shown in online courses where students were presented with different gendered names for the same instructor \citep{MacNell2015}, and it persists for questions not specific to the instructor, such as those about university policies or course expectations, in otherwise identical courses \citep{Mitchell.Martin.2018}.  Women also receive more comments about their personality and appearance in both formal and informal evaluations and are more likely to be referred to as the ``teacher'' rather than the ``professor'' compared to their male counterparts \citep{Mitchell.Martin.2018}. 

Successful women are also subject to hostility and to having their accomplishments downplayed; anecdotally, many have heard the remark that they only received an award or job because they are a woman. Implicit in that message is that expectations were lowered, requirements relaxed, or that they were otherwise unfairly favored as a result of remedial efforts to close the gender gap. In reality, however, studies show that women must often `overperform' to be afforded the same opportunities, and there is no evidence for preferential hiring of women in professional astrophysics \citep[e.g.,][]{Perley.2019}. On the contrary, cross-country studies suggest that male applicants may be substantially preferred in some contexts \citep[e.g.,][]{10.1093/astrogeo/atu080, Berne.Hilaire.2020}.
 
A recent study by \citet{Moss-racusin} provides strong evidence that bias against women contributes to their under-representation in academic science, refuting the notion that male and female students are treated equally.
This study focused on aspiring undergraduate scientists with some research experience but not yet firmly established careers.
In the study, a total of 127 science faculty from several universities were asked to evaluate an identical application for a lab manager position, differing only in the student applicant's name: half saw `John', half saw `Jennifer'.  
Using common scales, faculty rated the applicant's competence, hireability and proposed salary and the level of mentoring they would provide, under the belief that their feedback would be conveyed directly to the student.

The authors of this study found strong evidence that the female student was viewed as less competent ($P<0.001$) and less hireable ($P<0.001$) than the male student, despite the fact that the applications were identical. The female student was also offered less mentoring ($P<0.001$) and a significantly lower starting salary ($P<0.01$) than the male student. Interestingly, this result was true of both female and male professors, without a significant difference due to faculty gender, with the authors reasoning that professors (both male and female) are influenced by cultural messages that women are less competent in science. A subsequent study conducted at the postdoctoral level echoed the finding that female candidates were rated both significantly less competent and significantly less hireable than a male candidate with an identical CV \citep{Eaton2020}. These results strongly suggest that subconscious bias towards women continues to negatively impact their ability to compete for research and faculty positions or smoothly carry out their work. 

Similarly, \citet{Holleran.etal.2011} use intermittently sampled, anonymized recordings to examine the daily experiences of faculty. After controlling for rank, research productivity and impact, discipline, and department/institution to isolate gender effects, they find that women, in conversations with men, are perceived as less competent when discussing research. They also find that both men and women are less likely to discuss research when speaking with a woman. As stereotypes come to bear, women's professional satisfaction is diminished, leading to greater disengagement. Combined with studies showing that women's disproportionate attrition from the professoriate stems more from workplace climate and belonging than from caregiving responsibilities or ``work-life balance'' \citep{Spoon.etal.2023}, these results suggest that both implicit and explicit bias and professional perception play a central role in sustaining the academic gender gap.

In astronomy, women receive $\sim10\%$ fewer citations than men when correcting for paper characteristics \citep{Caplar.etal.2017} and men prefer male coauthors \citep{Joyce.etal.2022}, excluding women from collaboration networks. Women are also less likely to be awarded telescope time in non-anonymized competitions where the principal investigator's gender is visible \citep{Reid.2014, Patat.2016, Lonsdale.etal.2016, Spekkens.etal.2018, Carpenter.2020, Johnson.Kirk.2020, Hunt.etal.2021}, with female-led proposals being graded more harshly than comparable male-led ones. This bias persists regardless of the gender composition of the review panels \citep{Lonsdale.etal.2016}, though interventions such as increasing women's representation and openly acknowledging historic bias helped equalize the outcomes in \citet{Hunt.etal.2021}. Under dual-anonymous review, by contrast, women fare as well as men and do not receive systematically lower grades \citep{Johnson.Kirk.2020, Carpenter.etal.2022}. %Interestingly, women were even less likely to receive \textit{Hubble Space Telescope} time from more senior time allocation panels when the gender of the principle investigator was known or could be inferred \citep{Reid.2014}, indicating that entrenched biases and beliefs play a significant role. 

The pervasive nature of cultural messages about women's roles, competence, and capabilities suggests that combating bias requires broad engagement, with community-level interventions essential to addressing the persistent disparity in treatment.  In the remainder of this section we discuss recommendations for tackling anti-woman bias in science.

\subsection{Key Issues}

The results from this paper are gathered from the discussions during the Picture an Astronomer conference. In particular this discussion was centered around gender bias within astronomy. Gender bias remains a significant challenge \citep[see][for a short review]{Llorens2021}, and through this discussion and our recommendations, we aim to raise awareness and encourage collective efforts to create greater equity, inclusion, and fair opportunities for women in science. We highlight some key themes here.

Effective mentorship is crucial for the success of women in astronomy, yet current mentorship structures often fail to provide adequate guidance and advocacy. The lack of institutionalized mentorship programs and limited mentor training can leave women isolated in a male-dominated field, while the absence of formal mechanisms to address implicit bias in mentorship further compounds these challenges.

Gender bias in hiring and promotion remains widespread, with women often subject to unequal standards and subjective evaluations. One example is the use of gendered language in recommendation letters, where fewer standout descriptors are applied to female applicants \citep{Schmader2007, Dutt2016, Eberhardt.etal.2023}. Awareness of this bias may prompt some linguistic and evaluative self-correction \citep{Schmader2007}, yet women are still disproportionately praised for `soft-skills', such as science communication, team management, and event planning, at the expense of recognition for research excellence. They may also face gendered expectations, including being undermined or presumed less competent\footnote{\citet{doi:10.1126/sciadv.aba7814} found that the gender gap in perceived competency and offered compensation is driven by those who do not believe that gender bias is a problem in a particular field, reinforcing that awareness of bias is critical to mitigating it.} \citep{Leslie.etal.2015,Napp.Breda.2022}, which further impedes career progression.

Women are often expected to bear a disproportionate share of service tasks within academic departments \citep{Winslow.2010, Guarino2017, Jarvinen.Mik-Meyer.2025}. Explanations include gendered socialization toward agreeableness and collective responsibility \citep[e.g.,][]{Jarvinen.Mik-Meyer.2025}, perceptions that women should assume \textit{caregiving} roles in the academy \citep[e.g.,][]{Hanasono.etal.2019}, or the cultural assumption that they are more adept at conducting service \citep[e.g.,][]{Jarvinen.Mik-Meyer.2025}--and the simple fact that they are simply \textit{asked more often} \citep[e.g.,][]{OMeara.etal.2017}. These dynamics extend to informal contexts: students frequently approach female faculty with personal issues due to presumed ``mothering'' instincts \citep[e.g.,][]{Hanasono.etal.2019}. This unequal distribution of labor detracts from women's research productivity and professional advancement, reinforcing gender stereotypes about their roles in the academy. Even well-intentioned attempts at representational balance can  
in departments with few women, requiring their presence on every committee leaves them disproportionately overburdened with service not expected of men.  This dynamic extends to all underrepresented groups, underscoring the need for structural solutions. 

The pervasiveness of impostor syndrome among women in academia \citep{Muradoglu.etal.2022,PRICE2024100155} is exacerbated for fields where brilliance or natural talent is perceived as the key to success. This is further amplified in the academy, where this entrenched ``weed-out'' mentality further marginalizes those who do not conform to traditional norms of achievement. The up-or-out nature of academia, coupled with biased expectations of innate ability and uneven institutional support (or barriers such as disproportionate service demands), contributes to higher levels of stress and burnout, which, in turn, fuel attrition.

Though scientists strive for objectivity, science is ultimately done by humans who are equally susceptible to unconscious bias--and to acting on those biases, intentionally or not. Despite widespread acknowledgment, there is often no accountability or repercussion for those who perpetuate inequities or deny fair opportunities. This lack of oversight normalizes harmful behaviors and ensures its continuation in the absence of formal structures to prevent it.

\subsection{Recommendations}
Our group's discussions focused primarily on hiring practices, mentoring, service responsibilities, and student/researcher confidence. While some concrete recommendations are outlined below, the nebulous, persistent, and often unconscious nature of gender bias means the most effective interventions combine structural changes with cultural shifts, ensuring that bias is addressed both in formal processes and in everyday departmental life. For all recommendations, we emphasize the responsibility of leadership and senior-level faculty in fostering a fair environment and equitable departmental culture. 
\begin{itemize}
    \item \textbf{Hiring Practices}
        \begin{itemize}
            \item Leadership must be involved in trainings to help address bias in the hiring process within the department. While awareness of gender bias can \textit{begin} to improve fairness in evaluations \citep[e.g.,][]{Hunt.etal.2021}, the implicit and unconscious nature of bias is insidious and unlikely to be resolved by awareness campaigns alone. In-person trainings are generally more effective than online ones, as they encourage active participation.
            \item Standardized rubrics should be created and consistently enforced to ensure applicants are evaluated against the same criteria, reducing opportunities for bias in admissions and hiring.
            \item Adopt a phased hiring process that begins by first evaluating candidates' actual work, delaying consideration of recommendation letters, which often contain biased language.
            \item Incorporate bystander training to  encourage faculty and staff to intervene at the moment of the explicit bias, fostering a culture of accountability and allyship so individuals can respond at the moment of transgression.
            \end{itemize} 
        \item \textbf{Mentoring}
            \begin{itemize}
                \item Implementing yearly feedback or check-in forms can help align expectations between advisors and students while providing space for mutual feedback. Institutionalized mentorship policies further address implicit bias in advisor-trainee relationships and reduce variability across groups in the quality and frequency of mentorship.
                \item Students should have opportunities to practice written and oral communication both within and beyond the department, including talks and conference presentations, informal networking, collaborative writing, and the production of first-author publications.
                \item Women disproportionately shoulder service, teaching, and outreach responsibilities, yet these soft skills are often undervalued. Departments should provide opportunities and incentives for all students to practice teaching pedagogy through classroom experience and outreach events,  helping to close the gender gap in non-research academic work.
                \item Faculty should be strongly encouraged to attend mentoring and teaching workshops, such as those supported by the American Institute of Physics via the TEAM-UP report \citep{Teamup20} on academic support.
                \item Departments can proactively combat impostor syndrome by fostering an encouraging, safe environment where students feel comfortable asking questions, whether through closed sessions with guest speakers or facilitating anonymous question-asking.
                \item Community members should be equipped to identify and mitigate instances of bias. Departments can, for example, facilitate or encourage attendance at workshops on spotting and addressing gender bias in oral and written communication.
                \item The community must actively guard against bias toward nontraditional students who have experienced career gaps. Efforts may include avoiding evaluation metrics such as `time to degree' or career interruptions in hiring and awards, creating flexible leave policies, and establishing clear norms around research ownership during and after extended breaks.
            \end{itemize}
        \pagebreak
        \item \textbf{Building Student Confidence}
            \begin{itemize}
                \item Groups and departments should hold closed sessions that allow students to engage with guest speakers in discussions without senior faculty present.
                \item Individuals must challenge the assumption that avoiding jargon signifies disinterest or lack of credibility. In reality, clearly communicating complex topics demonstrates deeper understanding and expertise. Students should practice distilling their work and conveying ideas accessibly to an informed but non-expert audience.
                \item Promoting avenues for students to ask questions anonymously in classrooms and seminars may help close the gender gap in participation and ensure broader engagement.
                \item Across courses and graduate programs, the `weed out' mentality can be dismantled by clearly articulating expectations and emphasizing effort and growth over presumed natural ability.
            \end{itemize} 
    \end{itemize}

\begin{itemize}
    \item \textbf{Service Responsibilities}
    \begin{itemize}
        \item Values statements or codes of conduct must be consistently upheld and demonstrated by leadership at the group, department, collaboration, or institution levels. 
        \item Departments and institutions must ensure equitable distribution of service work, so that no faculty member is disproportionately burdened. Policies should recognize that, while well-intentioned, requiring \textit{representative} committees often removes choice in, or reprieve from, service work from underrepresented community members, while majority members are not similarly affected.
        \item Weaponized incompetence can be curtailed by combining voluntary choice with accountability: faculty may select committees, with preferred tasks assigned to early volunteers, but baseline service loads must still be tracked, rotated, and equitably distributed to prevent avoidance or overburdening.
        %Weaponized incompetence can be curtailed by allowing faculty to select their committees, with preferred tasks assigned to those who volunteer first.
        \item  Departments, collaborations, universities, and other institutions should maintain transparency around service responsibilities and guard against `backdoor deals' that trade or offload service duties.
    \end{itemize} 
\end{itemize}

\bibliography{references/bias}{}
\bibliographystyle{aasjournal}
\pagebreak
\section{Improving Belonging in Physics: The Role of
Representation and Stereotype Threat} \label{sec:representation}

\vspace{-40pt}
\begin{center}
    \includegraphics[width=0.4\linewidth]{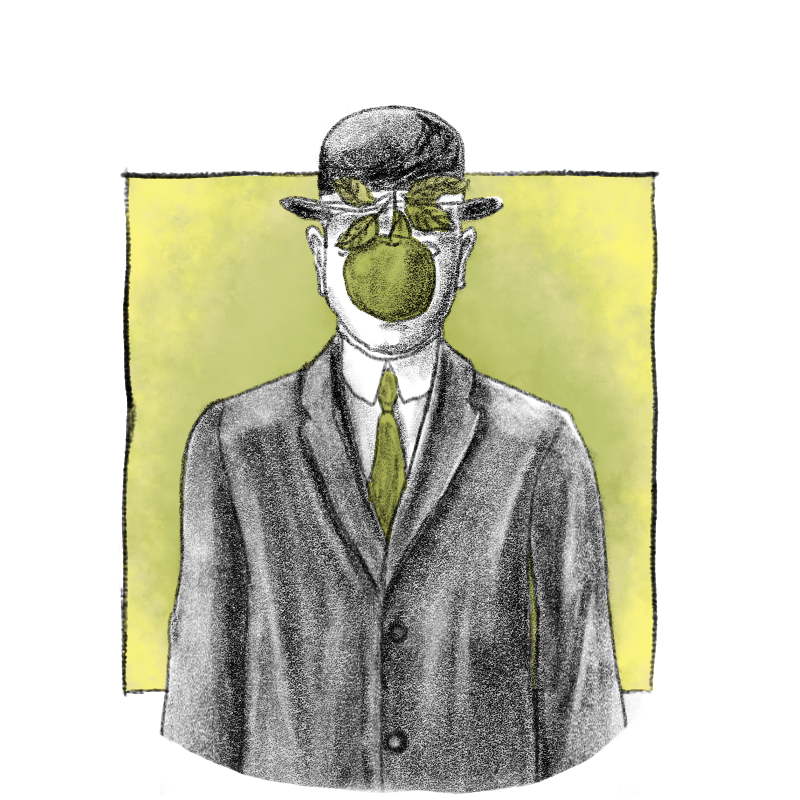}
\end{center}
\vspace{-30pt}

\begin{sectionauthor}
\normalsize{Ann-Marsha Alexis$^{1}$,
\href{https://orcid.org/0000-0002-5726-5216}{Andrea Gokus}$^{2}$, \href{https://orcid.org/0009-0002-6065-3292}{Lou Baya Ould Rouis}$^3$, \href{https://orcid.org/0009-0000-5561-9116}{Haile M. L. Perkins}$^4$, \href{https://orcid.org/0000-0002-5104-5263}{Pazit Rabinowitz}$^5$, Joseph Guzman$^{6}$, \href{https://orcid.org/0000-0003-3217-5967}{Sarah R. Loebman},$^{7}$ and \href{https://orcid.org/0000-0002-4186-6164}{Amanda Wasserman}$^{4,8}$\\}
\vspace{15pt}
\scriptsize{
$^1$ Department of Physics, Carnegie Mellon University, Pittsburgh, PA, USA\\
$^{2}$ Department of Physics, Washington University in St. Louis, St. Louis, MO, USA\\
$^3$ Department of Astronomy, Boston University, Boston, MA, USA\\
$^4$ Department of Astronomy, University of Illinois Urbana-Champaign, Urbana, IL, USA\\
$^5$ Department of Physics, Washington University in St. Louis, St. Louis, MO, USA\\
$^{6}$ Chicago Astronomer, chicagoastronomer.com, Chicago, IL, USA\\
$^{7}$ Department of Physics, University of California, Merced, Merced, CA, USA\\
$^{8}$ NSF-Simons AI Institute for the Sky (SkAI), Chicago, IL, USA\\
}
\normalsize{}
\end{sectionauthor}
\vspace{20pt}

\subsection{Introduction}
Physics, and by extension astronomy, is often perceived as a field where innate brilliance, rather than dedication or effort, is essential for success. This perception strongly shapes ideas about who belongs in the field.
Research shows that disciplines emphasizing raw intellectual talent tend to have the lowest representation of women, with physics being among the starkest examples \citep{leslie2015}. 
These perceptions are reinforced by persistent stereotypes about gender and intelligence and by the scarcity of visible role models from underrepresented groups. Together, these factors contribute to an environment where many--women, people of color, people with disabilities, LGBTQ+ individuals, and those from marginalized socioeconomic backgrounds--feel they do not belong.
In this chapter, we examine two interlinked consequences of this dynamic in physics and astronomy: underrepresentation and stereotype threat. In addition, we outline approaches to counter these effects and improve the retention of women across career stages.

\subsection{Lack of Representation} \label{sec:rep} % Andrea
Despite decades of efforts to attract more women into scientific careers, including astronomy, the field remains far from gender parity.
The number of women obtaining astronomy degrees has steadily increased since the American Astronomical Society's Committee on the Status of Women in Astronomy (CSWA) began conducting surveys in 1992, the same year C. Megan Urry organized a landmark series of conferences on the status of women in astronomy, leading to the publication of the `Baltimore Charter'\footnote{\url{https://www.stsci.edu/stsci/meetings/WiA/BaltoCharter.html}}.
Yet data from the 2013 CSWA survey showed that while the number of women in the field was rising, their proportional representation still declined at each successive career stage \citep{status2014jan}.

Persistent underrepresentation of women, people of color, and other marginalized groups is not simply a `pipeline' issue; it reflects deep-rooted structural and cultural barriers that determine who enters, stays, and succeeds in the field.
Many studies have sought to understand the obstacles that drive women away from academic science.
In this section, we focus on one major contributor: lack of representation.

Role models--individuals who share aspects of identity such as gender, race, upbringing, or other intersecting factors--are consistently shown to be critical for persistence \citep[e.g.,][]{Danielsson2012, keblbeck2024, shachnai2022walking}.
Yet for students from underrepresented backgrounds, especially those with intersecting marginalized identities, such identity-affirming environments remain rare.
External validation also reinforces women’s persistence, bolstering motivation and academic identity \citep{Nehmeh2021}.
Undermining retention and representation further, women are often given less credit for their scientific contributions compared to men \citep{ross2022women}.

Increasing representation requires long-term investment in recruitment and retention, but progress also depends on creating inclusive, supportive environments.
For example, secondary school teachers play a crucial role when they are active in addressing gender gaps in physics \citep{Masri2025}.

One effective approach is creating \textit{counterspaces}: environments where underrepresented individuals can gather, share experiences, and build community.
In \citet{hazari2024}, recurring undergraduate-focused conferences such as the Conference for Undergraduate Women in Physics (CUWiP) and Women in Physics Groups (WiPG) were shown to foster belonging, though strong intrinsic interest in physics remained essential for long-term resilience. 
Similarly, the \textit{Picture an Astronomer} conference served as a counterspace, welcoming participants across all career stages and genders. Anonyomous survey responses collected by the authors of this section highlighted its impact on fostering a sense of belonging, on emphasizing the value of shared experience, and on intersectionality: 

\begin{mdframed}[
  leftline=true,
  rightline=false,
  topline=false,
  bottomline=false,
  linecolor=gray,
  linewidth=2pt,
  innerleftmargin=10pt,
  innerrightmargin=5pt,
  innertopmargin=6pt,
  innerbottommargin=6pt
]
\textit{ ``Seeing so many other women at the symposium increased my sense of belonging. For once, I didn't feel like the outlier in a room full of men.''}
\end{mdframed}

\begin{mdframed}[
  leftline=true,
  rightline=false,
  topline=false,
  bottomline=false,
  linecolor=gray,
  linewidth=2pt,
  innerleftmargin=10pt,
  innerrightmargin=5pt,
  innertopmargin=6pt,
  innerbottommargin=6pt
]
\textit{``I felt more included in the field when hearing from folks who achieved brilliant careers in astronomy that they have had similar difficult experiences as me.'' }
\end{mdframed}

\begin{mdframed}[
  leftline=true,
  rightline=false,
  topline=false,
  bottomline=false,
  linecolor=gray,
  linewidth=2pt,
  innerleftmargin=10pt,
  innerrightmargin=5pt,
  innertopmargin=6pt,
  innerbottommargin=6pt
]
\textit{``Connecting with more brown women in physics and astronomy that are further along the academic track than I am strengthened my sense of belonging.'' } 
\end{mdframed}

Mentorship also emerged as a key factor: whether through a PI’s support, a postdoc advocating for a graduate student, or a high school teacher encouraging a career in science,  effective mentorship can sustain persistence.
The lack of representation can drive women from the field, as they navigate conflicting expectations--balancing societal notions of femininity with the dominant norms of the physics community \citep{Danielsson2012}. One participant reflected:
\begin{mdframed}[
  leftline=true,
  rightline=false,
  topline=false,
  bottomline=false,
  linecolor=gray,
  linewidth=2pt,
  innerleftmargin=10pt,
  innerrightmargin=5pt,
  innertopmargin=6pt,
  innerbottommargin=6pt
]
\textit{``I have considered leaving academia because I'm not [X] enough, where X is a quality of the over-represented groups.'' }
\end{mdframed}

A non-binary participant added: 

\begin{mdframed}[
  leftline=true,
  rightline=false,
  topline=false,
  bottomline=false,
  linecolor=gray,
  linewidth=2pt,
  innerleftmargin=10pt,
  innerrightmargin=5pt,
  innertopmargin=6pt,
  innerbottommargin=6pt
]
\textit{``Many of the statistics don't really include people like me, or group us with women, which, while understandable, can add to the sense of not belonging unless feminine--basically, not masculine enough to belong in the male-dominated, but not feminine enough to belong in the groups carved out for that.''}
\end{mdframed}

\subsection{Stereotype Threat}
Stereotype threat, which was first introduced by \citet{steele1995}, describes the fear of confirming a negative stereotype associated with a minority group to which one belongs, which can lead to underperformance.

Most studies on stereotype threat focus on pedagogical contexts, where the classroom serves as a controlled environment for understanding why students from racial, ethnic, or gender minorities may perform worse in tests \citep[e.g.,][]{steele1997,SPENCER1999,aronson2002,walton2007,walton2009, marchand2013}.
However, there have also been landmark studies examining the detrimental effect of stereotype threat at more advanced career stages. 
For example, \citet{Holleran.etal.2011} examined workplace conversations among STEM faculty and found that women face unique challenges, as they are often judged as less competent than men in research discussions.
For women of color, these challenges can be compounded by the effects of intersectionality \citep[e.g.,][see also Section \ref{sec:intersectional} for a more in-depth discussion]{clancy2017}.

A recent study by \citet{dehkordi2024} analyzed women's performance in sports under different conditions and concluded that while stereotype threat can impair performance, targeted interventions, such as communicating  counter-arguments to the stereotype, can help mitigate this effect.
In other contexts, especially in the workplace, counteracting stereotype threat is less straightforward and can hinder women's advancement to the highest leadership positions \citep{hoyt2016}. Even in fields where gender parity has nearly been achieved, such as internal medicine, women remain vulnerable to the negative gender stereotypes \citep{frank2024}.

In the field of physics, disparities in attainment between men and women are well documented \citep[e.g.,][]{Tai2001,mattern2002} and have been attributed to a variety of factors, including motivation, interest, and support from parents or teachers \citep[e.g.,][]{enman2000,she2001,tenenbaum2003,greene2004}. \cite{kost2009characterizing} found that roughly 70\% of the observed gender gap in introductory physics course grades can be explained by prior physics and mathematics preparation (e.g., having taken a high school physics course), as well as students' incoming attitudes and beliefs about their own competence in physics. These beliefs often interact with psychological factors such as stereotype threat.
A study by \cite{marchand2013} showed that stereotype threat affects women's performance regardless of whether the stereotype is voiced explicitly or remains implicit. However, \emph{nullifying} the stereotype, that is, by stating that men and women perform equally well on the given test, resulted in equivalent performance across genders.
Importantly, women in environments where they are severely underrepresented, such as introductory physics classrooms, and who endorse negative gender stereotypes, also underperform relative to their female peers who do not hold such beliefs \citep{maries2018}.

In the anonymous survey distributed by the authors of this section following the \textit{Picture an Astronomer} conference, attendees shared their experiences with stereotype threat. One participant described the tension she feels as a leader:

\begin{mdframed}[
  leftline=true,
  rightline=false,
  topline=false,
  bottomline=false,
  linecolor=gray,
  linewidth=2pt,
  innerleftmargin=10pt,
  innerrightmargin=5pt,
  innertopmargin=6pt,
  innerbottommargin=6pt
]
\textit{``Because I am very outspoken and driven to leadership, I am aware of and concerned about conforming to the `women are seen as bossy in professional situations' stereotype.''}
\end{mdframed} 

Given that women are underrepresented in physics, and women of color even more so, another participant remarked: 

\begin{mdframed}[
  leftline=true,
  rightline=false,
  topline=false,
  bottomline=false,
  linecolor=gray,
  linewidth=2pt,
  innerleftmargin=10pt,
  innerrightmargin=5pt,
  innertopmargin=6pt,
  innerbottommargin=6pt
]
\textit{``I've often been the only woman in the room, sometimes the only person of color, sometimes both. I feel pressure to both make my voice heard because I represent my
gender and my color, but also to make my voice heard while saying something smart. If I don't speak up, no female voice would be heard throughout the meeting.''}
\end{mdframed}

\subsection{Reshaping our communities for a better sense of belonging}

In the above sections, we listed out barriers to women's full participation in physics and astronomy. In the following paragraphs, we will make some suggestions that can be implemented at varying career stages to increase women's sense of belonging, a key factor influencing their interest in brilliance-oriented fields \citep{bian2018messages}. A greater sense of belonging decreases anxiety, which is a distraction from learning and work. Academic belonging refers to the feeling of being accepted and valued as a member of the given organizational space \citep{PhysRevPhysEducRes.12.020110}. A sense of belonging, which is an intrinsic human need, has been shown to predict the success and retention of female students and researchers. Bauer and colleagues’ Brilliance–Belonging Model (\citeyear{bauer2025brilliance}) highlights that some fields are perceived as requiring innate brilliance.  In these contexts, social stereotypes about who is believed to possess such ability undermine women’s and minoritized individuals’ sense of belonging through both self-doubt and others’ doubts about their fit. This dynamic creates a self-reinforcing feedback cycle of reduced motivation, lower self-efficacy, and underrepresentation.

As discussed in Section \ref{sec:rep}, retention, recruitment, and performance are also affected by the representation of women present at different organization levels. By addressing the lack of representation in STEM, it is possible to contribute to an increased sense of belonging among women.  At the classroom level, for example, studies have found that female students perform better on math tests when they are in majority-female settings compared to a majority-male settings \citep{promisingpractices}. Similarly, female students report lower interest and anticipate a reduced sense of belonging when attending conferences with a male majority, relative to conferences with greater gender balance \citep{promisingpractices}. However, it is important to note that increasing the number of women in physics and astronomy is neither the only nor the most reliable way to foster women's sense of belonging.
    
We suggest that \textbf{astronomy and physics educators emphasize a \emph{growth mindset} when approaching learning}. In this framework, learning outcomes and levels of success are not attributed to fixed ability, but rather to effort, practice, and persistence. One author recalls her undergraduate program, in which faculty fundamentally stressed that ``physics is a training''--a discipline centered on learning how to think about and solve problems. This framing made a dramatic difference in how students approached coursework, as they came to understand physics as a framework for reasoning rather than as something that should be innately intuitive. As a result, students developed a stronger sense of belonging, believing that success depended on their effort rather than on fixed attributes they may not have seen themselves as possessing. 
     
Consistent with this perspective, \citet{promisingpractices} note that adopting a growth mindset has been shown to improve students' grades across academic levels, encouraging students to value effort and learning more broadly. Concrete actions educators can take include: 1) praising effort and progress over innate talent, 2) being transparent about challenges and obstacles encountered in their own careers, as well as in those faced by well-known scientists\footnote{\url{https://growingupinscience.web.app/\#about}}, and 3) adopting grading practices that value improvement, such as allowing revisions and mastery-based assessment, rather than relying solely on test performance \citep{PhysRevPhysEducRes.12.020110, muradoglu2024culture}. 
     
On the last point, the author recalls that mastery-based learning and test revisions (with partial point recovery) were integral to her undergraduate program and proved highly motivating in her pursuit of a physics degree. Another intervention highlighted by \citet{promisingpractices} involves introducing dedicated workshops on the malleability of the brain and the flexible nature of intelligence. Such interventions have been shown to increase student motivation at the middle school level, but are likely applicable across a broader range of educational contexts.

\textbf{Active learning is a well-established strategy for increasing women's participation in astronomy and physics}. Research has shown that active learning decreases achievement gaps between students from disadvantaged backgrounds and their more advantaged peers \citep{haak2011increased}. One effective active-learning approach is peer-led team learning (PLTL), in which students work collaboratively to solve problems under the guidance of a peer mentor (typically a student with prior experience in the course). PLTL has been shown to increase women's persistence in STEM fields. 
     
More broadly, students report greater interest and engagement when learning in group-based settings. Working together in small groups also fosters social connection and peer support, which in turn strengthens women's sense of belonging in physics and astronomy learning environments \citep{promisingpractices}.  
     
\textbf{Social support plays an important role in retaining women in physics}. Although educators cannot directly influence students' social lives outside the classroom, they can facilitate the affirmation of students' value systems \citep{PhysRevPhysEducRes.12.020110}. Educators can also help students connect seemingly abstract physics concepts to concrete societal benefits or to their own daily lives \citep{PhysRevPhysEducRes.12.020110} and encourage reflection on why specific research questions are pursued and how they serve society. For some women, working with a faculty mentor who adopts a more communal approach may be particularly beneficial. Such approaches can increase participation and interest in STEM more broadly, suggesting that students who value helping others may be especially likely to persist in these fields. 

Alongside peer support, \textbf{mentorship during college and beyond is a strong predictor of women's success in physics and related fields} and is especially important for women of color, who remain even more underrepresented. The authors of \citet{promisingpractices} stress that effective mentorship depends less on matching mentors and mentees by race, gender, or LGBTQI+ identity--attributes for which mentors may be scarce--and more on pairing students with culturally responsive mentors whose values and approaches align with students' needs.
In addition, exposure to a female scientists, both through media representation and direct interaction, increases the likelihood that women will identify themselves as scientists \citep{promisingpractices}.

 Finally, \textbf{allies, particularly male allies, can contribute to improving women's sense of belonging} by showing interest in women's success, promoting less individualistic and masculine-centric academic cultures, and taking concrete actions in support of gender equity. However, further research is needed to better understand which behaviors and structures are most effective in fostering meaningful allyship.

\bibliography{references/representation}{}
\bibliographystyle{aasjournal}
\pagebreak
\section{Diminished Resources, Differential Praise, and
Unequal Workloads} \label{sec:resources}

\vspace{-40pt}
\begin{center}
    \includegraphics[width=0.4\linewidth]{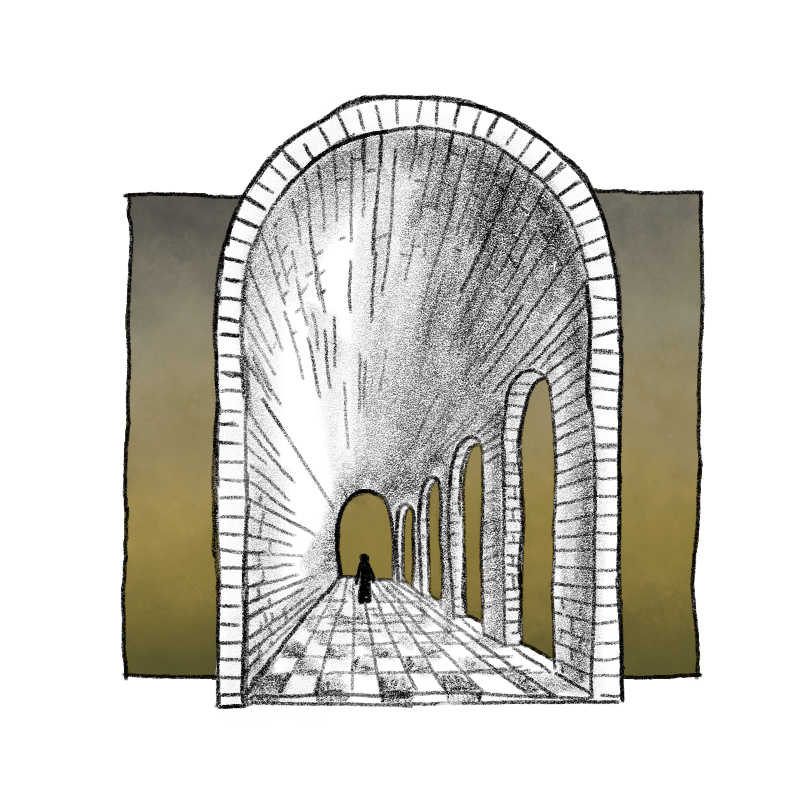}
\end{center}
\vspace{-40pt}

\begin{sectionauthor}
     \href{https://orcid.org/0000-0003-2404-2427}{Madison Brady}$^{1}$, \href{https://orcid.org/0000-0002-7155-679X}{Anirudh Chiti}$^2$, \href{https://orcid.org/0000-0002-5283-933X}{Ava Polzin}$^{3,4}$, \href{https://orcid.org/0000-0001-8023-4912}{Huei Sears}$^{5}$, \href{https://orcid.org/0000-0002-7759-0585}{Tony Wong}$^6$, \\\href{https://orcid.org/0000-0002-1819-0215}{Shanika Galaudage}$^{7,8}$, \href{https://orcid.org/0000-0003-3430-3889}{Jessica Speedie}$^{9}$, and \href{https://orcid.org/0000-0001-6746-9936}{Tanya Urrutia}$^{10}$

\vspace{15pt}
\scriptsize{

$^{1}$ Department of Physics and Astronomy, Michigan State University, East Lansing, MI, USA\\
$^2$ Kavli Institute for Particle Astrophysics \& Cosmology, Stanford University, Stanford, CA, USA\\
$^3$ Department of Astronomy \& Astrophysics, University of Chicago, Chicago, IL, USA\\
$^4$ Kavli Institute for Cosmological Physics, University of Chicago, Chicago, IL, USA\\
$^5$ Department of Physics and Astronomy, Rutgers, the State University of New Jersey, Piscataway, NJ, USA\\
$^6$ Department of Astronomy, University of Illinois Urbana-Champaign, Urbana, IL, USA\\
$^{7}$ Laboratoire Lagrange, Universit\'e C\^ote d'Azur, Observatoire de la C\^ote d'Azur, CNRS,  Nice, FR\\
$^{8}$ Laboratoire Artemis, Universit\'e C\^ote d'Azur, Observatoire de la C\^ote d'Azur, CNRS,  Nice, FR\\
$^9$ Department of Earth, Atmospheric, and Planetary Sciences, MIT, Cambridge, MA, USA\\
$^10$ Leibniz-Institut f\"{u}r Astrophysik, Potsdam, Brandenburg, DE

}
\normalsize
\end{sectionauthor}
\vspace{20pt}

\subsection{Introduction}

Despite many efforts to ``interest'' women and girls in science, far less attention is paid to ensuring that they remain in the field. At the most basic level, female scientists often do not have access to the same resources as their male colleagues. For instance, there is strong evidence for gender bias in the evaluation of telescope proposals, which can be fully mitigated by dual-anonymous peer review, in which the identities of both reviewers and proposers are concealed, meaning the gender of the principal investigator is not known \citep{Johnson.etal.2020}. Gender bias may also contribute to a preference for male proposers in the allocation of personal research grants \citep{doi:10.1073/pnas.1510159112}. Although in many programs women appear to win grants in proportion to their submission rates and representation in the field, women are less likely than men to submit proposals even when eligible \citep{RissleR2020-mm, Schmaling2023},
in part because they have less time available for research due to disproportionate service and teaching responsibilities \citep[e.g.,][]{o2019department}.  Moreover, the grants awarded to women are, on average, substantially smaller than those awarded to men \citep[e.g.,][]{Schmaling2023}.

Such resource disparities begin at hiring, which can already systematically favor men over comparably qualified women \citep[e.g.,][]{Moss-racusin,Eaton2020,Berne.Hilaire.2020}. Even after women have received an offer, they are penalized for negotiating \citep{BOWLES200784, Mazei2015-uw}, creating a disadvantage in commensurate resourcing. There is evidence to suggest that this is because women are judged more on their \textit{social skills} than their competence \citep{Phelan.etal.2008}, such that being perceived as less agreeable negatively affects assessments of their suitability for a role. This not only has an immediate deleterious effect, but it also disincentivizes future negotiations and contributes to a culture in which women are less likely to negotiate at all, let alone assertively.

Female scientists are asked and expected to do more service work \citep{OMeara2017}, and women are often tasked with less visible, undervalued service \citep{Hanasono.etal.2019}, such as administrative committee work and informal student mentoring or ``counseling'', whereas men are more often assigned highly visible (and sometimes compensated) leadership roles that materially advance their careers \citep{Jones.etal.2023}. Given the many demands on faculty time, disproportionate service responsibilities, particularly those that are not rewarded, pose a substantial disadvantage by leaving less time available for research \citep{toutkoushian1999faculty, link2008time, o2019department, foley2019national, eagly2020social}. Beyond the inequitable distribution of time as a resource in itself, reduced research time also has downstream effects, contributing to disparities in funding and perceived productivity. After accounting for field representation, women submit fewer proposals for which they are eligible than men across disciplines funded by the United States National Science Foundation, in part because female faculty spend more time on service and teaching and less time on research than their male counterparts \citep{RissleR2020-mm}.

Women are also \textit{tangibly} under-resourced. Female scientists have less physical space \citep{scripps}, and even after adjusting for career stage, department, research productivity, and other factors, there remains a substantial gender pay gap among academic faculty \citep[e.g.,][]{umbach2007gender, webber2015not, hatch2017gender, https://doi.org/10.1093/ajae/aaz017, kim24}. The pay gap appears in both base salary, likely reflecting disparities in recognition, and ``off-base'' compensation such as merit pay, supplementary funding, summer salary or grant top-ups, and compensation from external positions. The latter is not correlated with perceived productivity and instead reflects a combination of factors including bias in grant award and disbursement, women's more limited negotiating power, and gendered differences in self-promotion and advertisement \citep{kim24}.

Inequalities in recognition of similar work persist beyond pay and compensation. Women are substantially less likely to engage in self-promotion \citep[e.g.,][]{10.1093/qje/qjac003, Peng2025} and are less inclined to describe their work using promotional language that emphasizes importance and novelty \citep{Lerchenmuellerl6573, Qiu.etal.2024}. Women also receive less external promotion, despite evidence that gender-diverse collaborations \citep{Yang.etal.2022, doi:10.1073/pnas.1915378117} and woman-led projects \citep{Yang.etal.2024} are more creative, novel, and innovative. At the same time, men disproportionately favor male co-authors, meaning women are excluded from collaborations \citep[e.g.,][]{boschini2007team,Joyce.etal.2022,Pezzoni.etal.2023} and thus have reduced access to human and social capital. In astronomy specifically, female-led papers receive roughly 10\% fewer citations than comparable papers with a male first author \citep{Caplar.etal.2017}. Recognition is not only important for self-efficacy and job satisfaction; it is also critical for evaluation in hiring, tenure and promotion, and awards.

\subsection{Diminished Resources}\label{subsec:dimres}

\subsubsection{The Issue}

The discussion of available resources can be a difficult one and is often highly employer-specific, but studies spanning the past two decades \citep{mit1999,whoi2000,Scripps23} have shown that women academics and researchers are often offered fewer resources, including lower salaries \citep[e.g.,][]{umbach2007gender, webber2015not, hatch2017gender, https://doi.org/10.1093/ajae/aaz017, doi:10.1126/sciadv.aba7814, kim24} and less lab, office, and storage space \citep{scripps}.  
Historically, the lack of transparency in many of these metrics made identifying discrepancies difficult.  
However, salaries at all public universities in the United States are now publicly available, and recent studies indicate a persistent gender gap \citep[e.g.,][]{kim24}.

Salaries and resources are sometimes constrained by union-negotiated contracts, and, in general, the hiring of a faculty member involves negotiation between the department and the university administration. 
In group discussions, it was noted that male researchers and academics are often more likely to articulate preferences and expect them to be met.  Female researchers appeared less likely to do so and more likely to accept what is initially offered--perhaps because women are disproportionately penalized for negotiating \citep{BOWLES200784}.  
This can, and anecdotally does, lead to imbalanced working conditions and assignments for women researchers (as discussed in Section \ref{subsec:work}). Successful negotiation requires a combination of skills, including succinct self-advocacy; mutual goodwill between parties; and a robust understanding of what the employer (e.g., university) is--and has been--able to offer.

Astronomical researchers generally do not receive formal negotiation training (or even informal guidance through community groups), and are therefore left to learn these skills on their own--or not negotiate at all.  
While negotiation is often assumed to concern base salary (which may itself be constrained by union or administrative policies), negotiations can also include many other research resources across all academic stages. In astronomy, these may include office space, guaranteed conference travel funding  (e.g., for the winter AAS meeting), extended graduate student funding (e.g., a six-year rather than five-year PhD guarantee), and similar commitments.  
Anecdotally, students in engineering and other industry-aligned disciplines often receive training in negotiation and self-promotion from industry recruiters and academic societies, whereas such training is typically absent from traditional astronomy and physics communities (e.g., the Society of Physics Students). 

As discussed in greater detail below, we recommend the implementation of similar training for students at all levels, as well as more formal professional development for postdocs and faculty.  
In all cases, transparency and consistent standards--from students through faculty--regarding the resources required for success (e.g., a living document outlining available resources, request procedures, and teaching assignments), as well as clarity around appropriate negotiables (e.g., re-location allowances, equipment guarantees, and annual travel support), are strongly encouraged when positions are offered and accepted.

\subsubsection{Recommendations}

To promote better self-advocacy practices for women scientists, we suggest that \textbf{departments provide specialized training and/or resources focused on self-promotion practices and actions}.  Examples of direct action could include CV or personal statement revision workshops, or designated time at routine faculty meetings to advertise available awards and encourage women faculty to apply.  We also encourage departments, when legally permitted, to make salary levels and other key metrics (that is, negotiable resources such as discretionary research funding or office location) available to better enable transparency and empower women in job negotiations. \citet{Mazei2015-uw} found that offering negotiators a combination of formal training and a clear sense of negotiables and their historic ranges resulted in better outcomes.

To promote better mentorship practices for women scientists, we \textbf{encourage women to find and build communities with other women to provide mutual support}.  The exact implementation is likely to be individual-specific, but examples could include a large, university-unaffiliated messaging group or a weekly video-call check-in.  We also encourage employers (e.g., departments or research centers) to increase the visibility of women in senior positions in order to amplify role models for more junior women.  By seeing other women in the community speaking openly--even when views are controversial yet respectfully expressed--and having that be received positively, junior women may feel more confident contributing in similar ways.

\subsection{Differential Praise}

\subsubsection{The Issue}

Differential praise can be defined as the use of different descriptors, unequal emphasis on ability, and the reinforcement of stereotypes in areas of academia that directly affect career progression (e.g., award nominations, reference letters, talk introductions, mentoring, student feedback, and citations). Examples include men more often being described as ``standouts'', whereas women are more commonly described as hardworking but not necessarily exceptional \citep{Trix2003,Dutt2016,Madera2019,schmader2007linguistic}; information networks for awards potentially splitting along gender lines \citep{Lincoln2013}; and discrepancies in citation \citep{Wu2024} and speaking engagements \citep{Schroeder2013}. Anonymous student feedback has been found to be especially susceptible to bias against female instructors \citep{Aragon:23, storage2016frequency}, and  universities are increasingly reducing their reliance on student evaluations in promotion and tenure decisions for this reason.

In the area of reference letters in particular, a number of resources have emerged over the years to help letter writers avoid gendered language \citep{Urry1993,Pasadena05}, as broader awareness of this issue has grown. Some recent studies suggest that patterns of differential descriptors have begun to shift in certain fields \citep{Bernstein2022,Zhao2023}, whereas others still find evidence for persistent disparities \citep{Eberhardt2023}. Studies in this area vary widely in methodology (e.g., manual vs. algorithmic evaluations, or the specific classification of ``gendered'' or ``grindstone'' words; \citealt{Trix2003, Morgan2013, Madera2019, Zhao2023, Eberhardt2023}), which makes the degree of gender bias difficult to quantify. More broadly, as bias is likely more complex than the presence or absence of particular words, progress can be challenging to track. However, there is evidence that disparities in recognition--in references, award nominations, and speaker invitations--can be reduced when there is greater representation in senior roles and on selection committees \citep{Lincoln2013, Nittrouer2018, Zhao2023}. Given the salience of this issue in contributing to inequities in the field (e.g., the leaky-pipeline effect), we highlight issues and potential solutions in the contexts of reference letters and award nominations, self-advocacy (as discussed in Section \ref{subsec:dimres}), and representation.

\subsubsection{Recommendations}

There has been significant community effort over the past decade to address the commonly recognized problem of ``differential praise'' in promotional materials such as letters of recommendation and award nominations. More recent publications suggest that this issue may be less common now than in previous years \citep{Bernstein2022}.  Even so, anecdotally, experiences persist of letters of support that are poorly written, unhelpful, or even actively harmful.  It is widely recognized that letters of support serve different purposes at different academic stages and vary in how much influence they have on hiring and award decisions.

We recommend that, \textbf{at the level of faculty hiring and tenure, letters of reference be requested only in the final round of evaluation--or not at all}.  Letters at this stage are rarely the most informative component of the application package and often only confirm what is already clear from other elements of the file.  However, at earlier stages, especially for students and postdoctoral researchers, letters often play a more influential role.  Candidates at these levels may not yet fully recognize their strengths, or may be less confident in articulating them, resulting in application materials that do not reflect their full potential.  In such cases, strong letters of support from colleagues willing to speak openly and effusively about the candidate remain crucial assessment tools.  

We first, and most strongly, recommend that \textbf{employers require meaningful, ongoing anti-bias training for all those who mentor students and postdoctoral researchers}, which in practice includes nearly all faculty and many postdoctoral researchers.  We emphasize that such ``training'' should be an active and continuous practice within the department or research center, rather than a one-off workshop that may be skipped, receive limited engagement, or quickly be forgotten.  Examples of more active approaches include monthly collaborative, interdisciplinary seminars with departments such as Women and Gender Studies, or inviting advanced PhD students, postdoctoral researchers, and faculty from sociology-affiliated fields to observe and provide feedback on classroom or seminar dynamics.  The intention is for members of the astronomy community who hold positions of power to critically analyze the nuances of their behaviors and language--subtleties often overlooked in standard, standalone trainings.  

We also recommend that \textbf{employers provide resources to help women researchers become more effective advocates for themselves}. Women are, on average, less likely to engage in explicit self-promotion \citep[e.g.,][]{10.1093/qje/qjac003} and are often socially conditioned toward communal behavior. Women who do not conform to these expectations may be judged \textit{more} harshly \citep[e.g.,][]{McKinnon2020}, meaning that self-advocacy can sometimes carry additional risk.  Formal guidance can help women researchers more confidently, accurately, and safely represent their accomplishments in personal statements and other self-authored materials, which may, in turn, reduce the disproportionate influence of letters of support.

\subsection{Unequal Workloads} \label{subsec:work}

\subsubsection{The Issue}

Efforts to improve the representation of women in academia are often focused on the hiring stage, including how positions are advertised and how applications are solicited and reviewed.  However, gendered patterns in attrition and promotion present an important and distinct challenge that must also be addressed.  Women leave faculty positions at consistently higher rates than men at every career stage, with the differential impact being considerably worse in non-STEM fields \citep{Spoon2023}.  A feeling of being stuck and not progressing in one's career can be a major source of discouragement.  Studies indicate that women faculty take on heavier service loads, which are frequently skewed towards more time-consuming and less prestigious roles.  
Perhaps not coincidentally, women associate professors are much less likely to ever be promoted to the rank of full professor, and when they are promoted, the process takes significantly longer \citep{Misra2011}.

Among the reasons considered by \citet{OMeara2017} for gendered differences in service loads are the following:

\begin{enumerate}
    \item Women are more willing to take on service roles that relate to issues and groups they value.
    
    \item Women are {\it asked more often}, ostensibly to add diversity to committees or because of a perception that they will perform certain tasks well.

    \item Women may be less likely to decline requests because they tend to be more junior in status or because they do not want to appear selfish.
\end{enumerate}

While it can be difficult to establish which of these factors dominates in a particular context, the authors suggest that a common feature underlying inequity is a lack of clarity regarding expectations: faculty often do not know how much service is required of them and what (if any) consequences would result from not performing these duties well.  In situations where the ground rules are unclear, men negotiate more actively for preferred roles or additional resources.  Using detailed time diaries, \citet{OMeara2017} found that tenured women professors receive more work requests than men, that these requests tend to come from students, former students, or off-campus colleagues, and that they are less likely to be research-related.  Simply dealing with a larger number of requests is itself an additional time burden, highlighting the importance of creating a culture where professional service is a shared responsibility across the department or institute, and where women are supported in safely declining excess service work.

\subsubsection{Recommendations}

We recommend that \textbf{departments make service expectations explicit and transparent, and assign service loads deliberately rather than informally}, in order to prevent women from disproportionately assuming onerous service responsibilities. When expectations are unclear, women are more likely to take on heavier service burdens, so more conscious decision-making around service assignments may be an effective response.
Creating faculty workload dashboards that specify minimum, average, and high workloads for teaching, advising, and service can increase awareness of inequalities and improve accountability and fairness \citep{OMeara2016}.
Ironically, when women are ``asked more often,'' those requests usually come from other women \citep{OMeara2017}; by taking on heavier service loads, women may also be generating additional service work for other women.  Mentoring earlier-career women to more confidently decline requests and protect their scholarly work, while providing greater recognition for service work when it has broad impact, are additional strategies that can be considered.

\bibliography{references/resources}{}
\bibliographystyle{aasjournal}

\pagebreak
\section{Motherhood Penalty and Familial Expectations: Challenges and recommendations} \label{sec:family}

\vspace{-20pt}
\begin{center}
    \includegraphics[width=0.4\linewidth]{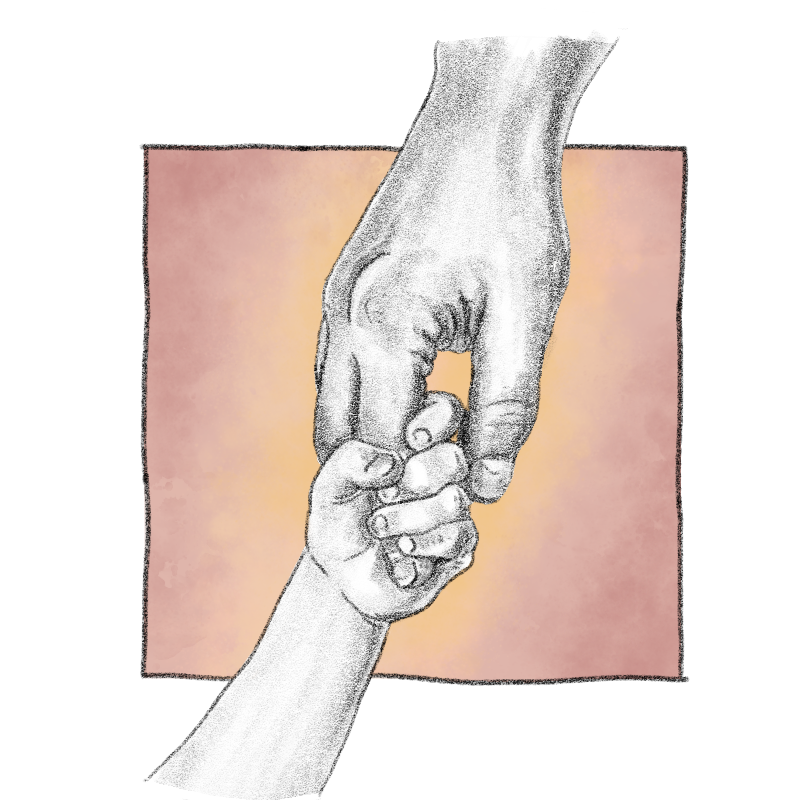}
\end{center}
\vspace{-30pt}

\begin{sectionauthor}
\normalsize{\href{https://orcid.org/0000-0002-0517-9842}{Melanie Archipley}$^{1,2}$, 
\href{https://orcid.org/0000-0001-7160-3632}{Katherine E. Whitaker}$^{3,4}$, 
\href{https://orcid.org/0009-0009-2685-4067}{Raagini Patki}$^{5}$, \\\href{https://orcid.org/0000-0003-2539-8206}{Tjitske Starkenburg}$^{6,7,8}$,
\href{https://orcid.org/0000-0002-5995-9692}{Mi Dai}$^{9,10,11}$, and \href{https://orcid.org/0009-0000-4830-1484}{Keerthi Kunnumkai}$^{12}$\\ 
}
\vspace{15pt}
\scriptsize{
$^{1}$ Department of Astronomy and Astrophysics, University of Chicago, Chicago, IL, USA\\
$^2$ Kavli Institute for Cosmological Physics, University of Chicago, Chicago, IL, USA\\
$^3$ Department of Astronomy, University of Massachusetts, Amherst, MA USA\\
$^{4}$ Cosmic Dawn Center (DAWN), Copenhagen, DK\\
$^5$ Department of Astronomy, Cornell University, Ithaca, NY, USA\\
$^{6}$ Department of Physics \& Astronomy, Northwestern University, Evanston, IL, USA\\
$^{7}$ Center for Interdisciplinary Exploration and Research in Astrophysics, Northwestern University, Evanston, IL, USA\\
$^{8}$ NSF-Simons AI Institute for the Sky (SkAI), Chicago, IL, USA\\
$^9$ LSST Interdisciplinary Network for Collaboration and Computing Frameworks, Tucson AZ, USA\\
$^{10}$ Pittsburgh Particle Physics, Astrophysics, and Cosmology Center, University of Pittsburgh, Pittsburgh, PA, USA\\
$^{11}$ Department of Physics \& Astronomy, University of Pittsburgh, Pittsburgh, PA, USA\\
$^{12}$ Department of Physics, Carnegie Mellon University, Pittsburgh, PA, USA\\

}
\normalsize{}
\end{sectionauthor}
\vspace{20pt}

\subsection{Introduction}
Despite advances in hiring women into astronomy roles over the last two decades \citep{porter2024} and no evidence of gender-based attrition through the postdoctoral stage of one's career, it has been shown that women are still more likely than men to leave the field at the faculty hiring stage \citep{flaherty2018}. Specifically, ``family formation'' including marriage and childbirth accounts for the majority of lost female\footnote{Here, we adopt binary gender classifications in the context of caregivers, though we acknowledge that the body of literature for non-binary parents is limited to date and thus inadequately addressed herein.} talent from the sciences \citep{goulden2011}. A broad literature spanning over a decade has uncovered the ways in which experiences of parenthood and childcare mediate the gender disparities in science academia \citep[e.g.,][]{fox_work_2011, thebaud_specter_2021, canetto_challenges_2017, cech_changing_2019, hong_parenthood_2025}, and studies specific to the field of astronomy confirm these findings \citep[e.g.,][]{barthelemy_experiences_2014, holbrook_astromoves_2019, pandey-pommier_status_2023}.

Familial caregiving responsibilities, which are still disproportionately shouldered by women \citep[e.g.,][]{pew2019, kenny2021, ervin2022}, are often at odds with a dominant social paradigm that assumes scientists must devote themselves entirely to their subject as a vocation without distraction or ``work-life balance.'' Studies show that in science, technology, engineering, and mathematics (STEM) fields, childbirth, motherhood, and other forms of caregiving are perceived as detracting from women's scientific excellence and devotion to their research, while the same standards are not applied to fatherhood \citep[e.g.,][]{national_academies_of_sciences_caregiving_2024}. In addition to parenting, familial responsibilities can include a variety of roles and relationships, including the care of elderly parents and disabled or other adult dependents \citep[e.g.,][]{zygouri_gendered_2021}. Women of color may additionally experience unique expectations and impacts of caregiving due to racialized cultural conceptions of family obligation and community \citep{national_academies_of_sciences_caregiving_2024}. 

While women in science graduate programs report that an academic career would be most fulfilling to their interests and aspirations, the academic career path is perceived as requiring them to give up on family life \citep[e.g.,][]{canetto_challenges_2017}. The gendered workloads of childcare lead to disproportionate loss of academic achievement \citep{hong_parenthood_2025}, and upon having their first child, mothers are more likely than fathers to leave STEM employment \citep{cech_changing_2019}. Although the effect has decreased over time, new parents who remain in academia continue to experience a gender productivity gap in their research publishing \citep{morgan_unequal_2021,bohm_impact_2023}.
Compared to White women, women of color, including Black, Latina, and Brown women \citep{gupta2022}, are subjected to additional bias and barriers as mothers in the workplace, which intersect with the pressures of the academic career path \citep{williams_beyond_2020}.
Women of color experience challenges prioritizing a lengthy career in astronomy and astrophysics, and these challenges stem in part from family roles, needing to financially support family, and upholding culturally female roles \citep[e.g.,][]{ko_narratives_2013}.

Furthermore, in astronomy and other STEM departments, women experience the gendered impact of the conflict between the traditional roles of women in families and expectations of scientists, {\it whether or not they are mothers or ever plan to be mothers}. Women are more likely than men to experience adverse effects when negotiating with a so-called ``two body problem'' in which they and their spouse are both searching for employment in their fields \citep{mcneil1999}. Early career scientists report that pregnancy, motherhood, and childbearing are a topic of public discussion, speculation, and judgment in their departments and professional spheres, while men's choices regarding fertility and fatherhood are not commented on, questioned, or policed \citep{thebaud_specter_2021}.
For young women, the ``specter of motherhood''--their status as even potential mothers--leads other scientists to question their status and legitimacy in their profession \citep{thebaud_specter_2021}. 

The example set by tenured faculty in a department is central to the perceptions of its junior members--women PhD students and postdocs report that they look to faculty members both for mentorship and to understand whether a suitable balance between work and family life is possible in their field \citep[e.g.,][]{barthelemy_experiences_2014,thebaud_specter_2021}. Given that women faculty report higher levels of interference of their family life on their work and vice versa than men \citep{fox_work_2011}, it is perhaps unsurprising that a survey of 750 participants ranked ``family and caregiving responsibility'' as the number one factor affecting women's careers in astronomy, above bias against women, social stereotypes, gender-based harassment, or other factors \citep{pandey-pommier_status_2023}. 

It is clear that women face unique and specific challenges when balancing their academic careers and family obligations. Thus, it is essential for academic institutions and departments to demonstrate institutional and cultural acceptance and support of members at all career stages as they navigate childbearing, childcare, and traditional family structures and expectations.

\subsubsection{Motivation for departments}

In this white paper, we advocate for recommendations to improve these systemic issues, some of which represent significant investment on the part of institutions.
We note that in astronomy academia, a resource-constrained environment focused on scientific progress and excellence, studies show that including all genders results in better science.
Gender-diverse teams have more novel and higher-impact scientific ideas \citep[e.g.,][]{zeng2016, yang2022}, produce work with more citations \citep[e.g.,][]{campbell2013, maddi2020}, and are more innovative and creative \citep[e.g.,][]{xie2020, vedres2023}. 

In the near future, adopting the policies that we recommend, and that accommodate families, may be essential to hire talent: the incoming workforce comprising ``Generation Z'' highlights several related workplace priorities in a recent report by \cite{deloitte2025}, ranking finances and cost-of-living as significant stressors. Institutional policies highlighting good work-life balance and flexibility for families are considered highly important by younger generations and early career faculty \citep[e.g.,][]{strong_worklife_2013,rodriguez-sanchez_investing_2020} and can help departments recruit the next wave of talented scientists. 

Departments should consider how programs that target recruits by supporting their families, such as spousal hiring policies at rural institutions, can result in a significant boost to their workforce rather than a detraction from their resources \citep{farrar2003}. As workers place higher value on how their job impacts their day-to-day lives, attractive policies related to wages, healthcare, parental leave, and a welcoming environment towards different family structures will have more influence on where people choose to work.

Finally, turnover of employees--especially those with highly specific and specialized training, as in astronomy--is expensive. Significant portions of the literature highlight how precarious the career paths of women and parents are, and retention matters for employers as much as employees. In academic departments, it is much more expensive to replace a faculty member than to retain one \citep{NASEM2007} and graduate students or postdocs who leave mid-appointment waste research funds and resources. It is in employers' best interests to devote attention to the training, recruiting, and retention of women in astronomy because institutional investment in such policies is an investment in research output.

\subsection{Context of these recommendations}
The following suggestions are based on discussions from the working group on traditional familial expectations at the ``Picture an Astronomer'' conference in March 2025. In our conversations, relevant experiences were shared among conference participants who were varied with respect to career stages in astronomy, marital and parental status, and caregiving duties. The astronomers involved in the discussions were primarily based at U.S. institutions: therefore we acknowledge that members of the international astronomy community may have different needs and concerns and encourage further engagement with this topic. 

A theme of discussion was the fact that specific solutions to the field of astronomy are needed in addition to those suggestions made in the broader literature on parenthood in academia and STEM. Thus, in this white paper we seek to provide policy recommendations specific to astronomy, narrowly focused on what is likely to help women astronomers, and all parents and caretakers, the most.

First, in academic astronomy, career progression is typically expected to follow a standard pathway, including a four-year bachelors degree, a six-year doctoral degree, and three to six years of postdoctoral positions before a person is on the job market for permanent positions \citep{polzin2023}. The ``standard academic track'' puts people at approximately age 33-35 before obtaining a permanent position whereas the average age of first-time mothers in the United States is 27.5 \citep{fitzpatrick2025}. While women with higher education are more likely to have children when they are older \citep{delbaere2020}, delaying childbearing can be medically detrimental to mother and child \citep{acog2022}; fertility rates sharply decline with increasing age and assisted reproductive technologies cannot fill the gap \citep{leridon2004}. As a result of these competing timelines, frequent relocation and career uncertainty occurs during the same period that many early career astronomers seek to start a family or have children \citep[e.g.,][]{holbrook_astromoves_2019}. Moreover, because of frequent relocations, academic parents are often living far from their support network of family and friends. This places additional pressure on these parents, who are predominantly mothers, and requires them to build a new local ``village'', a process that can take as long as a typical postdoctoral appointment. The lack of local support also makes it more challenging for parents in astronomy to travel for work: attend conferences, give colloquia and seminars, meet with collaborators, and travel for interviews; all crucial for networking, self-advertisement and effective collaboration. Particularly when both parents are academics or otherwise regularly travel for work, this can give rise to challenging planning exercises and add significant constraints. Lastly, although many colleagues and collaborators are supportive of caregivers, parents can still encounter gendered biases when traveling for work.  Fathers are generally not expected to limit or adapt their travel for their children, whereas mothers may be viewed critically for traveling at all, particularly when their children are young.

Additionally, this career path, with its frequent changes in employment, leads to repeated ``cycles'' of job applications and interviews during which parental bias may be experienced. Discussion participants shared concerns about encountering such bias during the job search process--for example, worrying that they might ``out'' themselves as parents simply by asking about parental leave policies. Securing childcare across successive research positions was also described as a major challenge. In the United States, daycare waitlists are often months or even years long, sometimes longer than the duration of a postdoctoral appointment. When the cost of daycare is compared with typical graduate student and postdoctoral salaries, parents at these career stages may find it particularly difficult to secure affordable childcare. 

Lastly, metrics of success in astronomy often center on first-author publications. Participants expressed concern that taking extended leave can create gaps in research productivity and disrupt the continuity of a research program.
Furthermore, astronomers whose research depends on grant funding may struggle with the reality that these projects cannot simply be paused, and they may worry that not applying for new grants while on parental leave will negatively affect future research prospects. 

The following recommendations reflect the themes that emerged from these discussions and the concerns raised. They are not intended to capture every astronomer’s experience with work or family life, but rather to highlight common challenges and potential responses.

\subsection{Recommendations}

\subsubsection{Hiring practices}

Some specific policies may help reduce bias against parents during the interview and hiring process at all career levels (undergraduate researchers, prospective graduate students, postdoctoral researchers, faculty, research scientists, etc.). \textbf{All candidates should be provided with a copy of the institution and departmental leave policies and benefits}, regardless of the candidate's gender, age, or career stage. \cite{lee2017} found that not only were postdocs uncertain about parental leave policies that may apply to them, but also that human resource (HR) offices were similarly uncertain about which policies applied to postdocs, resulting in widespread confusion. HR should be involved in the hiring process to answer questions, and the suggested policy of providing information prior to interviewing would allow parents and prospective parents to become informed about relevant policies without being forced to reveal their situation, and thus subject themselves to potential bias, during the hiring process.

While it should go without saying, during our discussions, some participants recalled instances of being asked unlawful or borderline\footnote{A seemingly casual question that came up during discussions was ``what neighborhoods are you looking to move to?'' that can be used as a proxy for inquiring about the candidate's family, since neighborhoods are associated with family status.} questions during interviews. \cite{mcneil1999} reported that potential academic employers often asked candidates prohibited questions both unintentionally and intentionally. All members of the department or institution that interact with a candidate (including human resources, faculty, postdocs, graduate students, etc.), as well as candidates themselves, should \textbf{be clearly informed on which questions are illegal to ask and steer clear of them}, such as marital status and whether or not a candidate is or plans to be a parent.

\subsubsection{Work hours and scheduling}

To improve working conditions and reduce bias for parents in astronomy departments and astronomical collaborations, we recommend several policies relating to the schedule of work. \textbf{Department meetings, colloquia, and other local or in-person obligations should be scheduled during the window of children's daycare and school to the extent possible}. Ideally, this would mean a high priority block for scheduled events from 9 a.m. to 2 p.m. local time and avoiding scheduled events outside of 9 a.m. and 5 p.m. altogether. As astronomy contains numerous large collaborations spanning the globe, it is often not feasible to restrict virtual meetings to any one timezone's ``work day,'' therefore we recommend \textbf{organizing virtual meetings in which people can participate asynchronously} by recording conference calls, taking notes, and making such meeting materials easy to find. Similarly, we emphasize the importance of giving as much notice as possible when setting up and canceling meetings. 

Accommodating the scheduling constraints of caregivers may be particularly difficult for the subfield of observational time domain astronomy, when alerts of cosmic phenomena may come at any hour of the day and require human attention and action. Thus we suggest that \textbf{research groups and collaborations in time domain astronomy adopt an equitable system where members are ``on call'' for limited time periods}, similar to the way that numerous other professions handle ``emergency'' work in a shift-like rotation without certain members being unduly burdened.

For those working in instrumentation or subfields that require field work in astronomy, such as deployment to telescope sites, \textbf{employees should feel empowered to request reasonable accommodations}: \cite{lee2017} reports that less than half of postdoc mothers request workplace accommodations, but of those requests, over 90\% are granted. Similarly, efforts should be made by principle investigators and group members where possible to accommodate employees with caregiving responsibilities, whether that means shifting the literal heavy lifting off of a pregnant colleague or training alternates for extensive field deployments in case of sudden need.

\subsubsection{Childcare}

Affordable and quality childcare represents a burden for parents in multiple ways. On-campus childcare typically prioritizes faculty and staff over postdocs and graduate worker parents, and financially, the cost of childcare can reach 50-100\% of typical early-career salaries \citep{lee2017}. \textbf{Departments that allow for flexible scheduling, remote work, and other accommodations can help alleviate employees' stress related to caring for family}. When employers allow for hybrid work, retention and job satisfaction of employees is improved without damaging performance \citep{bloom2024} and career opportunities for mothers are broadened \citep{harrington2023}.

We recommend that \textbf{departments reserve a certain number of guaranteed on-campus childcare spots for their members, including graduate students and postdocs}. While many universities offer some form of daycare on campus, waitlists can be months to years long, with an average of 90 children waiting for care at on-campus facilities prior to the pandemic \citep{reichlin_cruse_evaluating_2021}. A similar system to the priority model often used for university housing, where departments or units are allocated guaranteed places, could be particularly beneficial in helping academic parents secure childcare.

In recent years, it has become more common for astronomy conferences to offer childcare or childcare grants to participants. However, discussion participants with dependents reported mixed benefits when childcare is offered onsite. Bringing children to a conference site may be impractical, financially burdensome, and disruptive to children's schedules. Discussion participants reported experiencing a penalty to their networking and collaboration because they had dependent care obligations during many of the informal after-hours conference activities. We advocate instead that \textbf{conference organizers provide flexible childcare funding which can be used for care in a preferred location} \citep[such as those provided by][]{noauthor_dependent_nodate, princeton2025b}. Such funding should be prioritized by financial need so that parents in lower-income career stages or locales have meaningful access.

\subsubsection{Unionization}

Unions can retain critical benefits related to parental leave policies and childcare subsidies \citep{campbell2023}, but less than half of graduate workers \citep[38\%;][]{herbert2024}, 17\% of postdocs \citep{herbert2024, NSF2025}, and 27\% of faculty members \citep{nietzel2024} are represented by unions, leaving the vast majority of academic workers without union protections. For the academic sector in particular, unionization can reduce uncertainty and inconsistency between employees (such as those with different funding schemes even in the same department) among other benefits. In turn, the better working conditions that unions secure, including higher pay, improved healthcare benefits, and channels for resolving grievances, result in improved retention rates \citep{delery2000}. The positive effects of worker unionization are long-lasting in society: greater unionization has intergenerational benefits, as children of unionized parents are found to have fewer health issues and better eventual earnings and educational attainment \citep[e.g.,][]{freeman_how_2015}. For these reasons, we recommend \textbf{supporting unionization and advocating for family-friendly policies at the collective bargaining table}.

\subsubsection{Recommendations tailored to graduate students and postdoctoral scholars}

Because the early career stages are the most vulnerable for women and caregivers, we recommend several policies departments and institutions should adopt to protect their students and postdoctoral astronomers.

Firstly, universities should \textbf{include the option for all students and postdocs to be hired as employees,} even if they join a department with their own fellowship. Employee status allows for people to take advantage of any parental protections or leave available to employees at the institution, among other negotiated benefits.

Secondly, institutions should consider the effects of parental leave on the career of junior members in the context of the academic job cycle by \textbf{allowing students and postdocs to extend their positions for an additional year following parental leave}, rather than just the amount of time taken off (which is typically a few weeks to months). Such a policy lessens the impact of taking several weeks to months off from a research program, which can represent a significant fraction of time in a temporary appointment, shifting someone ``off'' of the job cycle. Many universities have already codified how family leave affects junior faculty members' path to tenure, ensuring that parents of all genders are not discriminately punished for taking leave during critical periods of employment (an example of this is Princeton's ``Tenure Clock Extension for New Parents'' policy, see \citealt{princeton2025a}).

\subsubsection{Departmental culture and further recommendations} 

We additionally caution that some studies show that parents are often reluctant to use supportive policies, because doing so violates social norms of science where productivity is prized \citep{national_academies_of_sciences_caregiving_2024}.
The reluctance is exacerbated for mothers, who are viewed as less productive when compared to equally productive fathers \citep{blair-loy_misconceiving_2022}. \textbf{Even if seemingly supportive policies are in place, the culture and norms must evolve alongside them, too,} or as \cite{lee2017} states, 
\begin{mdframed}[
  leftline=true,
  rightline=false,
  topline=false,
  bottomline=false,
  linecolor=gray,
  linewidth=2pt,
  innerleftmargin=10pt,
  innerrightmargin=5pt,
  innertopmargin=6pt,
  innerbottommargin=6pt
]
\textit{``Simple adherence to federal law would go a long way,''} because \textit{``[their] study found violations of the Family and Medical Leave Act, the Americans with Disabilities Act, Title VII [of the Civil Rights Act of 1964], and Title IX [of the Educational Amendments of 1972].''}
\end{mdframed}

When workplaces emphasize  ``true access'' to work-family policies (as opposed to those policies merely existing), performance and productivity see an increase \citep{medina-garrido_relationship_2017}. Therefore, it is paramount that institutions are \textbf{clear and vocal about their support for families} through both formal channels, such as adhering to helpful policies and the law, and informal channels, such as promoting role models, adjusting expectations to meet the reality of caregiving employees, and fostering a compassionate environment.

\bibliography{references/familial}{}
\bibliographystyle{aasjournal}
\pagebreak
\section{Mitigating Bullying and Harassment} \label{sec:bullying}

\vspace{-20pt}
\begin{center}
    \includegraphics[width=0.4\linewidth]{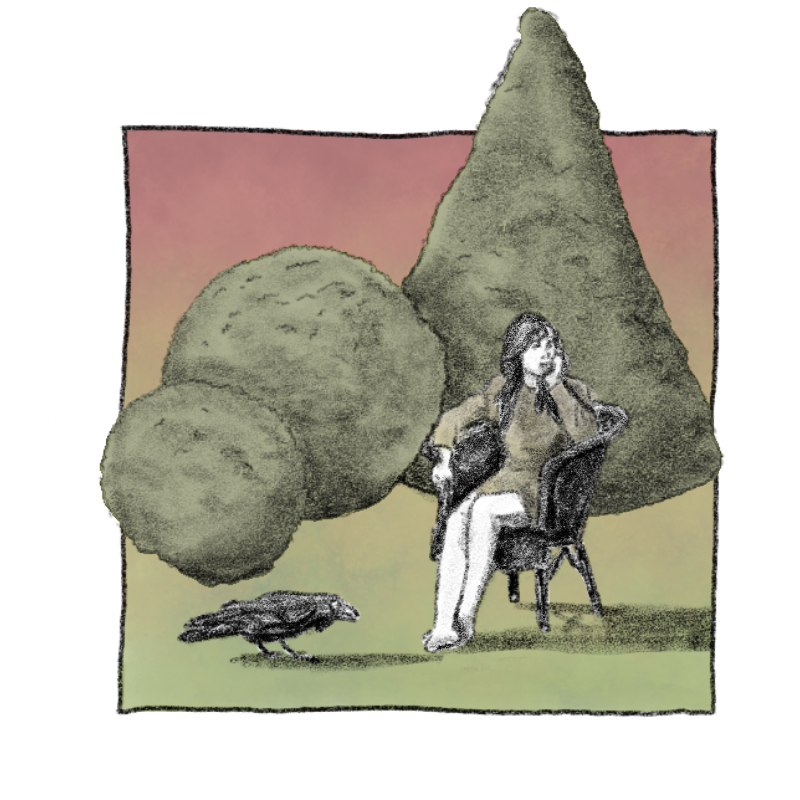}
\end{center}
\vspace{-30pt}

\begin{sectionauthor}
\normalsize{
\href{https://orcid.org/0000-0001-5228-6598}{Katelyn Breivik}$^{1,2}$, \href{https://orcid.org/0000-0002-0965-7864}{Ren\'{e}e Hlo\v{z}ek}$^{3,4}$, Molly Beth Jourdan$^{5}$,\\ \href{https://orcid.org/0000-0003-3953-1776}{Alexandra S. Rahlin}$^{6,7}$, and \href{https://orcid.org/0000-0003-0478-0473}{Chin Yi Tan}$^{7,8}$

}
\vspace{15pt}

\scriptsize{
$^1$ \hspace{1pt} Department of Physics, Carnegie Mellon University, Pittsburgh, PA, USA\\
$^{2}$ McWilliams Center for Cosmology and Astrophysics, Carnegie Mellon University, Pittsburgh, PA, USA\\
$^{3}$ Dunlap Institute for Astronomy and Astrophysics, University of Toronto, Toronto, ON, CA\\
$^{4}$ David A. Dunlap Department of Astronomy and  Astrophysics, University of Toronto, Toronto, ON, CA\\
$^{5}$ Kenwood Academy, Chicago Public Schools, Chicago, IL, USA\\
$^{6}$ Department of Astronomy \& Astrophysics, University of Chicago, Chicago, IL, USA\\
$^7$ Kavli Institute for Cosmological Physics, University of Chicago, Chicago, IL, USA\\
$^{8}$ Department of Physics, University of Chicago, Chicago, IL, USA\\

}
\end{sectionauthor}
\vspace{20pt}

\subsection{Introduction}
Academia is a resource-limited environment, one that can be `optimized' by principal investigators and heads of labs, groups and institutions. While academia, and science/STEM in particular, is often described as being `value-free', the scientific enterprise is generally framed as being guided by values such as openness, communalism, and objectivity, with reproducibility reinforcing norms of non-partisanship and skepticism \citep{merton1973}. However, \citet{slaughter2004} argue that growing connections to the market and an entrepreneurial framing have shifted the narrative.  Rather than emphasizing an abstract `public good', science is increasingly portrayed as a driver of economic prosperity, linking it more directly to the market and market values \citep{slaughter2004}. This narrative has strengthened as funding for traditional academic research positions has narrowed, data-driven alternative careers have ballooned, and interest in STEM degrees has risen dramatically over the past two decades.

The insertion of market values and priorities into academic spaces can intensify competition--both among labs working in similar areas and among trainees within the same group. It can also create conditions in which favoritism and workplace bullying are more likely to emerge \citep{salin2023,keashly/etal:2013}.

In parallel, \citet{tauber2022} summarize that bullying can function as a tool used by less-competent academics to eliminate competition while climbing the professional ladder. Although the narrative of scientific objectivity in \citet{merton1973} might lead one to assume that such behavior is atypical among scientists, \citet{tiidink2016} found otherwise. In a study of 1,833 participants across four medical centers, they reported that Machiavellianism--a personality trait associated with self-interest, the desire for power, and a lack of empathy--was positively associated with self-reported research misbehavior (including 22 forms of misconduct, from deleting or cherry-picking data to  deliberately omitting a collaborator from the author list or failing to publish null results).  %Machiavellianism can be seen as the tendency of someone to be manipulative and to focus on achievement while also being emotionally detached from morality. 

Another key finding from that study was that narcissistic and psychopathic traits appear to be more common among academics at higher ranks, suggesting that academic self-esteem does not simply increase with seniority. The authors suggest that this persistent insecurity at higher academic ranks could mean that narcissistic traits confer a kind of `survival benefit' in academia \citep{tiidink2016}. Such dynamics disproportionately harm those in lower academic ranks, including students and postdoctoral researchers, and risk perpetuating a culture of survival at all costs.

In a `Survivor's Guide to Academic Bullying,' \citet{mahmoudi2020} provide four key suggestions for individuals experiencing bullying: document all interactions, including abusive behavior; consult workplace mediation or ombudsperson offices and, if a formal complaint is filed, insist on receiving a summary of the investigation findings; seek out evidence of others who may be experiencing similar treatment, as collective reporting can strengthen a case;  and develop an exit strategy well in advance. The last recommendation is particularly relevant given that retaliation can follow reports of bullying--sometimes not directly from the accused, but from their collaborators--despite universities’ stated commitments to preventing such outcomes. 

In this white paper, however, we emphasize some \textit{community-based} recommendations to mitigate bullying and harassment that do not depend on individual reporting or on the investigation of specific incidents. If the academic community is to meaningfully reduce the prevalence of bullying, systemic, rather than solely individual, changes are required.

\subsection{Mitigation Recommendations}
Throughout the workshop, participants discussed a range of  bullying and harassment issues that can arise for both junior and senior members of the astronomy community. In parallel, they identified several strategies to help prevent or reduce bullying in different settings, as well as approaches for responding to bullying and supporting colleagues who are being targeted. These strategies are described below. 

\subsubsection{Codes of Conduct}
\label{ref:CoC}
Bullying can be a traumatic experience that leaves the victim feeling powerless. Codes of conduct can help mitigate this problem, and ideally prevent it, by \textbf{clearly defining expectations for behavior in collaborative environments, condemning unacceptable conduct, and outlining the accountability structures that will be used when violations are reported} \citep{aurora2018coc}. A key feature of strong codes of conduct is a clear and concise definition of unwanted behaviors, including (but not limited to) discrimination, harassment, and bullying. In addition, descriptions of the implementation, applicability, and limitations of the code of conduct (including those shaped by legal frameworks) are critical to ensure that victims understand the support structures available to them \citep{Nitsch2005}. 

When a clearly constructed code of conduct in place, \textbf{collaborations and professional organizations can also develop protected databases for reports}. These records can be used to identify repeated bullying behaviors that persist across leadership changes. Similarly, report logs can help track misuse of the code of conduct, such as repeated bad-faith complaints or reports filed in retaliation for unrelated conflicts.

Constructing and implementing these documents requires substantial time and effort. However, they serve as a public commitment to positive community behavior and, when developed in consultation with legal experts, can provide essential accountability structures, especially in large collaborations. Codes of conduct are thus a critical component of collaborative environments, including academic departments, science collaborations, and professional societies. Several publicly available examples may serve as models for future efforts, including those from the International Astronomical Union\footnote{IAU Code of Conduct: \url{https://www.iau.org/Iau/About/Statutes---Rules/Code-of-Conduct.aspx}}, American Astronomical Society (AAS)\footnote{AAS Code of Ethics: \url{https://aas.org/policies/ethics}}, and the Vera C. Rubin LSST Dark Energy Science Collaboration\footnote{LSST Dark Energy Science Collaboration Professional Conduct Policies: \url{https://lsstdesc.org/assets/pdf/policies/LSST_DESC_Professional_Conduct.pdf}}.

\subsubsection{Available Resources and Onboarding}

Lack of awareness about available resources, both within institutions and at the national or international levels, emerged as a common theme in discussions throughout the workshop. The working group discussed multiple past and ongoing efforts to recognize institutions that broadly support equity initiatives, including Project Juno\footnote{\url{https://www.iop.org/about/IOP-diversity-inclusion/project-juno}} and the Physics Inclusion Award\footnote{\url{https://www.iop.org/about/IOP-diversity-inclusion/physics-inclusion-award}}, the Athena Swan Charter\footnote{\url{https://www.advance-he.ac.uk/equality-charters/athena-swan-charter}}, and Sea Change\footnote{\url{https://seachange.aaas.org/}} run by the American Association for the Advancement of Science. Several members of the working group were unaware of one or more of these initiatives, highlighting the value of maintaining multiple access points to such information, potentially through professional societies such as the AAS or the IAU. 

In more local contexts, such as universities or physics and astronomy departments, the importance of standardized onboarding procedures was also emphasized. The level of institutional onboarding varied widely across participants' institutions: some offered no onboarding at any career stage, others focused primarily on students, while still others provided formalized materials for all career stages. Similarly, the extent to which onboarding materials included information about reporting bullying and/or harassment varied substantially. In some cases, working group members reported that departments provided robust logistical onboarding while omitting clear guidance on reporting bullying or harassment.

The working group defined \textbf{comprehensive onboarding materials, including explicit information on how to identify and report bullying and/or harassment, as a key way departments can directly help reduce these behaviors}. Because reporting structures and resources vary widely across academic institutions, the development of institution-specific onboarding materials\footnote{For example, the University of Chicago Department of Astronomy \& Astrophysics and Kavli Institute for Cosmological Physics maintain a ``Conflict Resolution Flowchart'', which shares details of possible interventions: \url{https://astrophysics.uchicago.edu/diversityinclusion/aa-kicp-conflict-resolution-flowchart}}
is essential to supporting new and junior colleagues. The working group further identified \textbf{a critical need for the creation of onboarding materials tailored specifically for postdocs and research scientists}, as these career stages are often overlooked in traditional university environments where most institutional structures are designed primarily for students and faculty.

\subsubsection{Letters of Recommendation}
Letters of recommendation remain an unavoidable (at present) component of career advancement within academia, despite several studies showing strong gender bias \cite[e.g.,][see also Section \ref{sec:resources}]{schmader2007linguistic,dutt2016gender, lin2019gender}. Beyond these biases, the working group identified letters of recommendation as a potential venue for bullying and harassment, as mentors hold immense power over the career trajectories of junior colleagues. Several members of the working group shared personal experiences, or those of close colleagues, where weak or negative letters appeared to influence career progression at the postdoctoral and faculty levels. 

Recognizing that letters of recommendation are unlikely to disappear from academic hiring and promotion processes, the working group discussed several strategies to mitigate the potential harm caused by negative or biased letters. An initial step identified was simply \textbf{increasing awareness that letters may be misused by bullies or harassers}--knowledge that some junior working group members did not previously have. The group also emphasized the importance of advising junior colleagues and mentees on how to seek confidential guidance from review committees or trusted senior colleagues when they suspect that a letter may be negatively affecting their career advancement.

Beyond sharing knowledge with junior colleagues, the working group discussed the potential value of \textbf{literacy training for review committees to help them identify signs of bullying or harassment, as well as cultural differences in letter-writing norms across countries}. This could include, but is not limited to, examining correlations (or lack thereof) between letters from direct mentors and close collaborators to identify anomalies, or contextualizing letters written in  cultures where `lukewarm' language by US standards may still signal strong support. 

Finally, \textbf{when an applicant has chosen not to request a letter from a direct mentor due to bullying and/or harassment, there should be a standardized mechanism for communicating this in the application process.} 
One possible implementation could be the addition of a field in job application systems allowing applicants to indicate that a direct mentor's letter was excluded for these reasons. Such an option could be incorporated into the AAS postdoctoral application guidelines\footnote{\url{https://aas.org/jobregister/postdoc-application-guidelines}} and, more broadly, into all jobs advertised on the AAS Job Register.

\subsubsection{Bystander trainings and creation of training materials on a wider scale}
\textbf{An important step in mitigating harassment in academic settings is training community members in how to respond when harassment occurs}--whether they experience it directly or witness it happening to a colleague. Academic working relationships (including mentor-mentee, professor-student, and peer-level interactions) introduce a variety of contexts in which witnessing or reporting harassment must be handled differently. As discussed above, access to clear information is essential for navigating this complex ecosystem. The working group therefore discussed a range of training materials that have either been helpful at their institutions or that they would like to see developed. 

Bystander trainings, in particular, were discussed extensively--both as resources that participants have found valuable and as trainings that others would strongly welcome.  \textbf{Group meetings, seminars, and conferences were identified as venues where explicit training on how to respond to inappropriate comments, microagressions, and direct attacks would be especially useful}. Indeed, \citet{HAYNESBARATZ2021100882} identify faculty bystander training as an effective tool for recognizing microagressions and building confidence in addressing them. While some academic institutions already provide such training, the working group also discussed the potential role of professional societies or large collaborations in organizing bystander training as part of workshops or conference programs. Training for session chairs--who may need to manage tense or uncomfortable situations during question-and-answer periods--was identified as particularly valuable. The development of a short video series based on realistic interactions was also discussed as a resource that could be widely shared with a single concentrated effort. Both the AAS and Astrobites were mentioned as potential venues for such an initiative. 

In addition to bystander training, participants discussed that \textbf{Title IX and mandated-reporter training is important for community members who hold relevant roles}. In some cases, training provided through Title IX offices helped working group members better understand when mandated reporting applies and what happens after a report is filed at their institution. Sharing this information with students and junior colleagues may be especially helpful, as individuals experiencing harassment are often reluctant to report unwanted behavior due to uncertainty about the consequences of doing so. The working group stressed that this information should ideally be communicated \emph{before} any incidents occur, so that students and junior researchers are not turned away, or surprised, at the moment where they most require assistance. Particularly important information includes the level of agency a reporter retains in the reporting process and how any reported information, including the reporter's identity, will be stored and used by the institution.

\subsubsection{Picture an Astronomer Hotline}
Building on the discussions above, the working group considered community-wide solutions to bullying and harassment that would allow the affected individuals to seek advice or assistance \textit{outside} their home institution or collaboration. These conversations led to the \textbf{proposed establishment of the `Picture an Astronomer Hotline' (PAH)}\footnote{name not binding}\textbf{: a confidential help desk staffed by trained volunteers who can provide resources or guidance to anyone who is, or believes they may be, experiencing bullying or harassment}. Such a resource could support concerns both small and large and, importantly, offer a point of contact who is \textit{not} a mandated reporter, allowing individuals to gather information and consider options safely. 

The working group recognized the significant effort required to develop such a resource but felt it worthwhile to outline a possible structure. A key priority is preserving the agency of the person seeking help, especially in cases where a PAH member may share an institutional affiliation with the requester. One possible implementation would be a website with a request form to select from a list of available PAH contacts, each linked to a brief profile. To reinforce confidentiality, the group suggested that PAH members use an  email account maintained specifically for the hotline, separate from any institutional affiliation. In addition, an automated reply confirming receipt of the request, and clearly setting expectations for how and when follow-up communication will occur, was identified as an important support feature.

Additional considerations included ensuring the PAH members receive training in navigating university, institute, and collaborative structures, as well as training related to secure data handling. While the group generally supported a principle of storing as little identifiable data as possible, it also recognized that clarity about what is (and is not) retained is essential. To help maintain continuity of support, the group suggested that PAH members could serve in scheduled shifts, recognizing the competing demands of academic travel and deadlines. 

Although an effort such as the PAH would require substantial resources, the group expressed the hope that the astronomical community would recognize the value. One possibility discussed was for the PAH to be housed as a multi-committee initiative within the AAS, where it could serve as a critical resource for the community. 

\bibliography{references/bullying}{}
\bibliographystyle{aasjournal}
\pagebreak
\urldef{\urlras}\url{https://ras.ac.uk/sites/default/files/2024-05/Final%20RAS%20Bullying%20and%20Harassment%20Report_digital_2024.pdf}

\section{Expanding the Universe of Opportunity: Building Diverse, Inclusive, and Accessible Communities in Astronomy and Physics} \label{sec:intersectional}

\vspace{-20pt}
\begin{center}
    \includegraphics[width=0.4\linewidth]{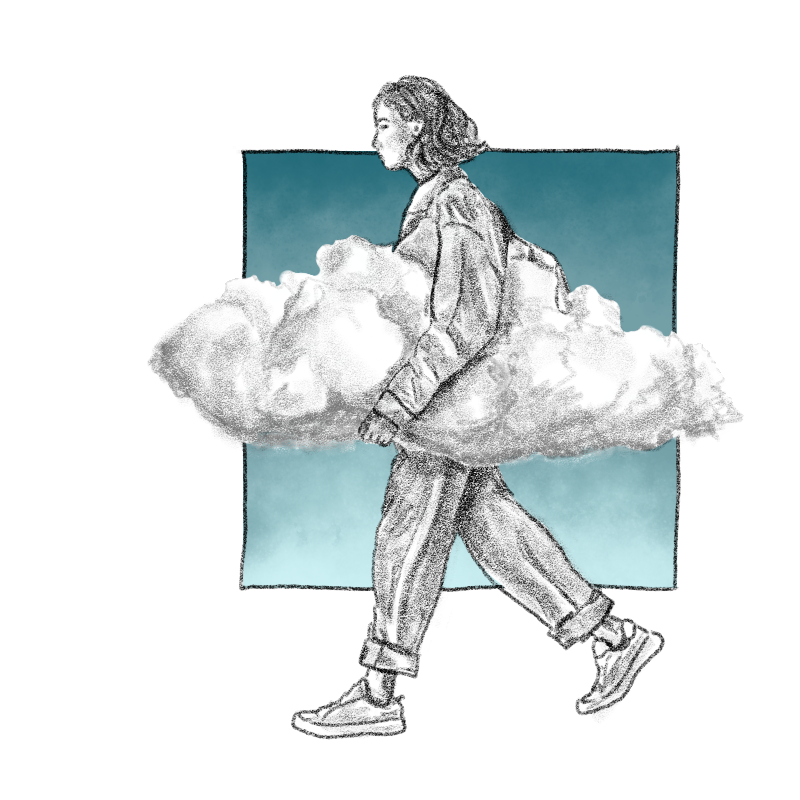}
\end{center}
\vspace{-30pt}

\begin{sectionauthor}
     \href{https://orcid.org/0009-0003-0415-404X}{Zuzanna Kocjan}$^1$, \href{https://orcid.org/0000-0002-4826-8642}{Sarah Biddle}$^{2}$, \href{https://orcid.org/0000-0002-1384-9949}{Tanvi Karwal}$^{3,4,5}$, and \href{https://orcid.org/0000-0002-8530-9765}{Lamiya A. Mowla}$^{6,7}$

\vspace{15pt}
\scriptsize{

$^{1}$ Department of Astronomy, University of Maryland, College Park, MD, USA\\
$^{2}$ Center for Astrophysics $|$ Harvard \& Smithsonian, Cambridge, MA, USA\\
$^3$ Department of Astronomy \& Astrophysics, University of Chicago, Chicago, IL, USA\\
$^4$ Kavli Institute for Cosmological Physics, University of Chicago, Chicago, IL, USA\\
$^{5}$ Enrico Fermi Institute, University of Chicago, Chicago, IL, USA \\
$^{6}$ Whitin Observatory, Department of Physics and Astronomy, Wellesley College, Wellesley, MA, USA\\
$^{7}$ Center for Astronomy, Space Science, and Astrophysics, Independent University Bangladesh, Dhaka, BD\\

}
\end{sectionauthor}
\vspace{20pt}

\subsection{Introduction}

It has long been established that creating inclusive and diverse spaces in STEM is not only a matter of equity but also essential for scientific progress and innovation; groups with greater diversity produce more creative and practical scientific and technical advances than homogeneous groups \citep{DiazGarcia_2012, Max_2013, Hofstra_2020}. However, in STEM--as in wider society--individuals continue to face discrimination based on gender, sexual orientation, race, socioeconomic background, disability, and other factors \citep{Richey_2020, noauthor_scourge_2024}.  Such experiences often lead to anxiety, burnout, depression, and decreased productivity. Research further shows that these harms are compounded for astronomers who hold multiple marginalized identities, with women of color facing the highest risks \citep{Clancy2017, prescodweinstein2017}. 

Discriminatory practices in STEM are not solely the result of individual actions; they are embedded in institutional structures, policies, and cultural norms. This structural dimension is reflected in large-scale survey data: a quantitative analysis conducted by researchers from the American Institute of Physics, using data from the Longitudinal Study of Astronomy Graduate Students (LSAGS) and the 2016 Survey of US AAS Members\footnote{\url{https://aip.brightspotcdn.com/a4/f3/22418fc159198396c5cb544e9dea/conferenceposter2017.pdf}}, found that underrepresented minority (URM) women are up to 20 times more likely than non-URM men to experience discrimination and harassment (20$\times$ among postdocs, 15$\times$ among non-postdocs). URM men and non-URM women also reported elevated risks compared with non-URM men. The study further notes that URM astronomers are more likely than Asian American or White astronomers to change advisors, and those who do so are more likely to leave the field. 

In 2022/23, the Royal Astronomical Society published results from a 2020 survey on bullying and harassment among astronomers, solar system scientists, and geophysicists\footnote{\url{https://ras.ac.uk/education-and-careers/bullying-and-harassment-astronomy-and-geophysics-report-and-recommendations}}. Within the two years prior to the survey, 44\% of the 661 respondents reported experienced bullying or harassment in the workplace. Scientists from traditionally underrepresented or marginalized groups--including women, Black, Indigenous, and People of Color (BIPOC), LGBTQ+\footnote{We use the term LGBTQ+ as an inclusive label that refers to lesbian, gay, bisexual, transgender, queer, questioning, intersex, asexual individuals, as well as others with marginalized orientations or gender identities.} individuals; and people with disabilities--were disproportionately affected and were 50\% more likely to report such incidents. 

Intersectionality highlights the fact that social categories such as gender, race, class, disability, and sexuality intersect to produce overlapping systems of disadvantage and discrimination \citep{crenshaw_mapping_1991}. Yet much of the existing STEM diversity research has focused narrowly on either gender--often within a binary framework--or race alone. To achieve meaningful inclusion and equity, our fields must embrace an intersectional approach; otherwise, policies risk leaving the most vulnerable groups unprotected. Moreover, adopting accessibility and inclusivity as guiding frameworks benefits everyone, not only those from traditionally marginalized groups, by creating environments in which all individuals have the opportunity to thrive.

The aim of this section is to offer evidence-based recommendations to physics and astronomy departments, highlight available resources to support these efforts, and compile relevant references from the rich existing literature on the topic. By translating research on discrimination and inequity into actionable strategies, departments can retain talented scientists and cultivate a more inclusive, supportive, and productive scientific community. While we recognize that there are areas of intersectionality we do not address here, due to limitations in authorship and perspective, we also acknowledge the significance of these experiences and honor the astronomers whose voices and perspectives are central to understanding them.

\subsection{Preventing racial discrimination in astronomy}

To date, multiple studies have documented the underrepresentation and marginalization of people of color in astronomy and physics, each highlighting different dimensions of systemic inequity. For example, Ko, Kachchaf, Ong, and Hodari’s ``Narratives of the Double Bind'' \citep{ko_narratives_2013} examined the life stories of 23 women of color in physics and astronomy by analyzing 41 written texts and conducting 10 oral interviews. These narratives reveal inequities that numbers alone cannot capture, such as the ``Double Bind'' experience--the exhausting tension of trying to discern whether negative remarks are motivated by racism, sexism, or both. One of their respondents, Evelynn Hammonds, framed it: 
\begin{mdframed}[
  leftline=true,
  rightline=false,
  topline=false,
  bottomline=false,
  linecolor=gray,
  linewidth=2pt,
  innerleftmargin=10pt,
  innerrightmargin=5pt,
  innertopmargin=6pt,
  innerbottommargin=6pt
]
\textit{``[Race and gender] aren’t separate in me. I am always black and female. I can’t say, `Well, that was just a sexist remark' without wondering would he have made the same sexist remark to a white woman. So, does that make it a racist, sexist remark? You know, I don’t know. And that takes a lot of energy to be constantly trying to figure out which one it is... somebody has some issues about me... being black, female, and wanting to do science and be taken seriously.''}
\end{mdframed}

The experiences of women of color in STEM cannot be reduced to either race or gender alone; instead, they are unique and intertwined, often creating heavier burdens than those encountered by white women or men of color. Through personal narratives, the paper also highlights the strategies of resilience that helped these women persist in physics, astrophysics, and astronomy--most notably activism and school/work-life balance. While these strategies often come with personal cost, both serve as a significant form of motivation, hope, and encouragement. Recognizing and valuing the labor that activism requires is crucial, as it both challenges systemic inequities and strengthens the community of women of color in STEM.

Between 2011 and 2015, \citet{Clancy2017} conducted an internet-based survey of 474 astronomers and planetary scientists and found that, across nearly every measure, women of color reported the highest rates of negative workplace experiences. Nearly 40\% said they felt unsafe at work because of their gender or sex, and 28\% felt unsafe because of their race. These experiences also led to exclusion from professional opportunities: 18\% of women of color and 12\% of white women skipped career-related events due to safety concerns. These findings reveal persistent systemic inequities at structural, cultural, and interpersonal levels, creating hostile environments that directly undermine career advancement. 

With this context in mind, the following actions can help faculty, students, and staff make astronomy and physics more inclusive and supportive for people of color:

\begin{itemize}

    \item \textbf{Focus on self-education \& accountability.} Change starts from within, so remember to address your own bias and actively educate yourself on equity, diversity and inclusion. An effective way to do so is by committing to attending training sessions, such as anti-bias or inclusive-mentorship workshops, and by reading work by underrepresented scholars. Regularly seek feedback from peers and colleagues about your behavior, mentoring practices, and classroom climate--it is important to be open to correction without defensiveness. 

    \item \textbf{Help with building students' physics identity and sense of belonging.} The American Institute of Physics' TEAM-UP Report \footnote{\url{https://aip.brightspotcdn.com/e3/c4/d4c805a50379da0014383538c204/teamup-full-report.pdf}} identifies structural barriers that affect African American physics and astronomy students. A crucial factor they highlight is physics identity, which is strongly shaped by role models and mentors: \textit{to persist, students must both see themselves--and be perceived by others--as future physicists and astronomers}. \citet{seyranian_longitudinal_2018} found that women reported lower sense of belonging and physics identification than men in an introductory physics course for STEM majors. \citet{rainey_race_2018} showed that women of color were the least likely to report a sense of belonging in STEM, with belonging shaped by interpersonal relationships, perceived competence, personal interest, and science identity. 
    These findings highlight that representation matters: role models sharing students’ racial or ethnic identities can be especially powerful supports.
    At the same time, in predominantly white departments, recognition from majority peers may unintentionally reinforce minority status. 
    Programs built by and for students demonstrate how community-driven recognition can better affirm belonging. Fostering diversity among faculty and empowering peer-led initiatives are therefore critical to cultivating physics identity and recognizing the success of underrepresented students.

    \item \textbf{Strengthen anti-harassment \& bullying policies.} The Committee on the Status of Women in Astronomy (CSWA) has published 26 recommended actions the AAS can take to address harassment and bullying in astronomy \citep{wexler_recommended_2025_II}. Their earlier survey \citep{wexler_recommended_2025_I} concluded that a major barrier to reducing harassment and bullying in astronomy is the limited understanding of how these behaviors manifest in practice.  
    Therefore, transparency must be prioritized; this includes regularly updating reporting procedures and follow-up protocols, clearly defining terms such as racial harassment, and ensuring that all claims are taken seriously.

     \item \textbf{Advocate for data transparency and accountability.} \citet{yang_liu_2025} argue that anti-discrimination efforts in the workspaces should be paired with employer transparency and accountability structures. Systematically collecting and publicly disclosing data on demographics, hiring practices, promotions, pay disparities, complaints, and retention can help identify inequitable patterns, track progress, and hold institutions accountable.

    \item \textbf{Support professional networks.} \citet{marin-spiotta_hostile_2020} highlight how accessible, active professional and social networks can transform scientific culture by providing mentorship, advocacy, and community--thereby reducing isolation caused by hostile environments. To maximize impact, such networks should prioritize racially diverse leadership and visibly recognize the contributions of leaders of color, ensuring students and early-career scientists see role models at the community level. 

    \item \textbf{Provide opportunities for systemic change within departments.} Lasting cultural transformation cannot depend solely on senior, and often less-diverse, faculty, who may hold more traditional views of support \citep{dancy_how_2023}. \citet{king_evading_2023} examined ``color-evasion'', the tendancy to avoid acknowledging race and racism, which can invalidate the lived experiences of students of color. They found that most instructors avoided discussing race even when it was clearly relevant.
    Without structures to elevate diverse voices, such as rotating leadership roles or empowered equity-focused committees, departments risk preserving exclusionary norms rather than enabling lasting changes. Departments should therefore establish mechanisms that allow students, postdocs, and junior faculty to meaningfully shape policy and  priorities. 
    \pagebreak
    \item \textbf{Distribute service fairly.} In ``The burden of service for faculty of color to achieve diversity and inclusion: the minority tax'', \citet{trejo_burden_2020} describes the disproportionate service loads assigned to underrepresented faculty, especially faculty of color, to support diversity and inclusion work. Such over-reliance often comes at the expense of research productivity and career advancement. As one anonymous astronomer noted, 
    \begin{mdframed}[
  leftline=true,
  rightline=false,
  topline=false,
  bottomline=false,
  linecolor=gray,
  linewidth=2pt,
  innerleftmargin=10pt,
  innerrightmargin=5pt,
  innertopmargin=6pt,
  innerbottommargin=6pt
]
\textit{``I have always felt that I have to do more to succeed in the field because I am a person of color. We are always seen as resilient, but there is little to no effort to make our work conditions better''}
\end{mdframed} 

This underscores how the emotional and professional toll of excessive service amplifies existing barriers. All faculty members--both underrepresented and well-represented--must share responsibility for building inclusive cultures.

    \item \textbf{Acknowledge the socioeconomic dimension of race.} Race and class intersect in ways that shape opportunity long before students arrive in our classrooms. \citet{whitcomb_not_2021} analyzed 10 years of institutional data on student GPAs 
    at a large public research university, categorized by gender, race/ethnicity, low-income status, and first-generation status. 
    They found that URM students earned lower overall and STEM GPAs than even the most disadvantaged non-URM peers, with additional penalties associated with socioeconomic disadvantage. Women in all demographic groups earned higher overall GPAs than men, except in STEM courses (aside from the most privileged groups; see extended discussion in Sections \ref{sec:bias}, \ref{sec:representation}, and \ref{sec:pedagogy}).  These results show how race, class, and gender intersect to shape academic outcomes. Mentorship, advising, and departmental policy must therefore account for socioeconomic diversity within racial groups--because guidance that seems straightforward from a privileged standpoint may be inaccessible or unsafe for students from working-class or first-generation backgrounds. 

\end{itemize}

While these recommendations address day-to-day inclusion work, it is important to remember that meaningful and lasting progress also requires coordinated institutional change and sustained attention to the lived experiences of underrepresented and marginalized astronomers.

\subsection{Ensuring the safety of LGBTQ+ astronomers}

Research shows that LGBTQ+ individuals in STEM fields regularly face increased risks of exclusionary and abusive behavior \citep{marosi_queer_2025}. \citet{Cech_2021} examined survey data from 21 STEM professional societies and found that LGBTQ+ professionals encounter higher rates of career limitations, harassment, and devaluation than their non-LGBTQ+ colleagues. These disparities, which appear consistently across disciplines and sectors, result in more frequent health difficulties and a higher likelihood of leaving the field. 

In 2016, the American Physical Society (APS) published a task force report on LGBTQ+ climate in physics \citep{APS_2016}, based on interviews, surveys, and focus groups. They found that a significant fraction of respondents reported uncomfortable or very uncomfortable departmental climates: 15\% of LGBT+ men, 25\% of LGBT+ women, 30\% of gender-nonconforming individuals, and 30\% of transgender respondents. Many individuals reported feeling isolated, lacking access to mentors or support networks. More than 20\% of respondents experienced exclusionary behavior in the past year (2015), while 40\% observed it; among transgender physicists, 49\% reported experiencing such behavior and 60\% reported observing it. Examples included sexual and verbal harassment, homophobic remarks, purposeful misgendering (calling a person by the incorrect pronouns), and exclusion from professional and social activities. Respondents with multiple marginalized identities faced higher levels of discrimination, with transgender people and gender minorities reporting the most hostile conditions.
Overall, more than one-third of respondents had considered leaving their positions during the previous year, often linked to bias they experienced or witnessed.

Reporting on the same \citet{Clancy2017} survey, \citet{Richey_2020} found that LGBTQ+ women and gender minorities experienced higher rates of verbal harassment related to sexual orientation and gender identity, and were twice as likely to face physical harassment based on gender or sex. Overall, queer women and gender minorities reported a more hostile workplace climate in astronomy and planetary science than their cisgender, heterosexual female peers.

These findings highlight that LGBTQ+ individuals face distinct systemic disadvantages in STEM, which highlights the need to examine the factors driving these outcomes. Faculty, students and staff can contribute to fostering safer and more supportive environments by taking the following practical steps: 

\begin{itemize}
    
    \item \textbf{Be an active bystander -- change starts with you.} Bystander intervention is proven to be a powerful and effective tool for addressing discrimination proactively and fighting against sexual harassment, bullying, and other forms of violence \citep{coker_rct_2017, inman_effectiveness_2018}. Yet \citet{lee_bystander_2024} found that only about 43\% of witnesses intervened during or after workplace sexual harassment. Decisions to intervene were strongly influenced by relationship to the victim and clarity of reporting systems. Clear, well-communicated reporting procedures therefore support bystander action. 

    \item \textbf{Respect individuals' chosen name and pronouns.} \citet{russell_chosen_2018} 
    found that allowing transgender youth to use their chosen names in multiple contexts significantly reduces depression, suicidal ideation, and suicide attempts, all of which are mental health risks known to be high in this group. Academic institutions should make it simple to update names and pronouns across records (e.g., diplomas, class rosters, department websites, etc) and ensure outdated data are not publicly displayed. 
    Instructors should collect students’ preferred names and pronouns at the start of the term. When unsure, colleagues or students can be asked respectfully, ``How would you like to be addressed?'' or ``What are your preferred pronouns?''.

    \item \textbf{Use gender-neutral and inclusive language.} \citet{he_debiasing_2025}
    found that replacing traditionally masculine wording in job advertisements with gender-neutral wording increases women's application rates. Using gender-neutral language affects the demographics of participation and leads to reduced bias in the workplace. Normalizing inclusive language, such as using `partner’, offering non-binary gender options on registration forms, and including pronouns in email signatures, helps reduce stigma and isolation. Providing gender-neutral bathrooms further supports safety and accessibility, particularly for transgender and non-binary people. Such actions can reduce anxiety, harassment, and health-related risks, while also signaling procedural fairness and fostering a more inclusive organizational climate \citep{chaney_gender_inclusive_2018, perales_access_2025}. However, as language evolves, it is important to stay open and receptive to feedback. 

    \item \textbf{Participate in and advocate for diversity trainings.} While findings on the effectiveness of anti-harassment or anti-bias trainings in workplaces are generally mixed, research shows that voluntary and solution-oriented initiatives with leadership engagement are likely to produce meaningful change \citep{dobbin_2025}. For example, \citet{perales_improving_2022} examined how employee trainings on gender and sexuality diversity and ally networks in workspace affect the comfort and safety of LGBTQ+ employees using data from a Australian large employer-employee survey of workplace inclusion. They found a strong, positive correlation between worker participation in these initiatives and increase in LGBTQ+ employee well-being index. To maximize effectiveness and promote systemic change, trainings should be recurring, emphasize skill-building (e.g., interrupting bias, inclusive communication), and be paired with structural reforms (e.g., transparent promotion pathways, inclusive hiring practices).

    \item \textbf{Create clear expectations for professional conduct.} \citet{foxx_evaluating_2019} reviewed 195 biology conferences in the US and Canada and found that only 24\% had codes of conduct. Among those, many were incomplete--43\% omitted sexual misconduct, 17\% omitted identity-based discrimination, 26\% lacked reporting mechanisms, and 35\% provided no consequences for violations. A lack of a clear code of conduct risks fostering environments where underrepresented participants feel unwelcome or unsafe, and in some cases, may even enable direct harm. Conference organizers, as well as departments and research groups alike, should establish explicit professional-conduct guidelines,  confidential reporting mechanisms, and clear accountability processes. This ensures that intervention is not left to individual discretion alone but is embedded in a transparent culture of accountability. Displaying codes of conduct publicly signals institutional commitment to inclusion.

    \item \textbf{Design inclusive mentoring and advising structures--and assess them regularly.} LGBTQ+ students in STEM often face isolation and may lack affirming mentors. \citet{charlton_development_2024}
    found that tailored mentor training significantly improved mentors’ skills in supporting mentees with identities different from their own. 
    Departments should provide identity-aware training, ensure multiple points of contact beyond primary advisors, and make LGBTQ+ resources visible. 
    Increasing the visibility of LGBTQ+ faculty and allies helps distribute ``cultural taxation''\footnote{Coined by Dr. Amado Padilla in 1994, cultural taxation describes the uncompensated, often unacknowledged labor expected of marginalized individuals, such as serving on diversity committees (see \citealt{padilla_ethnic_1994}).}. Regular climate surveys can provide feedback on these efforts, guiding continuous improvement and ensuring accountability.

    \item \textbf{Use tools and resources created by LGBTQ+ communities.} There are many online tools that can be used to create safer and more inclusive spaces in academia, examples include:
    
    \begin{itemize}
        \item The Astronomy and Astrophysics Outlist (maintained by the AAS Committee for Sexual-Orientation \& Gender Minorities in Astronomy, SGMA) contains a list of LGBTQ+ astronomers and allies who want to show their support for the community. Joining the list helps with increasing visibility, which helps normalize LGBTQ+ identities in academia and science, counteracting isolation and stigma.
        \item The Queer Speaker List (also maintained by SGMA) is a public list of LGBTQ+ astronomers who want to promote their research. It provides a resource for colloquia and seminar organizers to create a more diverse speaker list, increase visibility and normalize LGBTQ+ identities in astronomy, as well as help reducing the stigma.
        \item The Campus Pride Index\footnote{\url{https://www.campusprideindex.org}} is a US-only list of LGBTQ+ friendly colleges and universities, useful for LGBTQ+ students choosing institutions which they will apply. 
        \item The Social Acceptance of LGBTI People report: Using survey data from 175 different countries and locations, UCLA's School of Law produced a country list with Global Acceptance Index\footnote{\url{https://williamsinstitute.law.ucla.edu/publications/global-acceptance-index-lgbt/}}--a measure of the relative level of social acceptance of LGBTQ+ people and their rights. LGBTQ+ scientists may face additional safety concerns when traveling, particularly internationally. This tool can be used when deciding on or checking a conference location with LGBTQ+ protections in mind.
    \end{itemize}

\end{itemize}

For more comprehensive guidance, SGMA offers a more extensive guide, available online\footnote{\url{https://aas.org/sites/default/files/2019-09/LGBTInclusivityPhysicsAstronomy-BestPracticesGuide2ndEdn_small.pdf}}.

\subsection{Making the field accessible for astronomers with disabilities} 

\begin{figure}
    \centering
    \includegraphics[width=0.7\linewidth]{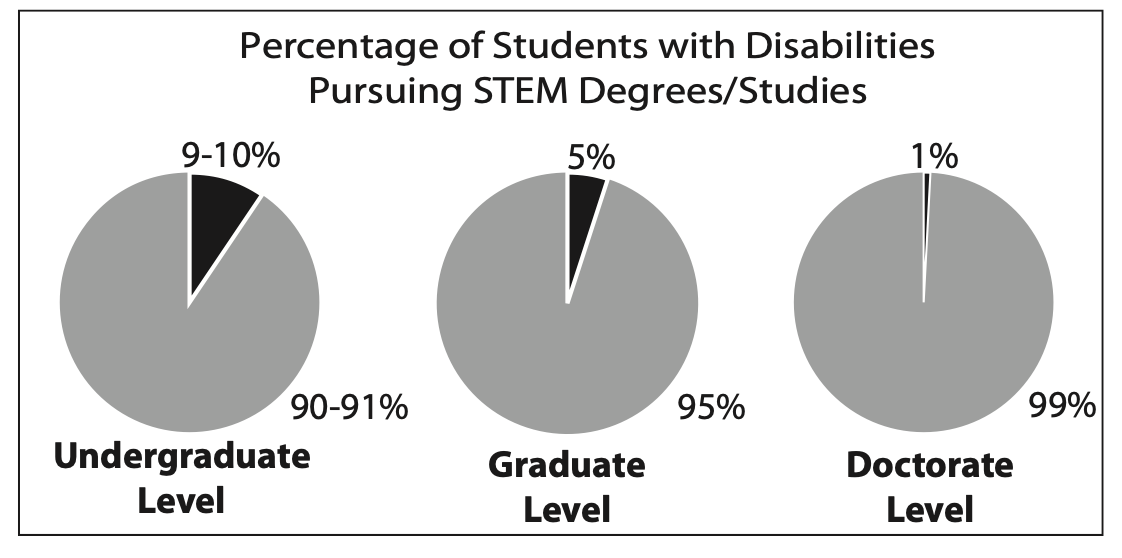}
    \caption{A pie chart showing the percentage of students with disabilities in STEM. At undergraduate level, the percentage varies from 9 to 10\%, at graduate level it is 5\%. Only 1\% of STEM doctorate students are disabled, showing that the fraction systematically decreases. Figure from \citet{moon_accommodating_nodate}, ``Accommodating Students with Disabilities in Science, Technology, Engineering, and Mathematics (STEM). Findings from Research and Practice for Middle Grades through University Education.'' SciTrain: Science and Math for All, Center for Assistive Technology and Environmental Access, George Institute of Technology, Atlanta, GA. }
    \label{fig:moon}
\end{figure}

On top of high rates of stigma and discrimination, disabled astronomers face persistent accessibility barriers and frequent failures of accommodation, which further contribute to exclusion. Obstacles occur across the field, from physical inaccessibility at conference venues to instructors being unprepared or insufficiently trained to recognize and support the needs of students with disabilities. As \citet{moon_accommodating_nodate} show, these barriers contribute to strikingly low representation of disabled students in STEM (Figure \ref{fig:moon}), with only about 1\% of PhD students identifying as disabled. 

In the article `Science Needs Neurodiversity' \citep{thorp_science_2024}, Herbert Holden Thorp emphasizes the importance of embracing neurodiversity within the scientific community. He argues that distinct ways of processing information and perceiving the world can spark novel scientific insights and provide alternative perspectives that might otherwise be inaccessible.  It is also important to recognize that many disabilities--including most forms of neurodiversity--are non-apparent, meaning they are not immediately visible to colleagues or instructors. This lack of outward visibility, combined with persistent stigma, contributes to the fact that many astronomers with mental or cognitive disabilities do not feel comfortable disclosing their disability status or seeking accommodations. Students, in particular, may hesitant to request accommodations because such requests are sometimes misinterpretted as attempts to gain an unfair advantage or as a sign of laziness \citep{shanahan_disability_2016}. 

Systematic changes focused on addressing discrimination and removing barriers are required to increase the participation of astronomers with both apparent and non-apparent disabilities and to create a consistently accessible and inclusive field. It is also crucial to remember that fostering a culture of inclusion in academia does not depend solely on large, long-term policy reforms--small, everyday practices play an equally important role, especially in the classroom. The Astro2020 white paper on accessible astronomy \citep{aarnio_astro2020_2019} identifies key obstacles faced by disabled astronomers and provides actionable strategies that can be implemented in both educational and professional spaces, including changes in teaching and mentoring methods, classroom organization, institutional culture, hiring practices, outreach, and workplace infrastructure, such as:

\begin{itemize}

    \item \textbf{Apply the rules of Universal Design.} The principles of Universal Design provide an overarching framework for creating environments, tools, and practices that are inclusive from the outset, rather than requiring later accommodations. Universal Design for Learning\footnote{\url{https://www.cast.org/what-we-do/universal-design-for-learning/}} (UDL) emphasizes providing multiple pathways for learners across three dimensions: representation (how information is represented), action and expression (how students can act and express their understanding), and engagement (how they engage with the material). For example, this could include offering varied forms of assessment, such as an exam, presentation, or project, so that students can select the option best aligned with their abilities, learning styles, and interests. By embedding UDL principles into teaching, research, and departmental culture, as well as treating Universal Design as a guiding framework rather than an afterthought, astronomers can promote inclusion of students and colleagues with disabilities and reduce barriers that may often otherwise go unnoticed by able-bodied or neurotypical astronomers \citep{quirk_applying_2025}. 

    \item \textbf{Design inclusive workspaces.} Physical and virtual spaces should anticipate diverse needs. This includes minimizing distractions in open-plan areas to support those with sensory sensitivities, offering meeting rooms that allow privacy, verifying that classrooms, conference venues, and departmental event spaces meet mobility, auditory, and visual access needs in advance. Similarly, it is crucial to ensure that public events, lectures, and outreach activities are accessible to individuals with disabilities, using accessible formats and providing necessary accommodations. Inclusivity is not a one-time achievement but an ongoing process, therefore these commitments should be reinforced through regular accessibility audits, with findings translated into actionable roadmaps that guide the removal of barriers over time.

    \item \textbf{Allow for hybrid participation.} Hybrid participation should be standard practice. Recording lectures and preparing comprehensive notes before or after class provides multiple entry points for students who cannot attend in person due to illness, disability, or other constraints. Department-wide events should offer virtual options (e.g., via Zoom, which includes transcription features) to support flexibility and broaden participation due to various circumstances. 
    
    \item \textbf{Ensure that class materials are accessible and transparent.} Course materials, papers, and presentations should be designed for readability and compatibility with assistive technologies. Use tools to check accessibility for individuals with color blindness or visual impairments (e.g., Color Oracle\footnote{\url{https://colororacle.org/}}). Ensure materials are compatible with screen readers and consider dyslexia-friendly formatting and fonts\footnote{e.g., recommendations at \url{https://www.weareteachers.com/best-fonts-for-dyslexia/}}. Provide text alternatives for figures, graphs, and videos to maximize accessibility. 
    
    \item \textbf{Design synchronous sessions with neurodiversity in mind.} Incorporate varied teaching strategies, such as combining short lectures, hands-on demonstrations, collaborative activities, and structured discussions. This approach helps accommodate different learning preferences and cognitive processing styles, supports sustained attention, and reduces cognitive overload. Providing clear agendas and predictable session structures can also help reduce anxiety. 

    \item \textbf{Establish clear learning goals.} Communicate learning objectives and expected outcomes at the start of the course and reiterate them at the beginning of each class. Clearly stated goals support self-regulation, provide a framework for note-taking, and help students (particularly those with executive function challenges) connect individual lessons to broader course outcomes. Making goals transparent also facilitates equitable assessment. 

    \item \textbf{Integrate accessibility into hiring and promotion criteria.} Departments should explicitly value demonstrated commitment to accessibility and inclusion in both hiring and promotion decisions. This can include experience with accessible teaching, mentorship of students with diverse needs, development of inclusive research tools, or leadership in accessibility initiatives. Job postings and review criteria should clearly signal that contributions to equity, diversity, and accessibility are considered alongside research and teaching accomplishments. Embedding these expectations into formal evaluation processes encourages faculty and staff to prioritize accessibility as a core, department-wide value.

\end{itemize}

\bibliography{references/intersectionality}{}
\bibliographystyle{aasjournal}
\pagebreak
\section{Building Global Community in Astronomy} \label{sec:global}

\vspace{-20pt}
\begin{center}
    \includegraphics[width=0.4\linewidth]{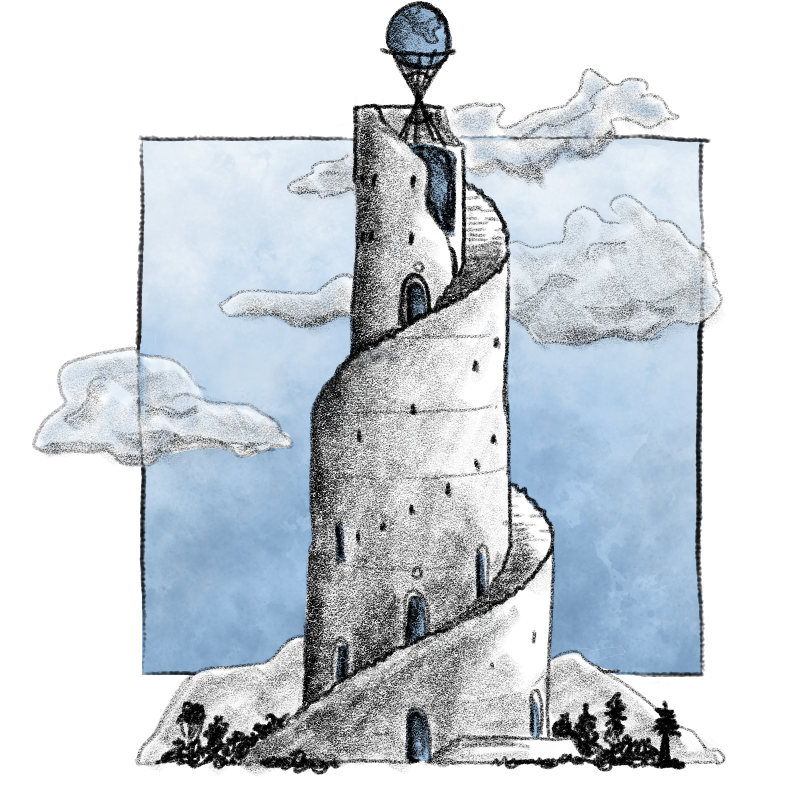}
\end{center}
\vspace{-30pt}

\begin{sectionauthor}
\normalsize{
\href{https://orcid.org/0000-0002-8530-9765}{Lamiya A. Mowla}$^{1,2}$, \href{https://orcid.org/0000-0003-0280-6617}{Aline Novais}$^3$, \href{https://orcid.org/0009-0007-8900-7178}{Camilla Nyhagen}$^{3}$, \href{https://orcid.org/0000-0002-0415-3077}{\'{A}lvaro Segovia Otero}$^{4}$, \\\href{https://orcid.org/0000-0002-4826-8642}{Sarah Biddle}$^{5}$, \href{https://orcid.org/0000-0002-1384-9949}{Tanvi Karwal}$^{6,7,8}$, and \href{https://orcid.org/0009-0003-0415-404X}{Zuzanna Kocjan}$^{9}$}
\\
\vspace{15pt}
\scriptsize{
$^{1}$ Whitin Observatory, Department of Physics and Astronomy, Wellesley College, Wellesley, MA, USA\\
$^{2}$ Center for Astronomy, Space Science, and Astrophysics, Independent University Bangladesh, Dhaka, BD\\
$^{3}$ Division of Astrophysics, Department of Physics, Lund University, Lund, SE\\
$^{4}$ Department of Astronomy, Tsinghua University, Beijing, PRC\\
$^{5}$ Center for Astrophysics $|$ Harvard \& Smithsonian, Cambridge, MA, USA\\
$^6$ Department of Astronomy \& Astrophysics, University of Chicago, Chicago, IL, USA\\
$^7$ Kavli Institute for Cosmological Physics, University of Chicago, Chicago, IL, USA\\
$^{8}$ Enrico Fermi Institute, University of Chicago, Chicago, IL, USA \\
$^{9}$ Department of Astronomy, University of Maryland, College Park, MD, USA\\
}
\end{sectionauthor}
\vspace{20pt}

\subsection{Introduction}

In astronomy, one of the oldest natural sciences, researchers from all over the world study the same Universe from different local contexts. Today, more than a century after its founding, the International Astronomical Union (IAU) brings together researchers from more than 90 countries to promote research through global collaborations\footnote{\url{https://www.iau.org/Iau/Iau/About/About.aspx}}. Many groundbreaking results in modern astronomy have only been possible through such global cooperation. One such discovery was of the accelerated expansion of the Universe \citep{RiessFilippenkoChallis.1998,PerlmutterAlderingGoldhaber.1999}, awarded the 2011 Nobel Prize in Physics, which was built on observations and analysis with contributions from teams on almost all continents. Global collaborations are essential for every aspect of astrophysics research, from building new observation facilities, performing large surveys to completing space missions.

However, participation in the global astronomy community is not equally accessible. Among the IAU member countries, the majority of countries do not have English as a primary language. Additionally, women are significantly underrepresented, with only 22\% of the IAU members being women\footnote{\url{https://www.iau.org/Iau/Iau/Geographical-and-Gender-Distribution.aspx}} and immense country-to-country variation. Latin American, Eastern European and South East Asian countries have overall higher proportions ($> 30\%$) of female astronomers compared to ``western'' countries.  Italy (31\%) and France (26\%) are exceptions within Western Europe and rank among the countries with the best overall gender balance in IAU membership. 

Even in countries often viewed as leaders in gender equality, such as Italy, France, the UK, and Germany, research documents persistent gender bias and elitism within astronomy. For instance, women remain underrepresented in permanent academic positions and are less likely than equally qualified men to be hired into these roles \citep{fohlmeister_careers_2014, berne_inequalities_2020}. 
Gender bias in the global astronomy community is also evident in scholarly recognition: a large study of publications from 1950–2015 across Astronomy \& Astrophysics, The Astrophysical Journal, Monthly Notices of the Royal Astronomical Society, Nature, and Science found that papers with women first authors receive approximately 10\% fewer citations than comparable papers with men first authors, independent of geographic origin \citep{caplar_quantitative_2017}. 

There are many challenges faced within the global astronomy community, some stemming from gender bias, others from cultural and structural inequities, and many emerging at the intersection of both.
While this chapter highlights several key challenges, it does so from the perspective of a small group of authors. We recognize that the experiences and perspectives represented here cannot fully capture the breadth and diversity faced by astronomers worldwide. The themes we discuss reflect only a subset of the many pressing global issues, and we acknowledge that important topics necessarily remain beyond the scope of this chapter. 

\subsection{Visas/Travel Considerations}

Maintaining immigration status is a critical and existential component of being an international scientist. 
For an appreciable fraction of international students and postdocs, returning to their home countries may mean the end of their academic careers, whether due to political instability, hostile relations between home and host countries, limited research opportunities, or discrimination based on gender or other identities
\citep[see e.g.,][for global statistics on gendered access to education]{unesco_gendered_atlas, unesco_heratlas}. 
These high stakes leave international scholars particularly vulnerable to exploitation, often with little recourse to protest. 
Greater visibility for international scholars and stronger international protections can help address some of these concerns. 

Universities and institutes should explicitly affirm their support for international scholars and protect them from retaliation if they participate in peaceful protests on university grounds. 
The cost of all immigration-related expenses, such as renewal of visas, should be reimbursed to all students, faculty and staff. Ideally, this should be codified in writing in offer letters. 
Because immigration paperwork is often complex, time-consuming, and highly consequential, institutions should employ trained staff whose explicit role is to assist scholars with documentation, filing, compliance, and communication with immigration authorities.

For astronomers from countries that require visas to enter North America, Europe, and Australia, attending conferences also poses a significant logistical and financial burden. The cost of obtaining a visa can range from \$100 to \$500 USD, depending on the country and visa type. In addition to application fees, many applicants must also travel to consulates for in-person interviews, incurring transportation and lodging expenses. Visa processing times vary widely, taking anywhere from a few weeks to several months, making it difficult to plan conference travel with certainty.

Visa applications typically require extensive documentation, including detailed five-year travel histories, bank statements, letters from employers and conference organizers, confirmed flight and hotel bookings, and, in some cases, proof of strong ties to the home country to demonstrate intent to return. For conferences in the Schengen area, visas are often granted only for the exact duration of the meeting, requiring repeated applications for each trip. Given that international collaborations are essential to astronomical research, the cumulative burden of visa applications fall disproportionately on early-career researchers from visa-restricted countries and can severely limit their ability to participate in the global scientific community.
Employing a `visa expert' at the university or department to assist in navigating visas for both employment and foreign travel can substantially reduce these barriers for international scholars.

\subsubsection{Recommendations for Conferences}
\label{subsec:conference_visas}

\begin{figure}
    \centering
    \includegraphics[width=\linewidth]{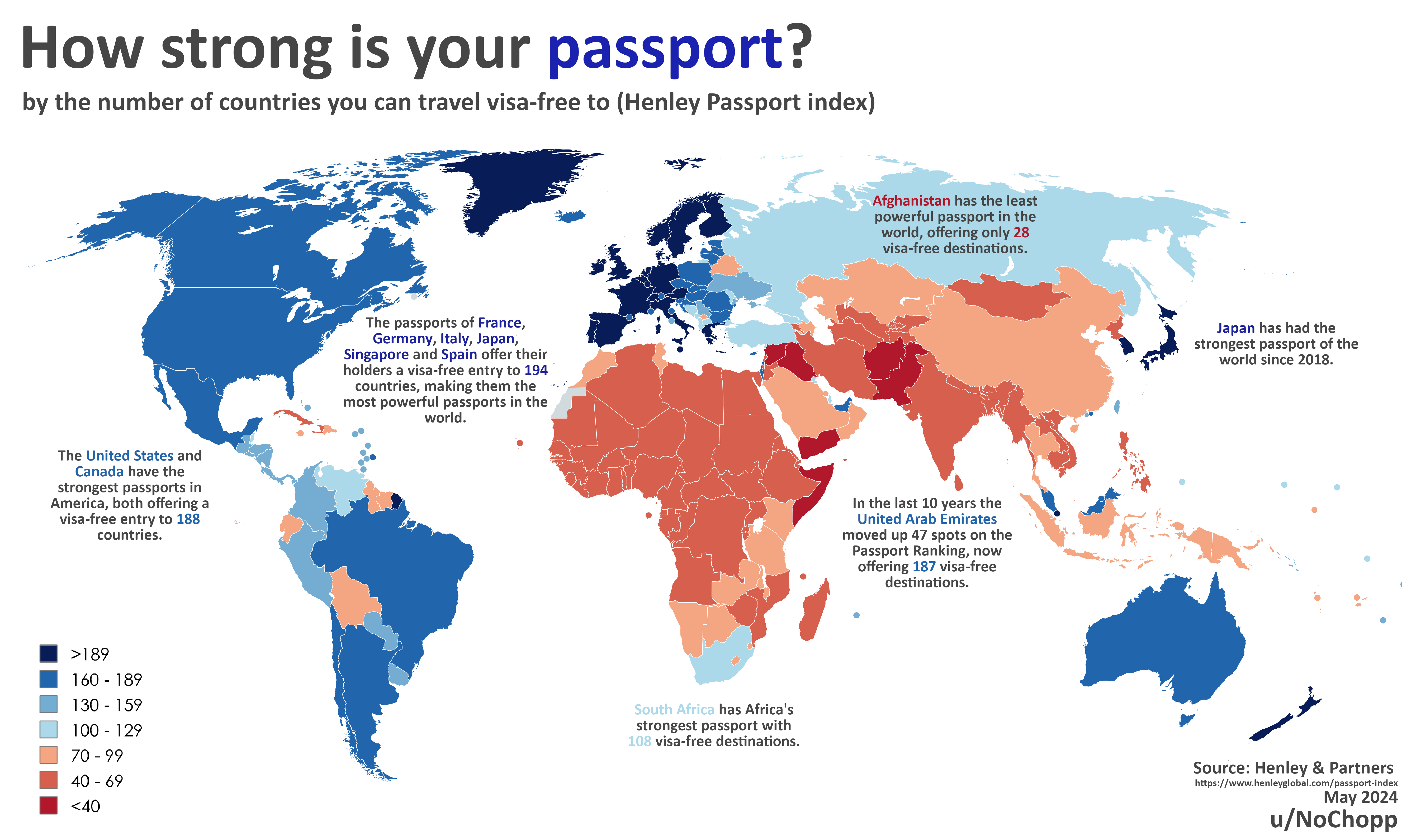}
    \caption{Ranking of passports by the number of countries the citizens can travel to visa free. Source: Henley Global, \url{https://www.henleyglobal.com/passport-index/ranking}}
    \label{fig:enter-label}
\end{figure}

The global center of science remains concentrated in the Northern and Western regions of the world, which can create structural exclusion for scientists from the Global South and East due to visa and travel barriers. Conference organizers can reduce these inequities by implementing the following best practices:

\begin{enumerate}
    \item \textbf{Early Invitation Letters} – Provide official invitation letters as early as possible, ideally six months in advance, to allow sufficient time for visa applications. The letter should explicitly state the applicant's role in the conference and any financial support offered, as well as contact information for the organizers. Enforcing an early deadline for responding to conference-talk applicants will also help internationals decide whether to undergo the visa application process which can be costly and time-consuming. 

    \item \textbf{Visa Support Desk} – Establish a dedicated visa support team to assist applicants. This could include offering guidance on required documents, providing letters addressed to embassies, and advocating for urgent processing if necessary. This can be shared across a few departments to reduce costs for creating such a resource. 

    \item \textbf{Flexible Refund and Virtual Participation Policies} – Given the uncertainty of visa approvals, conferences should offer full refunds for registration fees if a visa application is denied. Providing an option for virtual participation for those unable to attend in person ensures that visa-related issues do not entirely exclude researchers from engaging with the community.

    \item \textbf{Funding for Visa Costs} – Where possible, conference travel grants should include visa fees and travel expenses for visa appointments. This is particularly important for early-career researchers and students with limited financial resources. Visa costs should alternatively be covered by travel funds of scientists. 

    \item \textbf{Multi-Year Visa Facilitation} – Encourage institutions and funding agencies to support applications for longer-term visas (where available), reducing the frequency of visa applications. Some embassies consider letters from conferences when issuing long-term visas, so explicit support in this regard could be helpful. 

    \item \textbf{Embassy Engagement} – Work with embassies in key locations to advocate for smoother visa processes for academic travelers. Some large conferences have successfully coordinated with embassies to expedite applications or secure special consideration for attendees. 

    \item \textbf{Host Institution Support} – For conferences held at universities, the host institution can play a role in visa applications by issuing official letters, offering local support, and, where possible, helping arrange accommodations that meet visa requirements. 

    \item \textbf{Choosing Conference Locations with Fewer Visa Barriers} – When possible, organizers should consider hosting conferences in countries with more accessible visa policies for international attendees. Some regions have more flexible visa-on-arrival or e-visa options, which can significantly reduce the administrative and financial burden on participants. Rotating conferences among different continents can also improve time-averaged accessibility for researchers from diverse regions.

\end{enumerate}

\subsection{Admission and Hiring Advice for Applicants}

The astronomy and astrophysics community is global, and with that comes persistent biases concerning gender, ethnicity, class, religion, and other identities. These biases act at every level of academia--from individual interactions and group dynamics to the institutional structures that shape academic careers around the world. We therefore have the collective responsibility to address these inequities so that participation in the field is more accessible and sustainable for everyone. Globally, men continue to be favored in hiring (and even earlier in the process, since men are more frequently invited to apply for positions). This trend holds even in countries that are viewed as having better gender equity \citep{Moss-racusin, fohlmeister_careers_2014, Eaton2020, berne_inequalities_2020}. This section aims to address challenges that arise during the hiring process of academic jobs in science in the context of the field of astronomy and astrophysics, with a specific focus on demystifying implicit cultural considerations. 

A recent study carried out by Nature in collaboration with the research consultancy Thinks Insights \& Strategy\footnote{\url{https://www.thinksinsight.com/}} targeted over 1100 recruiters, over half of whom are in academia, with questions relating to hiring for STEM positions. To gain insights on current hiring practices worldwide, 77 countries were involved in answering the survey: 63\% of the respondents come from Europe or Central and North America, 24\% from Asia, and 12\% from Africa, South America, or Australasia. Far from unpacking all the findings from this survey, we focus on the main takeaways ordered according to the hiring process\footnote{A more comprehensive summary can, along with the full survey data, be found here: \url{https://figshare.com/s/7aea1f12aaca90a1030d?file=49646184}}, noting that only 11\% of respondents were in the physical sciences, so qualitative rather than quantitative conclusions are of the most use here:

\begin{itemize}
    \item \textbf{Before applying:} It is critical to build a professional network even before sending out job applications. In academia, 65\% of the recruiters rely on their own professional network, emphasizing interactions where the candidate has met the hirer face-to-face either by being invited to give a seminar at the host institution or during a conference \citep{linda_nordling_17oct24}. Instead of depending on your own or your supervisor's professional network, unsolicited job requests or \textit{cold emails} are likely to be considered by 37\% of the recruiters in academic science if they are appropriately tailored, e.g., whether messages feel personalized to the principal investigator and the research topic of the group, or are timed with the resolution of grants, other funding sources, or job vacancies. This already highlights one of the main issues of the \textit{leaky pipeline}, where recruitment based on personal networks and reputation amplifies and favors those already visible and connected \citep{brinkdragon2012}. Critically, women and people of color are less likely to receive response to their informal inquiries about potential opportunities \citep{Milkman2015-mh} even in organizations with better existing representation for these groups.

    \item \textbf{Screening applications:} The main key points considered when reviewing applications in science encompass experience, publications, and expertise in the field of research. For candidates with the same level of technical ability, soft skills such as communication, team work, personality, or passion are underlined as characteristics that ``tip the scale'' \citep{linda_nordling_21oct24}. As such evaluation is prone to subjectivity, some soft skills are highlighted more than others in different countries. For example, the UK and Germany particularly value time management and organizational skills. 

    \item \textbf{Interviewing:} Standing out and performing in a formal interview is the most important factor influencing hiring decisions. It is at this point that the interviewer can directly test if the applicant’s performance matches the stated experience \citep{linda_nordling_18nov24}. Around 45\% of hirers in science identify key mistakes during interviews and are twice as likely to say their overconfident interviewees lack creative thinking or knowledge about their field of research \citep{linda_nordling_17oct24}. It is also during interviews when soft skill gaps often reveal themselves. Nowadays, the nature of the scientific method is based on research teams with a larger number of, and more diverse,  members, so candidates must be flexible and adapt to different contexts and working dynamics (seminars, informal discussions, working for long hours). Women's productivity metrics are often considered decisive as caregiving and childbirth are expected to cause interruptions in their long-term careers, papers and citations, and grant applications \citep{nicolo_romano_2021}.
    
    \item \textbf{Job offer:} Nature reports a trend in ever-growing frustration among early-career researchers who no longer see themselves pursuing a long term career in science \citep{nature_postdocs,linda_nordling_17nov25}. Even if they accept the offer, it has become a pressing issue to retain talent. Academia loses qualified candidates to industry, who offer more attractive salaries, reduce living-cost concerns, and help administrative processes including visa applications or renewals \citep{linda_nordling_4dec24}. The fact that academia is less predictable than industry directly impacts women; as they are often in the position of being primary caregivers, the absence of childcare support, parental leave, and, in cases, dual-career constraints will continue reinforcing long-term gender disparities, specially in countries with a poor welfare state \citep{gronlund10}. In addition, restrictive visa systems may disproportionately affect women researchers \citep{SeattleTimes, awis}.

\end{itemize}

\subsection{Disparities in Proposal Evaluations}

Observations are a fundamental part of studying the Universe.
To obtain telescope time at an observatory, astronomers must submit observing proposals that clearly motivate and justify their scientific goals. These proposals are then evaluated through a peer-review process, typically carried out by volunteer members of the astronomy community \cite{carpenter_systematics_2020}. Because successful proposals determine the scientific directions pursued in 
the near future, it is essential that the review process is as fair and unbiased as possible \citep{carpenter_update_2022}. 

In the last decade, multiple facilities have released studies showing disparities in the evaluation of telescope proposals. For the Hubble Space Telescope (HST), for example, \cite{reid_gender-correlated_2014} found that female principal investigators (PIs) had a lower acceptance rate than those led by men. In response, the Space Telescope Science Institute introduced new procedures for the HST proposal review, including dual-anonymous peer review, n which both the identities of the proposal team are concealed from reviewers and the identities of the reviewers are not shared with the proposers. 
Following the adoption of dual-anonymous review, proposals led by women PIs  have been accepted at significantly higher rates, effectively removing the previously observed gender gap \citep{johnson_dual-anonymization_2020}. 

After the HST study, European Southern Observatory (ESO) and National Radio Astronomy Observatory (NRAO) published statistics from their proposal review processes. At ESO, \cite{patat_gender_2016} found that observing proposals submitted by women had a lower acceptance rate than those submitted by men. Although some of this difference was attributed to the lower proportion of senior women PIs, a gender gap remained even after accounting for seniority.
For the NRAO-operated facilities, including Jansky Very Large Array, the Very Long Baseline Array, the Green Bank Telescope, and the Atacama Large Millimeter/submillimeter Array (ALMA), \cite{lonsdale_gender-related_2016} reported that proposals led by men were more likely to receive higher rankings than those led by women. In response, NRAO began informing review committees about gender bias and increased female representation among reviewers.  

Beyond gender, \cite{carpenter_systematics_2020} examined proposal outcomes at ALMA with respect to regional affiliation. They found that the proposals from PIs based in European and North American institutions were ranked higher, on average, than those from PIs working in Chile, East Asia, and other regions. Although the source of the disparity is unclear, the authors speculated that language may play a role, since most affected PIs are not native English speakers. 
They also found that, across regions and even after correcting for seniority, proposals from women tended to receive lower grades than expected, while those from men tended to receive higher grades, although this effect was not statistically significant (see their Table 5 and Figure 14). 

Following the findings by \cite{carpenter_systematics_2020}, ALMA implemented a series of changes to reduce potential bias in the proposal review process, including dual-anonymous review and randomization of the investigator list to minimize name recognition effects. An updated analysis of ALMA Cycles 7 and 8 by \cite{carpenter_update_2022} showed evidence that name recognition contributed to bias in earlier cycles. The study also confirmed that proposals from PIs based outside Europe and North America continued to receive lower rankings, 
though the disparity between Chile-based and North American PIs narrowed somewhat. Variations in acceptance rates between men and women from cycle to cycle were not statistically significant, making it difficult to attribute trends solely to the changes in the review processes. However, in Cycles 7 and 8, after dual-anonymous review and PI randomization were introduced, proposals led by women PIs were accepted at higher rates consistent with expectations based on proposal grading, suggesting progress toward equity \citep[see Figure 13 in][]{carpenter_update_2022}. 

Although there is still much work to be done to fully eliminate bias in proposal review, these studies collectively demonstrate that review processes which emphasize the proposed science--while reducing emphasis on the identities and reputations of proposers--can improve equity in proposal outcomes. 

\subsection{Astronomy in a Second Language}

Within the astronomy community, communication happens at many different levels: informal discussions with colleagues over morning coffee, email exchanges with collaborators across institutions and countries, journal articles, grant proposals, and conference presentations. While the different forms of communication can be carried out in different languages, effective collaboration requires a shared language \citep{Mahoney2000}. 
In astronomy, as in much of science, that common language, or \textit{lingua franca}\footnote{Lingua franca refers to a common language used by a group of people speaking different languages.}, is English. It is the dominant language of major astrophysics journals, and by the early twenty-first century, the majority of doctoral theses in astronomy were also being written in English \citep{tenn_astrogen_2023}.

However, many countries with large and active astronomy communities do not have English as a primary national language. The IAU currently includes members from more than 90 countries, most of which are non-English-speaking. Some countries, such as the Netherlands, Sweden, and Germany, have high overall English proficiency. Others, including India, China and Japan, also have strong astronomy programs but lower average English proficiency\footnote{\url{https://www.ef.com/assetscdn/WIBIwq6RdJvcD9bc8RMd/cefcom-epi-site/reports/2024/ef-epi-2024-english.pdf}}. Even within individual countries, the language of instruction varies, with some universities teaching undergraduate courses in English and others in national languages. As a result, English fluency among astronomers worldwide spans a very wide range. 

This variation does not affect all astronomers equally.  Women and researchers from countries where English is not the primary language face additional barriers. Recent survey data from \cite{Amano2023, amano_who_2025}, including 908 environmental scientists from eight countries, show that women publish up to 45\% fewer English‑language papers than men\footnote{Since the survey by \cite{Amano2023} focused on environmental scientists, the numbers do not necessarily represent the exact situation in the astronomy community. We therefore use their numbers as guidelines and assume there are \textit{parallels} from which we can draw inspiration and conclusions.}. The discrepancy is larger when considering women who are non-native English speakers, who publish up to 60\% fewer English-language papers than male native English speakers from high‑income contexts \citep{amano_who_2025}. Formally, gendered disparities in publishing behaviors are often referred to as the ``productivity gap'' and are primarily attributed to differences in expectations around institutional service and family caregiving, as well as overt sexism \citep[for instance, women are less likely to be offered authorship as recognition for their work than their male counterparts;][]{Ross2022}, though there are also indications that in spite of these effects women have more consistent performance over a career and the gap is instead attributable to increased stochasticity in year-to-year research output \citep{ZHENG2025} and shorter careers overall \citep{doi:10.1073/pnas.1914221117}.

\citet{Bohm2023} examined in detail the impact of COVID-19 on gendered research productivity in astronomy, finding that the ``productivity gap'' widened \textit{globally} during the pandemic, with little apparent correlation with country GDP or spoken language, likely due to a loss of easy access to professional networks and increased caregiving responsibilities \citep[e.g.,][]{Peetz02092023}. Individual institutes were able to retain more parity in productivity \citep{Tran2025Achieving}, suggesting that institutional and community support can play a key role in facilitating tangible research output.

\subsubsection{Challenges for Non-native English Speakers and Recommendations for the Global Community}
While having a common language opens up a world of possibilities for collaborations across borders, there are challenges that occur for researchers that are from countries where English is not the primary language. We will refer to these researchers as non-native English speakers for the remainder of this section.
\begin{itemize}
    \item \textbf{Communicating your science at conferences:} Giving talks in English as a non-native English speaker adds another layer of insecurity to whether you speak ``good enough'' English. Additionally, the allocated time for talks at conferences are often the same for every researcher, regardless of English proficiency, meaning that a speaker with lower proficiency in English effectively get less time to present their work because they need more time when talking in English \citep{Lenharo2023}. These barriers can potentially discourage non-native English speakers from presenting their work, asking questions, or even participating at conferences altogether.

    In preparation for conference talks and networking, individual research groups and collaborators can offer support by helping English editing and preparation of presentations. At conferences, promoting multilingual presentations can reduce the pressure to perform in perfect English \citep{pavon_recommendations_2025}. Additionally, ensuring gender‑balanced panels and speaker lists can help mitigate ``stereotype threat'' (see Section \ref{sec:representation}) among presenters that adds to the feeling of pressure.
    \item \textbf{Peer review and publishing:} Few astronomers receive any formal training in how to write academic papers, and for non-native English speakers this challenge is compounded by the fact that most leading astrophysics journals require manuscripts to be written in English. \cite{Amano2023} performed a survey to investigate the barriers faced by non-native English speakers in science when it comes to reading and publishing research. In addition to their results showing that significantly more time is needed for reading, writing, and publishing papers, it became clear that researchers from countries with lower proficiency in English had their papers rejected 2.5 times more often due to language-specific problems. When gender is taken into account, these barriers become even more pronounced: as noted above, women publish up to 45\% fewer English‑language papers than men, and women who are non‑native English speakers from lower‑income countries publish up to 60–70\% fewer English‑language papers than male native English speakers from high‑income communities \citep{amano_who_2025}.
    
    It is therefore important that research groups and the global astronomy community explicitly acknowledge the extra time and effort required from non‑native English speakers to publish papers written in English and take steps to mitigate the largely community-driven ``productivity gap'' for female researchers. On the journal level, double‑blinded reviews helps reduce bias linked to gender and nationality, and there should also be clear guidelines on separating the language and scientific content. Where possible, editing and language support should be offered to authors. This may be especially effective when targeted toward early‑career women from non-native English‑speaking countries, who face additional systemic barriers to both funding and publication. Offering complementary in-house editing services through journals, or subsidizing the cost of external proofreading, can facilitate publication without placing an additional financial burden on authors.

    \item \textbf{Science communication in native languages:} For researchers whose first language is not English, conducting research solely entirely in English can make it difficult to translate technical ideas back into their native language, especially when key concepts and technical terms were first learned and practiced in English. Surveys of non‑native English‑speaking researchers show that they invest substantial extra effort not only in reading, writing, and presenting in English, but also in sharing research in multiple languages, which includes translating their own work for local audiences \citep{Amano2023}. This added burden can discourage some researchers from engaging in public communication in their native language, which in turn makes astronomy less accessible to the broader public. 
    For female non‑native English speakers, these challenges intersect with gendered expectations around outreach and public engagement, as studies have shown that more women than men participate in outreach activities and science communication \citep{wilkinson_roles_2022, eizmendi-iraola_gender_2025}.  Supporting a more equitable distribution of outreach responsibilities is therefore important. In parallel, creating resources such as astrophysics glossaries, translated summaries of scientific findings, and public talks in multiple languages can meaningfully assist non-native English-speaking researchers. 
\end{itemize}

\subsection{Publication Fees and Access to Financial Support}

A common challenge faced by astronomy researchers based in countries outside Europe and North America is the extremely high publication fees charged by the main high-impact, peer-reviewed astronomy and astrophysics journals. This challenge is exacerbated for female researchers, who are both less likely to receive research funding and are awarded funding in smaller amounts than their male counterparts \citep[e.g.,][across disciplines, career stage, and country of work]{BORNMANN2007226, doi:10.1073/pnas.1510159112, guardian, 10.1001/jama.2018.21944, Ma2019-gj, Schmaling2023, 10.1162/QSS.a.18}. Stretching available resources farther can become a necessity. The costs to publish a paper may be charged per page or quanta or as Article Processing Charges (APCs), the latter introduced with the rise of open-access (OA) publishing \citep{SegadoBoj+22}. Table \ref{table:publicationfees} summarizes the publication costs of major journals in astrophysics.\\

\begin{table}[!h]
\resizebox{\textwidth}{!}{%

\begin{tabular}{|c|l|c|}
\hline
\textbf{Journal} & \textbf{Charge (as of December 2025)} & \textbf{Reference} \\ \hline
\begin{tabular}[c]{@{}c@{}}ApJ, AJ,\\ and ApJS\end{tabular} & \begin{tabular}[c]{@{}l@{}}1\,357 USD ($\leq$ 30 quanta)\\ 3\,011 USD (31–50 quanta)\\ 5\,315 USD (51–100 quanta)\\ 250 USD (extra quanta $>$ 100)\end{tabular} & {[1]} \\ \hline
ApJ Letters & \begin{tabular}[c]{@{}l@{}}2\,836 USD ($\leq$ 40 quanta)\\ 450 USD (extra quanta $>$ 40)\end{tabular} & {[1]} \\ \hline
MNRAS & 2\,356 GBP & {[2]} \\ \hline
MNRAS Letters & 1\,122 GBP & {[2]} \\ \hline
A\&A & \begin{tabular}[c]{@{}l@{}}150 EUR per page (up to 12 pages)\\ 200 EUR per extra page\end{tabular} & {[3]} \\ \hline
A\&A Letters & \begin{tabular}[c]{@{}l@{}}150 EUR per page (up to 4 pages)\\ 200 EUR per extra page\end{tabular} & {[3]} \\ \hline
\begin{tabular}[c]{@{}c@{}}Nature and\\ Nature Astronomy\end{tabular} & \begin{tabular}[c]{@{}l@{}}Free if not OA\\9\,190 GBP\,/\,12\,690 USD\,/\,10\,690 EUR (OA)\end{tabular} & \begin{tabular}[c]{@{}l@{}}{[4]} \\ {[5]} \end{tabular} \\ \hline
Nature Communications & 5\,290 GBP\,/\,6\,990 USD\,/\,5\,890 EUR (OA only) & {[6]} \\ \hline
OJAp & Free & {[7]} \\ \hline
PASA, PASA Letters & Free & {[8]} \\ \hline
PASP & \begin{tabular}[c]{@{}l@{}} 129 USD per page\\ 3\,325 USD (OA)\end{tabular} & {[9]} \\ \hline
PRD & \begin{tabular}[c]{@{}l@{}} Free if not OA\\ 2\,840 USD (OA)\end{tabular} & {[10]} \\ \hline
PRL & \begin{tabular}[c]{@{}l@{}} Free (optional 830 USD charge) if not OA\\ 3\,980 USD (OA)\end{tabular} & {[10]} \\ \hline
\end{tabular}
}
{[1]}: \hspace{1pt} \url{https://journals.aas.org/article-charges-and-copyright}\\
{[2]}: \hspace{1pt} \url{https://academic.oup.com/rasjournals/pages/read-and-publish}\\
{[3]}: \hspace{1pt} \url{https://www.aanda.org/for-authors/author-information/page-charges}\\
{[4]}: \hspace{1pt} \url{https://www.nature.com/nature/for-authors/publishing-options}\\
{[5]}: \hspace{1pt} \url{https://www.nature.com/natastron/submission-guidelines/publishing-options}\\
{[6]}: \hspace{1pt} \url{https://www.nature.com/ncomms/open-access}\\
{[7]}: \hspace{1pt} \url{https://astro.theoj.org}\\
{[8]}: \hspace{1pt} \url{https://www.cambridge.org/core/journals/publications-of-the-astronomical-society-of-australia}\\
{[9]}: \hspace{1pt} \url{https://iopscience.iop.org/journal/1538-3873/page/publication-charges}\\
{[10]}: \url{https://journals.aps.org/authors/apcs}\\
\caption{Publication charges of the main astronomy and astrophysics peer-reviewed journals, as of 14 December 2025. ``OA'' refers to open access publication.}
\label{table:publicationfees}
\end{table}

\begin{itemize}
    \item \textbf{ApJ, AJ, ApJS, and ApJ Letters:} These are full open access journals. The Astrophysical Journal (including Letters and Supplements) and Astronomical Journal offer a 15\% discount\footnote{\url{https://aas.org/join/author-discounts}} on the APC if the first author or any of the co-authors is a member of the American Astronomical Society (AAS). AAS Publishing also has a generous fee waiver policy\footnote{\url{https://journals.aas.org/support}}, and priority is given to authors with no funding, including those from countries or institutions with limited resources. Financial support decisions are made concurrently with editorial decisions.
    \item \textbf{MNRAS and MNRAS Letters:} These are full open access journals. The Monthly Notices of the Royal Astronomical Society (and its Letters section) offer a 20\% discount to the APC\footnote{\url{https://academic.oup.com/mnras/pages/mnras-open-access}} if the corresponding author is a member of the Royal Astronomical Society. In case the primary affiliation of the corresponding author is a university or research institution that has an agreement\footnote{\url{https://academic.oup.com/pages/open-research/read-and-publish-agreements\#Institutions}} with the Oxford University Press, the APC may be fully payed by the institution. In addition, they also offer a full APC waiver which is automatically applied if the corresponding author is based in countries considered ``low and middle income countries\footnote{\url{https://academic.oup.com/pages/purchasing/low-and-middle-income-countries-initiative/participating-countries-and-regions}}''.
    \item \textbf{A\&A and A\&A Letters:} These are hybrid open access journals, published under a Subscribe to Open (S2O) model, which means that the content becomes open access if enough subscriptions to the journals are renewed each year. Astronomy \& Astrophysics and Astronomy \& Astrophysics Letters offer a full waiver to the page charges if the primary affiliation of the first author is a university or research institution located in an A\&A sponsoring country\footnote{\url{https://www.aanda.org/about-aa/copyright/184-about-aa}}. They also offer a full waiver if the first author is affiliated with one of the countries in Group A of Research4Life\footnote{\url{https://www.research4life.org/access/eligibility}}. In addition, a discount of 50 EUR to the cost per page and the cost per extra page is offered if the first author is not in a A\&A sponsoring country but is in a subscribing institute [3].
    \item \textbf{Nature and Nature Astronomy:} These are hybrid journals. Publications are are free of charge if the publishing option is not open access. If the author wishes to publish open access, a full APC waiver\footnote{\url{https://www.springernature.com/gp/open-science/policies/journal-policies/apc-waiver-countries},\url{https://datahelpdesk.worldbank.org/knowledgebase/articles/906519}} is offered if the corresponding author is based in a country classified as ``low-income economy''. If the corresponding author is based in a country classified as ``lower-middle-income economy'', they are eligible for a 50\% discount\footnote{\url{https://datacatalog.worldbank.org/search/dataset/0038130}}. In case the primary affiliation of the corresponding author is a university or research institution that has an agreement with Springer Nature, the institution may fully or partially cover the APC.
    \item \textbf{Nature Communications:} This journal offers the same discounts, waivers or institution agreements as Nature and Nature Astronomy: However, it is fully open access, with no ``free publication'' option (as available in Nature and Nature Astronomy). 
    \item \textbf{OJAp:} The Open Journal of Astrophysics is a ``diamond'' open access arXiv overlay journal, with no associated charges to authors or readers. The whole operating budget of OJAp is comparable to a single society journal APC.
    \item \textbf{PASA and PASA Letters:} The Publications of the Astronomical Society of Australia and PASA Letters are open access publications and do not have page charges.
    \item \textbf{PASP:} The Publications of the Astronomical Society of the Pacific is hybrid open access with APCs charged for open access publication and page charges for all other publication. APCs are waived for authors at institutions in Transformative Agreements\footnote{\url{https://publishingsupport.iopscience.iop.org/questions/institutional-open-access-agreements}} with the IOP Science, and waivers and discounts are guaranteed for authors from low-income and lower-middle income countries\footnote{\url{https://publishingsupport.iopscience.iop.org/questions/paying-for-open-access}} respectively.
    \item \textbf{PRD and PRL:} The Physical Review D and Physical Review Letters are hybrid open access journals with APCs for open access publication. These APCs may be waived\footnote{\url{https://www.aps.org/publications/less-resourced-countries}} for ``less-resourced countries''. There is an additional charge associated with the publication of color figures in the print journal, but no such charge is incurred for online-only publication.
\end{itemize}

Woman-led papers receive fewer citations when corrected for manuscript content and quality \citep[e.g.,][]{caplar_quantitative_2017,TeichKimLynn.2022} and garner less publicity and media recognition \citep[e.g.,][]{doi:10.1073/pnas.2102945118, Arabi2025}. Notably, publishing research open access leads to greater visibility for, and engagement with, the scholarly work \citep[e.g.,][]{Lawrence2001, Adie2014, Piwowar2018-ru, Huang2024}. Relying on a free tier of closed access publication to broaden who publishes in a particular journal may only serve to widen the gap in who receives recognition and citations.

\subsubsection{Challenges}

All journals enumerated above offer the possibility to request a full waiver or a discount to publication costs at submission if the first/corresponding author is based in an low-income or lower-middle-income country, often referred to as ``developing'' countries. This may also be requested in case the first author is able to demonstrate financial hardship. While this initiative seems helpful at first, waivers or discounts are still not an effective solution to reduce the inequality within scientific publishing on their own.

Waivers are evaluated on a case-by-case basis, which does not guarantee the discount will be granted. In fact, this decision can depend on many factors such as the available budget, and/or the number of waiver requests that journal has recently received or approved. Furthermore, these waivers should be requested by the corresponding author at the time of submission, yet they are often only granted \textit{after} the manuscript is accepted for publication. This means that the manuscript will go through revision with the uncertainty that the waiver will be granted and so that the manuscript will even be published if accepted. It is important to note that these waivers may be automatically granted to authors in lower-middle-income countries, while many other middle-income countries may not have access to the same support.

Another challenge is that the journals may request a declaration of financial hardship in order to consider the possibility to grant a waiver or discount. This means that corresponding authors must explain the lack of funding and likely provide a declaration from their primary institution, which may feel intrusive, embarrassing, and/or humiliating for the authors, potentially discouraging waiver requests. Even when in need of financial support, some institutions may not want to offer a declaration of financial hardship, instead discouraging authors from publishing their research in expensive high-impact, OA journals. These implicit limitations create a further structural barrier and can discourage authors from submitting to these main journals, disadvantaging underfunded researchers and, particularly, early career scientists whose career outcomes rely on publication in recognized outlets.

\subsubsection{Recommendations for Professional Journals}

Studies show that simply allowing authors to request waivers for article-processing charges does not fully eliminate inequities in access to high-impact journals.  In practice, APC-based publishing models often reinforce disparities, since well-resourced researchers remain better positioned to publish than those with limited funding \citep{KlebelRossHellauer23}.

Effective solutions to solve these inequities would involve a major change in the policies of high-impact journals, which would include:

\begin{itemize}
    \item Remove the requirement for authors to individually cover publication costs, allowing institutions to fully cover publication charges;
    \item Remove the requirement for authors to request waivers individually, providing automatic waivers;
    \item Create a system where well-funded institutions continue to pay, allowing publications free of charge from underfunded institutions;
    \item Become a fully open-access journal, funded by societies, institutions, consortia, or even public funding.
\end{itemize}

\subsubsection{Recommendations for Authors}

Since a structural change in the policies of high-impact journals is necessary to raise the number of publications from underfunded authors, immediate strategies are suggested, considering the current status of these journals. Practical recommendation for authors include:

\begin{itemize}
    \item Choose the desired journal based on the publication costs, on any agreement that your institution or country may have to cover these publication costs, and on the possibility to request a waiver;
    \item If possible, request waivers;
    \item If possible, consider the possibility of publishing letters or short papers, which often require lower publication costs;
    \item Apply for publication grants or funding that cover publication costs;
    \item If possible, consider the possibility to designate a co-author from sponsoring countries or institutions as the corresponding author.
\end{itemize}

Additionally, (more senior) authors from more privileged countries and institutions can push for structural change by submitting primarily to ``diamond'' OA journals, thereby broadening the perception of what constitutes a main venue for publication in astrophysics.

\bibliography{references/globalchallenges}{}
\bibliographystyle{aasjournal}
\pagebreak
\section{Patching the Leaky Pipeline Through Pedagogy: How Teaching and Learning Affect the Retention of Underrepresented Persons in Astronomy and Physics} \label{sec:pedagogy}
\vspace{-30pt}
\begin{center}
    \includegraphics[width=0.4\linewidth]{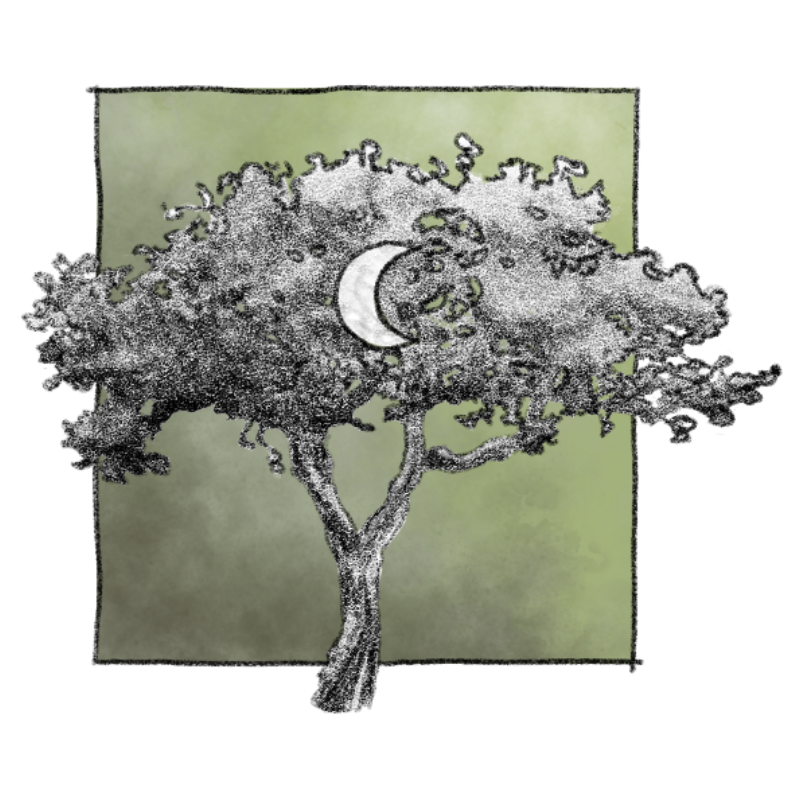}
\end{center}
\vspace{-40pt}
\begin{sectionauthor}

\normalsize{\href{https://orcid.org/0009-0009-2307-2350}{Katherine Myers}$^{1}$, 
\href{https://orcid.org/0000-0001-9348-9738}{Sydney Andersen}$^{2}$, 
\href{https://orcid.org/0009-0006-4036-4919}{Mikel Charles}$^{3,4}$, 
\href{https://orcid.org/0009-0007-8812-1630}{Ryn Grutkoski}$^{5,6}$, 
\\Disha Chakraborty$^{7}$, \href{https://orcid.org/0000-0001-8638-2780}{Padmavathi Venkatraman}$^{8}$, \href{https://orcid.org/0009-0000-7177-9697}{Henna Abunemeh}$^{8}$, \href{https://orcid.org/0009-0008-1965-9012}{Aadya Agrawal}$^{\,8}$, Anika Goel$^{\,9}$,   \href{https://orcid.org/0009-0006-5765-1607}{Lillian N. Joseph}$^{10}$, Varun Kore$^{7}$, \href{https://orcid.org/0000-0003-4307-634X}{Andrey Kravtsov}$^{5,6,11}$, \\\href{https://orcid.org/0009-0007-8974-9016}{Shalini Kurinchi-Vendhan}$^{12}$, \href{https://orcid.org/0000-0002-6187-4866}{Rebecca McClain}$^{13,4}$, \href{https://orcid.org/0000-0003-4248-6128}{Stephen McKay}$^{14}$, \\\href{https://orcid.org/0000-0003-3585-3356}{Cassidy Metzger}$^{15}$, \href{https://orcid.org/0000-0003-0343-0121}{Janiris Rodriguez-Bueno}$^{8}$, \href{https://orcid.org/0009-0005-6146-7151}{Laura Stiles-Clarke}$^{16}$, \\and \href{https://orcid.org/0000-0003-1535-4277}{Margaret E. Verrico}$^{8,17}$\\}
\pagebreak
\scriptsize{
$^{1}$ Department of Physics and Astronomy, York University, Toronto, ON, CA\\
$^2$ Department of Astronomy, University of Washington, Seattle, WA, USA\\
$^3$ Department of Physics, The Ohio State University, Columbus, OH, USA\\
$^{4}$ Center for Cosmology and AstroParticle Physics, The Ohio State University, OH, USA\\
$^5$ Department of Astronomy \& Astrophysics, University of Chicago, Chicago, IL, USA\\
$^6$ Kavli Institute for Cosmological Physics, University of Chicago, Chicago, IL, USA\\
$^7$ Department of Physics \& Astronomy, University of Kansas, Lawrence, KS, USA\\
$^8$ Department of Astronomy, University of Illinois Urbana-Champaign, Urbana, IL, USA\\
$^9$ Department of Astronomy, Indiana University Bloomington, Bloomington, IN, USA\\
$^{10}$ College of Science and Health, Benedictine University, Lisle, IL, USA\\
$^{11}$ Enrico Fermi Institute, University of Chicago, Chicago, IL, USA\\
$^{12}$ Max Planck Institute for Astronomy, Heidelberg University, Heidelberg, Baden-W\"{u}rttemberg, DE\\
$^{13}$ Department of Astronomy, The Ohio State University, Columbus, OH, USA\\
$^{14}$ Department of Physics, University of Wisconsin-Madison, Madison, WI, USA\\
$^{15}$ Department of Physics and Astronomy, Dartmouth College, Hanover, NH, USA\\
$^{16}$ Department of Astronomy and Physics, Saint Mary's University, Halifax, NS, CA\\
$^{17}$ Center for AstroPhysical Surveys, National Center for Supercomputing Applications, Urbana, IL, USA\\

}
\normalsize{}
\end{sectionauthor}
\vspace{20pt}

\begin{abstract}
The relationship between pedagogy (teaching and learning) and professional development is profoundly important for those in science, technology, engineering, and math (STEM) careers. There is a direct link between having a welcoming, nurturing, and principled space in which to learn and grow, and the likelihood of staying in STEM and academia. This link is especially crucial for supporting students and young professionals whose demographic is underrepresented in their field. In this chapter, it is demonstrated through examples and recommended `best practices' how \textbf{lowering systemic barriers, providing social support, and implementing equitable practices} can contribute to higher success and retention rates for underrepresented persons in astronomy and physics. 
\end{abstract}

\pagebreak
\subsection{Introduction}
Teaching and learning are necessarily intertwined with the academic pipeline followed by any astronomer or physicist. As students in childhood science education, then post-secondary and graduate education, and potentially as instructors (whether teaching assistants, lecturers, mentors, or professors), the various pedagogies encountered will have a significant role in shaping careers, relationships, and lives. However, the diminished number of underrepresented persons in astronomy and physics (and broader STEM fields) at each successive stage of academia has been described as a \textit{leaky} pipeline \citep{JacksonMaryAnne2023TLPi}, where representation and numbers drop as one rises through the ranks. 

It is well documented that  women and racialized persons are underrepresented in physics and astronomy, and encounter systemic barriers and challenges in physics and broader STEM fields. A series of highly-cited figures from the American Physical Society (2022)\footnote{\url{https://www.aps.org/apsnews/2022/10/mixed-progress-women-marginalized}} show that only 21\% of physics doctoral degrees were awarded to women in 2020, and students of colour made up only 6\% of doctorates awarded, with neither number showing any substantial increase over the last decade. The diversity gap in physics is among the worst of all science disciplines \citep{XuYonghong2015FoWi, LeslieSarah-Jane2015Eobu}.

This chapter focuses on best practices for pedagogical interactions designed to improve the retention of women in astronomy (and physics), in line with Picture an Astronomer's objectives. Many of the aspects of retention through pedagogy are also relevant to other underrepresented students, including Black, Indigenous, and People of Color (BIPOC) students, international students, first generation students, disabled students, and transgender, 2SLGBTQI+, and gender non-conforming students. ``Underrepresented'' is not a monolith, there are many best practices that are applicable to the needs of a diverse student base, and as such this section will refer to the collective ``underrepresented students''. Additionally, the use of research-based pedagogical practices has positive effects for \textit{all} students including the white, male majority \citep{LiYangqiuting2020HPoB, AIP2019}.

Pedagogy impacts the experience of underrepresented persons in physics in numerous ways, providing us with many avenues for improving the integrity of the pipeline. This section refers to \textbf{three core values} that are pillars for healing the leaky pipeline and supporting the systemic inclusion of historically underrepresented groups in physics and astronomy. They are outlined as follows:

\begin{enumerate}
\item \textbf{Lower systemic barriers.}

Within academic spaces, many hurdles exist that may increase the difficulties faced by marginalized groups. First and foremost in the movement towards equal representation and experience in the field is removing these barriers, which are diverse in nature, where they exist. One example is the lack of support in balancing family responsibilities (see Section \ref{sec:family} for more in-depth discussion): studies show that 40\% of women compared to 23\% of men leave science or reduce their working contributions upon starting a family \citep{CechErinA.2019Tcct}. There is also systemic discrimination at play, demonstrated by phenomena such as ``white empiricism'', which describes the underlying belief that ``only white people (particularly white men) are read as having a fundamental capacity for objectivity'' \citep{Prescod-WeinsteinChanda2020MBWS}, and the ``equity tax'', where underrepresented scholars face larger workloads \citep{HenryFrances2017TEMR}. Another 
barrier can be power imbalances and a fixed ``status quo'' that is resistant to change. \citet{DancyMelissa2023Hwwm} demonstrate how even the most well-intentioned physicists of majority demographics who self-identify as progressives are still contributing to status, power, and privilege imbalances through patterns of inaction. Marginalized groups in physics and astronomy also fall victim to prejudicial practices when it comes to hiring, mentoring, assessing applications \citep{WittemanHollyO2019Aggd}, or citing work (women tend to accumulate fewer lifetime citations than men, and are less likely to receive reciprocal citations \citep{LermanKristina2022Gcpa}).

When barriers are removed or overcome, retention of underrepresented groups in physics improves. For example, many grant applications and observing proposals are now anonymized of proposer personal information, which is shown to reduce gender bias and result in female applicants being more fairly rated \citep{JohnsonStefanieK.2020DYPR, RadebaughJani2021TVoa}. To reduce the number of women leaving science to begin families, funding institutions and universities can allow new parents to take paid parental leave without the risk of losing their funding or position \citep{cdi_proquest_journals_2126986311}. When gender parity and diversity is achieved in research teams, it is shown to improve collective problem solving and impact of results \citep{NielsenMathiasWullum2017Gdlt, strength}. Adopting diverse curricula, inclusive teaching, and emphasizing development of students' identities all directly contribute to improving retention and success in physics \citep{BlueJennifer2018Gm}. Working to remove or decrease systemic barriers is an important activity for all astronomers and physicists because it improves educational climates and spreads awareness for systemic issues. These in turn have positive effects in classrooms, reaching current and future students. 

\item \textbf{Provide social support}

A supportive and welcoming learning environment is imperative for students' success as they are introduced to academia. However, not everyone has such a sense of community; underrepresented students report a reduced feeling of acceptance and kinship \citep{ItoTiffanyA.2018FIHS}, and these ``early disparities in belonging may have downstream effects on the retention of women and other groups underrepresented in STEM'' \citep{FinkAngela2020Bigc}. Students from minority groups are less likely to have role models in their field (see Section \ref{sec:representation}), but when they do, they perform better, are more motivated, and participate more frequently \citep{BarabinoGilda2020StGB}. Supporting students can help ``the transition from new student to contributing physicist'' as well as increasing students' ``sense of belonging to the physics community'', according to \citet{RethmanCallie2021Ioip}. This sense of belonging is directly linked to underrepresented student success \citep{FinkAngela2020Bigc,EdwardsJoshuaD.2023TEoS,HausmannLeslieR.M.2007SOBA,TrujilloGloriana2014CtRo,LewisKarynL.2016Fioo}. Having a sense of belonging is linked to ``intentions to persist'', and such sentiments are stronger and more lasting among students in intervention groups aimed at fostering their sense of belonging \citep{HausmannLeslieR.M.2007SOBA}. 

As highlighted in her essay in \textit{Blazing the Trail: Essays by Leading Women in Science}, Jarita C. \cite{HolbrookBtT} emphasizes:

    \begin{mdframed}[
  leftline=true,
  rightline=false,
  topline=false,
  bottomline=false,
  linecolor=gray,
  linewidth=2pt,
  innerleftmargin=10pt,
  innerrightmargin=5pt,
  innertopmargin=6pt,
  innerbottommargin=6pt
]
\textit{``Don't even think about asking about ``quality'' and ``qualified'': anyone who gets into a Ph.D. physics or astronomy program is clearly qualified! The rest is good or bad projects, good or bad advisors, publishing, time, and knowing about the right opportunities. However, a hostile environment can make all of this far more difficult, and thus mentoring is so important in giving students extra support.''}
\end{mdframed}  
    
Developing these senses of self-efficacy---a belief in their ability to succeed as scientists and researchers---and belonging is crucial; social support and integration are significant indicators of students' intention to persist in undergraduate and graduate programs \citep{SachmpazidiDiana2021ItRb}. Therefore, fixing the leaky pipeline requires cultivating a supportive social environment.

\item \textbf{Implement equitable practices}

Implementing equitable practices into teaching and learning pedagogies is a cornerstone for progress towards retention of underrepresented students in physics and astronomy. Equitable practices can be defined as initiatives and structures that aim to ensure all students are valued, respected, and have the same chances of succeeding. Despite efforts to make the field more accessible, subtle acts of discrimination remain an issue within physics and astronomy. \citet{BarthelemyRamónS.2016Gdip} captures the breadth of these experiences with discrimination, having surveyed female students about facing microaggressions, hostile sexism, or gendered differential treatment. In situations like these, we see how equitable structures like affinity groups and societies both provide support and aid to dismantle unwelcoming hierarchies within physics departments \citep{GonsalvesAllisonJ.2020NoSI}. For example, traditional teaching methods lead to larger achievement gaps for underrepresented students compared to active learning \citep{TheobaldElliJ.2020Alna}. Adopting flexible and equitable practices in grading and student evaluations is directly linked to retention of underrepresented students \citep[][ see their section 2 for more]{vanderSluisHendrik2013Fast}. Learning physics material can be particularly challenging for disabled students in ways that go beyond standard accommodations. For example, students with ADHD may struggle with information uptake and processing new information \citep{JamesWestley2020Dbeb} while students with physical disabilities may be unable to participate in traditional lab courses \citep{LannanAmanda2021RfSS}. Though this chapter does not always address disability specifically, many of our recommendations align with \citet{AarnioAlicia2019AAWP}, which focuses on policies aimed at improving the participation of disabled astronomers. Overall, one of the strongest ways the leaky pipeline can be mended is by actively building up support systems for those who need them most. 

\end{enumerate}

Within the larger white paper focused on addressing various stages of the leaky pipeline, this section confronts the relationship between pedagogy and the retention of underrepresented groups. We have assessed the relationship between pedagogy and retention in three post-secondary stages of academia: How in-classroom (Section \ref{subsec:inclass}) and out-of-classroom (Section \ref{subsec:outclass}) interactions, environments, and pedagogies affect retention, and how pedagogy within the teaching assistant and graduate student experience affects retention (Section \ref{subsec:ta}).

\subsection{The in-classroom relationship to retention \label{subsec:inclass}}
Within the classroom there are many student-instructor relationship elements that have a direct effect on the student experience and the likelihood of retention within academia --- especially for students belonging to underrepresented groups. Everyone in and around academic spaces can play a role in improving this relationship. \\

\subsubsection{Incorporate Representation in Class Content}\textbf{Lowering systemic barriers in the classroom} is fundamental to success --- feelings of belonging and seeing representation in academic spaces enhances retention of underrepresented students \citep{BowmanNicholas2023TRoM, IjomaJennyN.2022VbWB}. Promoting a diverse learning environment encourages diversity of thought, and it is crucial to give a platform to results and innovation from underrepresented scientists. The `Diversity-Innovation' paradox highlights that while diverse scientists are often responsible for more novel research, they are also more likely to be discredited or devalued \citep{diversity}. Making a conscious effort to acknowledge the contributions of underrepresented persons to science in lecture content and to teach about both historical and modern trailblazing figures in the field can also have a monumental impact on students. Doing so can be as simple as creating a portfolio of profiles of women in physics, a calendar of diverse science-related holidays, or creating a list of resources, scholarships, and internships available to students and for circulation among the teaching faculty. Consider making resources like the Mirror Project\footnote{\url{https://blogs.wellesley.edu/mirror/}} by Wellesley College (PI: Mowla), which showcases women in physics, or related examples highlighted by the American Society of Plant Biologists\footnote{\url{https://plantae.org/fostering-inclusivity-in-stem-teaching-tips-and-best-practices/}}, available as well. Emphasize how all students are deserving of recognition and opportunity in physics and astronomy. Promoting institutional support, with information about internships or mental health resources works to break down the hurdle of entering academic spaces for students who may be underprivileged or otherwise lack their own support system. Creating spaces of support and representation increases a feeling of belonging in STEM \citep{DostGulsah2024Spot}. Consider the following examples of how to lower systemic barriers in the classroom: 

\begin{itemize}
    \item Announce to students the International Day of Women and Girls in Science\footnote{\url{https://www.un.org/en/observances/women-and-girls-in-science-day}} (February 11), and 2SLGBTQI+ STEM Day\footnote{\url{https://prideinstem.org/lgbtstemday/}} (November 18). This lowers the barrier that is lack of representation in academic spaces.
    \item During your lectures, acknowledge the contribution of women or underrepresented persons to relevant scientific discoveries.
    \item Post scholarship or internship opportunities online on the course webpage, especially those targeting underrepresented groups. This reduces barriers relating to finding research positions and financial support. 
    \item Add institution-specific resources for accessing mental health support to the course syllabus; dealing with unsupported mental health issues can present significant barriers.
\end{itemize}

These small actions, among others, have a significant impact if widely adopted, and many of these resources can be created/assembled by anyone - department leadership, faculty, or interested student groups.

% \underline{A socially supportive, and inclusive learning environment}. 
\subsubsection{Apply Inclusive and Active Learning Teaching Methods}\textbf{Providing social support in the classroom} and an inclusive learning environment improves the academic experience for underrepresented students. Physics education research has shown that interactive teaching and learning strategies that move away from a traditional ``lecture'' format result in more peer collaboration, comfort in the classroom, and deeper conceptual understanding \citep{PhysRevSTPER.6.020123, TunggyshbayMeiirbek2023Fcsa}. Having access to equitable opportunities and an inclusive learning environment allows underrepresented students to engage with learning efficiently and successfully \citep{CwikSonja2022RoIo}, and provides the social support essential for success. 

An inclusive learning environment can look like in-class discussions, Q\&As, group work, and other interactive learning strategies like those outlined in the Inclusive STEM Teaching Project\footnote{\url{https://www.inclusivestemteaching.org/}}. This should be done using informed, and socially conscious teaching methods -- especially when dividing students into small groups. Taking any opportunity to structure groups intentionally based on demographic and performance has been shown to foster the greatest problem-solving performance for group activities \citep{HellerPatricia1992Tpst}. Engaged students are more likely to retain course material, and collaboration and peer learning are valuable skills. These methods also encourage communication among students and professors, which can provide confidence-building opportunities \citep{DowningVirginiaR.2020FoNE}. 

Workshops to support teaching best practices can also reduce disparity in STEM achievement among underrepresented groups \citep{OLearyErinSanders2020Cicb}. \citet{APS_AAS_RBT2024} describe the importance of a department chair's role in implementing research-based teaching practices. Departments should provide resources and training; one possibility is an early-in-the-year colloquium filled by a physics education research expert, for teaching assistants, instructors, and professors to encourage research-supported interactive and inclusive teaching techniques, which have been shown to improve instruction quality \cite[see][]{JohnsonDavidW2014CLIU}. Information and resources concerning proven successful teaching strategies will facilitate their introduction in the classroom. 

% \underline{Engaging in pedagogy by implementing inclusive and equitable practices in the classroom} 
\subsubsection{Facilitate an Equitable Classroom Environment }\textbf{Implementing equitable practices in the classroom} and engaging in inclusive pedagogy supports retention of underrepresented students and mends the leaky pipeline. Faculty involvement has a direct effect on institutional transformation in favor of diversity \citep{WhittakerJosephA.2014CITa}. Instructors themselves play a fundamental role in ``establishing and maintaining safe, equitable and inclusive environments for students'', which has a profoundly positive impact on retaining underrepresented students in physics and astronomy \citep{PhysRevPhysEducRes.20.020142}. To build a department culture around equitable and supportive classroom practices, departments may choose to assess willingness to engage in pedagogy, learning reflection, and educational collaboration as desirable skills during the instructor/faculty hiring process, as well as providing resources and training as mentioned in the previous section. For departments and teaching faculty, we propose three equitable classroom practices to improve retention of underrepresented students. 

\begin{enumerate}
\item \textbf{Rolling feedback models for course delivery.}

When instructors are willing to implement and consider student feedback on course delivery content and methods, it opens doors for increased student understanding and meaningful learning. Implementing the results of teaching method surveys has been shown to support multidimensional learning and inclusive teaching \citep{Johnson-OjedaVanessa2025MSIL}. Creating the opportunity for course delivery feedback will encourage adaptability in teaching practices to support students. Potential surveys may allow students to rate their own performance and comprehension, the effectiveness of course delivery, and the helpfulness of graded materials for accomplishing learning goals on a weekly or monthly basis. This feedback should be implemented with careful attention to the gender bias that can be introduced through performance surveys \citep{MacNellLillian2015WiaN}. Instructors should consider exercising an open mind to adjusting their teaching practices based on student feedback. Choosing to request and act on feedback on an ongoing basis (rather than at the end of term, as is typical) creates an opportunity to make mid-term modifications to teaching strategies, as to best support student needs. 

\item \textbf{Distinct, universal, and flexible policies for graded material.}

Consider department-wide policies for grading exceptions such as late assignments, missed work, and sick/emergency leaves/extensions, instead of instructor-by-instructor policies. Having uniform, enforceable policies reduces opportunity for biased judgment \citep[as documented by][]{DelToroJuan2022TRoS} otherwise introduced while responding to student situations on a case-by-case basis. It also reduces students' need to self-advocate for extenuating circumstances, which disproportionately disadvantages underrepresented students \citep{alma99161797315301452}. These department-wide policies could include universalizing the penalty for late work across courses, or standardizing the extension policy for extenuating circumstances. It is also recommended that grading and attendance policies have a degree of freedom, with examples such as dropping the lowest quiz or assignment mark, or allowing a set number of missed days without penalty. Flexible grading and assessment techniques have been shown to support retention of underrepresented students \citep{vanderSluisHendrik2013Fast}, who may have  financial or disability needs, undeniable familial responsibilities, health issues, or other potential barriers affecting their participation. 

\item \textbf{Grading transparency through rubrics and explicit learning goals.}

Explicit instruction contributes to positive learning outcomes and student achievement \citep{AusEdReAs2023}. This can take the form of creating rubrics, preparing detailed answer keys, and writing out explicit learning goals for each course. This guides student focus towards desired learning objectives, and reduces the opportunity for bias when assessing academic performance \citep{BloodhartBrittany2020OyuU}. It has been shown that students are more confident and engage in more meaningful learning with clear rubrics and grading criteria \citep{ReyndersGil2020Rtac}. This practice is beneficial for teaching assistants as well; providing marking guidance alleviates stress caused by unclear expectations or conflicts related to grading: teaching assistants \textit{and} students benefit from rubrics indicating clear learning outcomes \citep{ReyndersGil2020Rtac}.
\end{enumerate}

\subsection{Supporting retention outside of the classroom \label{subsec:outclass}}
Outside of the classroom, the educational role between departments, faculty, and students becomes less well-defined and more dependent on institutional resources. We identified key areas of interest where additional support and instruction would be the most beneficial, including improvement of department culture, access to research experiences, and preparation for post-graduation opportunities. There are many ways that departments can facilitate these areas of learning and professional development, as well as many actions individual students and faculty can take to make information more accessible.

\subsubsection{Department Climate and Culture}
Department climate and culture shape the experiences of students throughout their educational career and are influenced by many factors. Experiences of sexism, as discussed in \citet{BarthelemyRamónS.2016Gdip} and \citet{AycockLaurenM.2019Shrb}, directly influence department culture and create barriers. Improving and maintaining this culture is not the role of any individual alone, but there are incremental steps individuals can take \citep{Stiles-ClarkeLaura2025Mtbi}. Because the challenges students with underrepresented identities face are not just academic, but also environmental in nature \citep{KeblbeckD.K.2024Upse}, it is important that these concerns are addressed at an institutional level.

\paragraph{Support Professional Societies and Affinity Groups.}
One key action to address systemic problems within department culture is to establish and support the efforts of affinity groups and professional society chapters. Student engagement with these organizations will allow them to meet peers and share professional resources, creating a strong foundation for department culture. An anonymized study of a Woman Physicist Group \citep{GonsalvesAllisonJ.2020NoSI} demonstrated how an affinity group can positively impact its minoritized members. Observed impacts included disruption of traditional hierarchies and positionality and shared resources for information and support. Notably, this group had active faculty engagement and department backing, demonstrating how external support can help in developing these networks. Professional societies can also affect change by making recommendations on how to improve inclusivity within astronomy \citep{DesirSteve2025Cpsc}. These recommendations, such as the ones released by the American Astronomical Society's Committee on the Status of Women in Astronomy \citep{WexlerRachel2025RAft} --- especially the need to prioritize anti-harassment training and update reporting procedures  --- can kickstart important dialogues within departments on how best to make positive change. It is important to consider that these affinity groups do not inherently meet the needs of all community members and that they can be co-opted or turned against intersecting identities \citep{GutzwaJustinA.2024Hwal}, highlighting the need to actively develop affinity networks in ways that deter hierarchical authority and incorporate feedback.

Providing the opportunity for student affinity groups to engage with professors, research staff, and visiting speakers allows them to ask questions and connect with colleagues. Involvement of professors with these organizations helps students feel supported \citep{GuzzardoMarianaT.2021“OtC} while individualized interactions allow for more networking opportunities. Affinity groups can contribute to a sense of belonging and community for BIPOC students \citep{Adefiyiju-MonwubaTaiwo2023TLEo}. In her thesis, Adefiyiju-Monwuba makes many suggestions for how institutions can support students, including developing ``affinity within an affinity'' mentorship programs to connect BIPOC students and campus leaders. She also highlights that ``implementing affinity groups within PWIs (primarily white institutions) can help students and staff recognize their own unconscious biases while equipping them with real-world strategies, techniques, and pedagogy to engage in meaningful conversations and promote diversity, equity, and inclusion on campus.'' The student-professor pedagogical connection in these affinity group environments facilitates a partnership that can contribute to an increased sense of belonging and identity for Black students \citep{Cook-SatherAlison2021'wia}. Similarly, participation in Women in Physics affinity groups, as well as events like the annual Conferences for Undergraduate Women in Physics\footnote{\url{https://www.aps.org/initiatives/inclusion/gender-inclusive/undergraduate-women-minorities}} (CU*iP) supported by the American Physical Society, increases the sense of belonging for participants \citep{HazariZahra2024Etrb}. The same principles apply to professional societies more broadly. \citet{HuledeIreneV.2018PSfS} highlights that, to be successful, professional societies must support diversity and inclusion, actively recruit and train underrepresented minority members, and develop mentoring networks in order to prepare and retain these students. 

Department administrations play an important role in preventing power dynamics that are harmful to underrepresented students \citep{TäuberSusanne2024Upds}. \citet{PadillaAmadoM.1994EMSR} describes the ``cultural taxation'' ethnic minority students face, where they must use their own time to act as experts in diversity to educate their peers and teachers. This additional labor can take away from their research and contribute to burnout, as ``mentoring is a first step in setting up the mechanics of professional development, but more is required since the mentor is frequently in as much need of support as is the mentee'' \citep{PadillaAmadoM.1994EMSR}.

It is important to acknowledge that faculty members do not have unlimited time to provide additional mentorship and advising outside of coursework settings, and that women faculty disproportionately perform more service and mentoring work \citep{MisraJoya2011TICo}. It is clear that mentorship is critical for students' development and self-efficacy, therefore it is important that administrations acknowledge the disparity in service and consider this mentoring work when evaluating promotions, raises, and tenure \citep{SmithElizabethE2019HaFR}. 

To ensure the success of affinity groups, departments should financially and materially support these groups where possible, including providing meeting spaces and faculty advising and participation. Group leadership should utilize institution resources to facilitate presenters and opportunities. This can take the form of panels, presentations or interactive methods, such as brief interviews where students can ask research- and career-related questions of more senior academics. These frameworks do not need to be birthed from scratch; The article `Making Space: Affinity groups offer a platform for voices often relegated to the margins' and the accompanying toolkit by \citet{Bell2015MakingSpace} provides steps on how to establish effective affinity groups, highlighting that these are not just spaces for pride, but for conversation and action. 

\paragraph{Provide Access to Peer Mentorship.} Departments can encourage social support and improve culture by facilitating mentorship relationships between students. Mentoring ensures students feel supported academically as well as professionally by providing access to organizational and study skills and a setting to discuss stress and social issues \citep{ZaniewskiAnnaM.2016ISsa}. If some faculty are unable or unwilling to mentor, we suggest peer mentorship to distribute the workload while still maintaining the benefits of mentoring relationships. \citet{RethmanCallie2021Ioip} discuss the impact of peer-to-peer mentorship within the Department of Physics and Astronomy at Texas A\&M. Their Discover, Explore, and Enjoy Physics and Engineering\footnote{\url{https://artsci.tamu.edu/physics-astronomy/outreach/deep/index.html}} (DEEP) program facilitates direct graduate to undergraduate mentorship. Student responses to being surveyed about the program indicated an improvement in their communication skills as well as a reinforced understanding of physics concepts. Similar programs exist at other universities (for instance, Yale University's Astro Sibs\footnote{\url{https://astronomy.yale.edu/academics/undergraduate-program/astro-sibs-undergraduate-mentorship-program}}), and these established mentorship networks can serve as a model for broader ``near-peer'' mentorship programs. To this end, departments can leverage the knowledge of post-doctoral researchers, graduate students, and higher-level undergraduate students to provide support and mentorship for their most junior members. An initial approach could include surveying students to gauge interest and developing pilot programs to mentor first and second year undergraduate students. Eventually, these programs could be improved by calling on examples like DEEP or Astro Sibs and expanded to include additional levels of ``near-peer'' mentoring for more senior department members.

\paragraph{Address Student Health and Sense of Belonging.} Both mental and physical health as well as feelings of isolation can compound inequities underrepresented students face. With the rise in mental health concerns for graduate students \citep{EvansTeresaM2018Efam}, it is crucial that departments connect students and faculty in supporting, and working towards destigmatizing concerns about, mental health and improving student sense of belonging. Feelings of inadequacy do not begin at the graduate level, as students enter their undergraduate institutions with perceptions about the role of gender and work-life balance \citep{Tan-WilsonAnna2015CSVo} which can contribute to stress and mental health struggles later in their careers. Addressing mental health and student belonging through panel discussions and interaction with established scientists can reassure students of their STEM career goals and create a precedent for prioritizing self-care. At the administrative level, mental health services should be well-funded and easily accessible, but increasing the visibility of existing resources also goes a long way \citep{PriestleyMichael2022SPoi}. When compassionate professors direct students towards existing resources, it leaves a positive impact and improves the students' impression of the student-faculty relationship \citep{GuzzardoMarianaT.2021“OtC}. 

Many factors can impact a student's sense of belonging within the department. Dealing with name mispronunciation \citep{NajjarRana2023SaIE} and incorrect pronoun usage \citep{WoodsChris2024Tmad} can heighten feelings of isolation and ostracization. Having all students document their name pronunciations and pronouns through a centralized learning management system can contribute to an inclusive learning environment and gentle reminders to do so are effective in improving the climate \citep{ArmstrongCherylL.H.PhDRDNLDMBA2023ITSC}. This approach and others contribute to safety and inclusion for many students. In particular, 2SLGBTQI+ STEM professionals face many challenges, including social exclusion, and are more likely to consider leaving their fields \citep{CechE.A.2021SifL}. Feelings of invalidation and isolation are also common for 2SLGBTQI+ and disabled physics students \citep{KeblbeckD.K.2024Upse}.

Several simple steps can be taken to address these concerns and create a more equitable environment. Departments and instructors can ensure students are connected to campus mental health resources, including counseling, emergency contacts, and wind-down or self-care activities, by promoting these resources via course orientations, department websites, and in public meeting spaces. Accessing these resources can also be further encouraged in the course syllabus with specific efforts to target first year classes. It is also necessary to encourage healthy work-life balance across the department -- including faculty. Students will struggle to maintain this balance if they do not see healthy examples among faculty \citep{AyresZoëJ2022MYMH}. Further institutional recommendations include discouraging working outside of established hours, taking time off for illnesses, and encouraging students to set aside time for self-care. To facilitate student belonging, set an example of inclusivity by displaying pronouns, correct name pronunciations, and other information in email sign offs and messaging profiles. These steps may not address the deeper, societal issues that impact students' mental health issues and social isolation, but can help individuals feel more safe and accepted. Receiving direct support from peers and faculty is central to supporting 2SLGBTQI+ students in challenging academic environments \citep{GutzwaJustinA.2024Hwal}.

\subsubsection{Research Experiences}
Research is a critical part of career progression in astronomy, but not all students have access to summer or department research programs. In addition to being competitive, systemic barriers exist that may prevent students from accessing these opportunities, such as citizenship requirements or financial limitations. While some students may have access to classes on research skills and pedagogy, this is not standard across all programs. 

\paragraph{Access to Research and Related Skills.}Students need access to research opportunities in order to kickstart research careers, but systemic barriers exist that make this more difficult for underrepresented students. A study of gender-based biases in lab management hiring points to faculty members of all genders perceiving female students as less competent, when compared against identical applications with masculine names attached \citep{Moss-RacusinCorinneA2012Sfsg}. These hiring biases, where female applicants are perceived as less capable, affect female students as they navigate landing research roles. This presents a significant barrier, as undergraduate research experience has been shown to aid in retaining female students \citep{BrainardSuzanneGage1995RFUS} and to play a significant role in their interest in pursuing science careers \citep{HarshJosephA2012APoG}. Partaking in hands-on opportunities such as working directly with telescopes increases students' self-efficacy and feeling of competence as researchers  \citep{FreedRachel2022Davo}. Early access to research reinforces coursework and ultimately increases retention \citep{GheeMedeva2016FSRP}. Furthermore, \citet{SeymourElaine2004Etbo} document the many areas in which students experienced benefits following undergraduate research experiences. Broadly, undergraduate research experience increases interest in STEM research and facilitates future career planning and professional preparedness. Access to undergraduate research opportunities is deeply important for training and retaining underrepresented astronomers and physicists. In an effort to focus on strengthening retention and identity through research experience, the following areas of focus are presented:

\begin{enumerate}

\item \textbf{Transparent and equitable hiring practices.}

To overcome these systemic biases, it is vital that departments not only provide research opportunities to their undergraduate students, but standardize their application processes to account for unconscious biases in hiring. Providing clear expectations for each project, including expectations for prerequisite coursework and technical skills (like familiarity with programming languages or other specialized software) ensures fairness. Departments should facilitate opportunities for students with all levels of experience to pursue research by designating a fraction of opportunities to be ``low barrier to entry'', meaning not requiring previous research or programming experience. This will make research projects more accessible and allow more students to develop strong, foundational skills. Faculty should also be trained specifically on equitable hiring practices, following similar suggestions as recommended for faculty searchers in \citet{Cosgriff-HernandezElizabethM.2023Ehst}. 

\item \textbf{Variety in available research-based opportunities.}

Where possible, departments should provide publishable research opportunities to students, as external research may not be accessible. This can take many forms, including undergraduate research courses, guided independent study, final research projects and undergraduate theses, or practicum options. Allowing students to contribute to papers, practice scientific writing, and develop their presentation skills will shape their future careers. Formally structure research groups in ways that benefit students' professional development and sense of belonging, such as those described by the Affinity Research Group (ARG) Framework. The ARG model provides more equitable access to research, development of research skills and expertise, as well as social development and community trust \citep{VillaElsaQ.2013ARGi}. It is also important for students to understand the context of their research and why individual activities are meaningful \citep{QuanGinaM.2018Ibdp}. Research groups are important avenues for addressing student belonging and, as a result, should provide transparent expectations, recognition, and encourage social connections \citep[as in][]{NidiaRuedas-Gracia2022Tsrf}.

\item \textbf{Resolved skill and knowledge gaps.}

New undergraduate students should have the opportunity to learn about career paths and expectations in academia. Students should be presented with general timeline outlines for a typical academic career trajectory, a high level overview of the crucial skills they will need to develop (such as data analysis and programming), as well as information on how to apply for funding and write proposals for instrument time.

Presentation of this information could take many forms, whether in a department-sponsored colloquium or through a dedicated course or seminar. Some career information can also be included in introductory courses where appropriate. For example, covering proposal and paper-writing in a class on observational astronomy techniques or experimental design. To ensure they understand how to read research papers, create or expand existing journal clubs and courses to allow undergraduate students to observe or participate when appropriate. This will expose them to broader research areas and allow them to develop important scientific reading and critical thinking skills. When possible, connect students to useful external resources, such as `hackathons' or introductory seminars for programming, like Penn State's Astrostatistics Summer School\footnote{\url{https://sites.psu.edu/astrostatistics/su25/}}, Northwestern University's Code/Astro workshop\footnote{\url{https://semaphorep.github.io/codeastro/}}, and other similar summer opportunities.
\end{enumerate}

\subsubsection{Connect Students to External Opportunities and Information} It is beneficial to connect students with external programs such as conferences\footnote{For those not already familiar, the Canadian Astronomy Data Centre maintains a list of astronomy meetings and conferences: \url{https://www.cadc-ccda.hia-iha.nrc-cnrc.gc.ca/en/meetings/}}, workshops, and summer schools, expanding their social connectedness and professional networks. Providing advance information about these opportunities should be proactive: making announcements in relevant courses, having mailing lists, or other methods of directly delivering information are likely the most effective. Helping students navigate the process of finding opportunities will enable them to put forward competitive applications, explore research careers, and pursue further education. This will be valuable for many, including first-generation students, who may be less familiar with the academic career track \citep{GardnerSusanK.2011"iba}, and international students, who may have restricted opportunities to conduct research due to funding and visa limitations. Notably, international students face additional stressors and may have different expectations of pedagogy \citep{ZhouYuefang2008Tmoc}. 

Bridge and post-baccalaureate programs like the American Physical Society Bridge Program\footnote{\url{https://www.aps.org/initiatives/inclusion/bridge-program}} or the Fisk-Vanderbilt Master's-to-PhD Bridge Program\footnote{\url{https://www.fisk-vanderbilt-bridge.org/}}, designed to help participants transition to higher education, have been shown to increase diversity within graduate programs \citep{RobertsSoniaF.2021RoRE} and be of value to students of underrepresented groups who may not have had previous access to research. In addition to career development, these external structures provide opportunities for networking and connection as students build their professional support systems.

Student awareness of these career-growing opportunities will increase participation. Awareness is generated by facilitating sessions (colloquia, panels) on how to apply for research opportunities and postdoctoral fellowships. These sessions should address what opportunities are available to international students -- for example, in the United States, National Science Foundation Research Experiences for Undergraduates (REU) programs are not open to non-citizens\footnote{\url{https://www.nsf.gov/funding/undergraduates\#research-experiences-for-undergraduates-reu-3f1}}. Information about relevant (and ideally vetted) external opportunities should also be on the department's website year-round. Departments should also take steps to ensure students can attend conferences and present research, regardless of personal financial ability either by providing funding for research-related travel or directing students to sources of external funding. It is important to facilitate conference participation, which helps in developing both research-adjacent skills that will be crucial as students progress through academia and professional clarity and identity \citep{FLAHERTYELIZABETHA.2019MCES}. While student conference attendance should be encouraged in general, special emphasis may be placed on conferences for underrepresented students in physics and astronomy (e.g. CU*iP, Black in Astro\footnote{\url{https://www.blackinastro.com/}}).

\paragraph{Update Information on Department Research.} To ensure equitable access to opportunities, it is important that departments provide accurate, up-to-date information online. Website updates have been shown to improve recruiting for fellowships and residencies in the medical field \citep{ShaathM.Kareem2020IFRH, HashmiAsraMD2017HIat} so it follows that the same lessons can be applied to research and graduate positions. By providing the most up-to-date information, departments can encourage the professional development of current students by highlighting ongoing work and assist incoming students in understanding the program offerings and allow them to reference updated information in their applications. In turn, this will aid departments in attracting and retaining students. To this end, department web pages should be maintained, on a yearly basis at minimum. Websites with accurate information on current opportunities and requirements help students make informed decisions. This does not necessarily have to be an added faculty responsibility (although many choose to have personal webpages that deserve the same amount of care and upkeep), but could potentially be done by paid students or administrative staff. This suggestion aligns with the recent recommendations released by the AAS Graduate Admissions Taskforce \citep{LevesqueEmilyM2025AGAT}, which highlight the need for centralized, accessible information about graduate school application processes.

Giving already matriculated students an updated resource for navigating the department and its varied research is also beneficial. Undergraduate research opportunities have been shown to increase graduate school participation for underrepresented minority students, and students who interact with faculty through research are more likely to receive letters of recommendation \citep{HathawayRusselS2002TRoU}. Strong letters of recommendation remain one of the most critical pieces of admissions criteria for physics programs \citep{PotvinGeoff2017Iatd}. As these often come from research supervisors and PIs, research helps students' future career progress while allowing them to develop a better understanding of their interests.

\subsubsection{Post-Graduation Preparation}
Just as there are structures in place to ensure students' academic success in classroom learning environments, it is as important to support their plans following graduation. This post-graduate training represents an important junction in the academic pipeline, as it will inform student's access to post-secondary education or a career beyond academia.

\paragraph{Support for Graduate School Applications.} Systemic barriers make navigating the graduate application process more difficult for underrepresented students. Some students may enter undergraduate studies with clear expectations of their career trajectory but others may not have been exposed to the world that is graduate school, postdoctoral research, and tenure-track planning. First-generation students have less access to social, cultural, and financial capital, and as a result, may be less aware of pathways to higher education and financial aid \citep{GardnerSusanK.2011"iba}. Without pedagogical structures in place to share this knowledge, many students will remain at a disadvantage in planning their futures. Other barriers exist that may prevent retention of women, especially those with intersecting identities, from pursuing graduate studies. Assessing the validity of Graduate Record Examination (GRE) standardized test scores in astronomy and physics admissions is beyond the scope of this paper, but scores have been found to correlate strongly with identity, and it has been further demonstrated that the heavily weighting scores from the GRE Physics Subject Exam (PGRE) in admissions can negatively impact women and other minority applicants \citep{GlanzJames1996HNtP, MillerCaseyW2013ACaD}. Additionally, application fees discourage underrepresented students from applying to graduate programs \citep{RobertsSoniaF.2021RoRE, HoxbyCaroline2013ECO}. Although fewer schools are now requiring the GRE and PGRE\footnote{The American Astronomical Society recommends that the role of the (P)GRE be limited in graduate admissions: \url{https://aas.org/about/governance/society-resolutions/GRE-scores}}, the tests may remain important to facilitate stronger applications for those who choose or need to take them \citep{PosseltJulieR.2019MFDL, PotvinGeoff2017Iatd}. 

\paragraph{Connections with Alternate Post-Graduation Opportunities.}By establishing connections with community resources, students can have additional social support as they explore careers outside of academia. Not all undergraduate and graduate students in astronomy will pursue research careers. According to a report by the American Institute of Physics regarding the combined undergraduate classes of 2018, 2019, and 2020\footnote{While this does include a number of students whose undergraduate experience and post-graduation plans were impacted by the COVID-19 pandemic, these survey results do not differ significantly from the same analysis performed on the combined 2014, 2015, and 2016 classes \citep{AIP2019}.} in the United States \citep{AIP2023}, only 47\% of astronomy majors who responded were enrolled in a graduate program a year after matriculation (22\% in Astronomy and Astrophysics, 13\% in Physics, and 12\% in other fields). While this does not consider students who take a year off or who follow a non-traditional path, it does suggest that nearly half of US-based undergraduate students who earn a bachelor's degree in astronomy will enter industry positions. As many astronomy majors will go on to jobs in private industry, it is valuable for departments to train students for these careers in addition to the academic career track. Options outside of academia are varied, as captured in the AAS Astronomy-Powered Careers webpage\footnote{\url{https://aas.org/careers/astronomy-powered-careers}}.

Outreach opportunities are another key area for students to explore leading up to graduation. For those pursuing research, this presents an early opportunity to explore interests in teaching. Surveyed students suggested participation in outreach impacted their identity as educators and astronomers \citep{MatthewsAllison2022GSPi}, with benefits to their professional development as graduate students. Those same students suggested that they simultaneously held the perception that these identities were in conflict; they felt that because spending time on outreach meant that they were not researching, outreach, education, and astronomy were separate pursuits. A different study concluded that participation in informal physics outreach programs enabled students to develop their identities and see themselves as part of the physics community \citep{PrefontaineBrean2021Ippa}. The apparent disparity between these results is compelling and suggests that, possibly, the devaluation or lack of emphasis on outreach is not universal and can be overcome. Both studies highlighted that outreach allows students to develop important teamwork and communication skills beyond what would develop in classroom settings.

As students leave academia or otherwise interact with the public via outreach, it is important for departments to establish and maintain a presence in, and connection with, local communities. This may include active involvement in cultural institutions such as planetariums, museums, observatories and non-profits like astronomy clubs, as well as building a rapport with local companies to present students with a variety of potential current-and-future roles and positions. Community relationships can be developed by leveraging connections of current students and alumni, including by compiling lists of opportunities/groups/institutions students have found valuable or by encouraging clubs and affinity groups to develop ongoing relationships with local organizations. In addition to benefiting students as they look for summer and post-graduation positions, these connections will improve outreach opportunities and community engagement while developing critical networking and social support resources for students.

\subsection{Retaining Teaching Assistants \label{subsec:ta}}

Many physics and astronomy programs rely on graduate students (and sometimes undergraduate students) to assist in teaching and managing courses. What these roles look like varies drastically from institution to institution, but there are consistent areas for improvement and support that have been identified by teaching assistants (TAs). This is important to consider, as the pedagogical dynamics between TAs and their students as well as TAs and their departments are significant intersections of learning and retention. Being taught by a well-trained and effective TA will influence a student's undergraduate experience, just as that training will inform the TA's future career in education. Both undergraduate and graduate students have their own coursework and life demands, and TA roles can be demanding of both time and effort \citep{GeptsThomas2024TTTB}. The focus of this section will be on graduate students, as they are more commonly in TA roles compared to undergraduate students, but the recommendations herein are applicable to all TAs. We have identified the following areas of the TA experience as having particular importance for the retention of graduate students: feeling and being supported, having a work/life balance to reduce burnout, appropriate grievance networks, and clear expectations. Improving the skills of TAs is also a way to improve undergraduate student retention, due to the importance of TA roles in instruction and the frequency of contact between TAs and undergraduate students \citep{GardnerGrantE2011PPot}.
\\

\subsubsection{Comprehensive Training Course or Workshop}As many students come into graduate school with no prior experience teaching \citep{Alicea-MuñozEmily2021Ttpo}, providing adequate preparation for this role is a critical step to decrease barriers to success. Some new TAs may have been tutors in a one-on-one setting, but it is relatively rare for them to have experience teaching in a classroom setting. Students entering graduate school are often navigating many difficulties, such as a shifting support network as the result of a move/relocation, the transition from undergraduate to graduate coursework, and learning to balance a different kind of workload that now additionally includes teaching responsibilities. This shift can be especially difficult for women and gender minorities who are also face societal barriers \citep{LiYangqiuting2020HPoB, Porter2020AIP}. One way to mitigate the stress of this transition is to institute a mandatory comprehensive TA training course or workshop  for first-time TAs \citep{DoucetteDanny2020Pdcc, Alicea-MuñozEmily2018AaGp}. Such trainings do not have to be reinvented at the department level. Instead, departments can encourage or mandate attendance at existing short form introductory programs\footnote{For example: Teaching@Yale day (\url{https://poorvucenter.yale.edu/teaching/teaching-programs-and-events/teaching-at-yale-day}) or Teaching@UChicago (\url{https://teaching.uchicago.edu/programs/teachinguchicago})} run through institutional centers focused on teaching and learning, which can be augmented by both discipline-specific instruction and longer term, more in-depth pedagogical training.

Various methods of delivery for TA training and information have been explored, but this paper will focus on the type of information valuable for TAs prior to their first day on the job. TAs should understand what is expected of them. Examples of information that should be covered in TA training are as follows: 

\begin{enumerate}
\item Description of TA duties both inside and outside the classroom -- with clear expectations. 
\item Successful practices to prepare for the first day of teaching/lab instruction.
\item Tips for interacting with students and facilitating group work as well as instructions for dealing with challenging classroom dynamics.
\item Outline of the chain of command for grievances and conflict resolution (with specifics about who to contact for which issue).
\item Information about existing labor and union laws as they pertain to the university and the TAs. 
\end{enumerate}

\pagebreak
Hands-on experience for TAs can help build confidence and teach strategies for answering student questions as well as effectively communicating course material \citep{DoucetteDanny2020Pdcc}. Training for TAs should include information about research-backed teaching and learning practices aligned with the educational goals of the university \citep{GardnerGrantE2011PPot}. Below are examples of \textit{specific} questions and topics that should be addressed during TA training: 
\begin{itemize}
    \item \textbf{Lab}: How is the lab structured for this course? What is the policy if a group finishes the lab early? What is the TA supposed to do during the lab? What does the TA do if lab equipment breaks? 
    \item \textbf{Recitations (or equivalent)}: What is the recitation format? What is the TA responsible for during the recitation? How should the TA prepare for the first day of classes? What should the TA do on the first day of class? How should they introduce themselves and the course? What student resources are available for TAs to promote (tutoring, disability services, etc.)?
    \item \textbf{Office Hours}: What are office hours? Should TAs have office hours? How many office hours per week are required? Can office hours be held online, and what are appropriate times to hold them? What are some recommendations for choosing a time/format for office hours? 
    \item \textbf{Grading}: Are solutions or rubrics provided? How should an answer key or rubric be created (if necessary)? How should grades, including those for the course, be released to students? How should various types of assignments be graded?
    \item \textbf{Quizzes/Exams}: What does effective proctoring look like in the course? What should a TA do if a student is suspected of an academic integrity violation?
    \item \textbf{TA Resources}: How does the TA gain access to the instructor version of the textbook? How does the TA use any tools/technology required for the course? Who should the TA contact with student issues or concerns? How much freedom do TAs have in addressing student concerns?
\end{itemize} 

It is also important for TAs to understand how to handle emergencies and what concerns should be reported directly to the professor or the department. Having a clear understanding of expectations and how to meet them boosts the confidence of graduate students undertaking teaching assistantships and makes their experience more positive, supporting retention.

\subsubsection{Clear and Accessible Grievance Processes} TAs typically report to a faculty or staff member, whether it is directly to the instructor of the course, to a course manager, to a TA supervisor, or directly to the department chair. However, the process for reporting grievances is not always clear and accessible to TAs, creating a barrier that can make navigating disputes difficult for underrepresented students. Departments can consider implementing a paid teaching assistant position: someone to act as a liaison between the physics/astronomy TAs and those with the power to make changes for them, similar to the ``head TA'' position at some universities\footnote{For an example, see \url{https://carter.faculty.pstat.ucsb.edu/TAHandbook/Part_1_TA_Expectations/head_tas.html}} or a TA ombudsperson. When possible, we advise that the person in this role not have a course load, have previously experience as a TA, be post-candidacy, and not hold a TA position at the time of taking the role. TAs are often overworked and their well-being is, unfortunately, frequently ignored \citep{HodgkinsAngela2024TARE}. Having this position would ensure that TA concerns are heard and addressed by those in a position to make changes. It would also likely lead to TAs feeling more comfortable speaking to a dedicated support person, rather than directly to a more senior member of the department. Lumen Learning \footnote{\url{https://courses.lumenlearning.com/hrmanagement/chapter/12-3-administration-of-the-collective-bargaining-agreement/}} shows an example grievance process for unionized TAs. 

\subsubsection{Transparency of TA Workload} Every university and department has different requirements and expectations for TAs. These requirements are often not clearly communicated to the TAs, which can represent a significant barrier to success. This lack of clarity on responsibilities and time expectations can cause unnecessary stress. Departments should outline a clear definition of TA duties, as well as estimated times for each genre of responsibility. An example of a clearly-outlined TA position would look like: 

\begin{quotation}
   \noindent The TA is responsible for instructing two weekly recitations per section (4 hrs/week), one lab per section (4 hrs/week), grading student assignments (3 hrs/week), proctoring exams, holding one office hour per week per section (2 hrs/week), and responding to student emails (0.5 hrs/week).
\end{quotation}

Such lists of responsibilities should be public and accessible, allowing prospective graduate students and TAs to develop a clear picture of their work expectations, which is extremely useful for requesting specific TA assignments or choosing between graduate programs. Testimonials can be collected (and ideally similarly made publicly available and accessible to students) from previous TAs about their work experience to help assess the workload for the position. Throughout the semester, there should also be ``check-ins'' to make sure the amount of time TAs are spending on different tasks is reasonable and in accordance with contracted expectations, and if not, adjustments should be made.

TAs are typically graduate students who are balancing their own coursework, research, and personal lives in addition to teaching \citep{GeptsThomas2024TTTB}. Ensuring that TAs have clear knowledge of their responsibilities and then checking in to make sure they are managing the work provides a more consistent experience for students across courses. This especially benefits women and gender minorities, who face similarly heightened expectations of emotional labor \citep{BartosAnnE.2019'tro} as female faculty \citep{MisraJoya2011TICo}. In turn, TAs will feel that their well-being is prioritized, and they will be less frequently overworked, leading to an improved balance of priorities between teaching, research, and coursework. 

\subsubsection{Department Undergraduate Tutoring} Peer tutoring represents an opportunity to lower systemic barriers for both TAs and their students by alleviating some TA workload and providing additional resources for student support. Undergraduate students are frequently looking for extra support in their physics and astronomy courses. When students are unaware of extra support, or when that extra support is not offered by the department, this often falls on the TA or course instructor, by means of office hours or increased demand for one-on-one attention. Departments are invited to coordinate paid tutoring positions for undergraduate physics and astronomy courses, where the tutors can be senior undergraduate students or graduate students. 

Depending on available resources, departments or faculties could offer an hourly wage for tutors to work in a designated ``Physics Help Room''. Tutors can sign up to work in the room at specific times and students can drop in for free tutoring. The schedule with available tutors and at what time/location should be easily accessible on the department webpage and in high-traffic areas such as meeting spaces. An example of a ``Physics Help Room'' is being implemented at the University of Michigan\footnote{\url{https://lsa.umich.edu/physics/undergraduate-students/introductory-physics-courses/tutoring/}}, and the SNAP Centre at Saint Mary's University\footnote{\url{https://www.smu.ca/faculty-of-science/science-snap-centre.html}} is an example of a functioning peer-tutoring program. The department should ensure there are dedicated spaces reserved for said tutoring, with a sign-up process for graduate/senior students. If funding is not available for paid positions, departments should consider connecting students to existing tutoring resources within the university or beyond -- such as coordinating or advertising private tutoring databases.

Providing tutoring is beneficial for both students and TAs. It can provide a free resource for students to get help, which can be especially beneficial for students who do not have the financial means to get this support otherwise. It also helps to lighten the ``extra-help'' workload of TAs and instructors and can provide supplemental income for those students who act as tutors. Tutoring also provides lower-stakes teaching experience for those who may be interested in teaching or teaching assistantships in the future. 

\subsubsection{Track TA and Award Recipient Demographics} Disparities in assigned TA roles can indicate systemic issues within a department. Being a TA adds a significant workload to students. This can put TAs at a disadvantage in comparison to students who are supported exclusively by fellowships or research assistant (RA) positions. Departments should track the demographics of TAs, proportionally to student demographics, as well as award recipients (for department/university provided scholarships, fellowships, etc.) to see if there are trends in gender or race. This holds departments accountable and can provide insight into whether the TA roles are disproportionately being held by certain groups as a discrepancy has been observed in other sources of funding such as research grants \citep{genderlanguage}, and is documented in similar workplaces \citep{babcock}. Being aware of patterns can help reduce future biases.

\subsubsection{Acknowledge TA Achievement} By acknowledging TA achievement, departments can provide social incentives for TAs to excel in their role. Ensuring quality of TA work is a multi-dimensional effort that leverages personal motivation to succeed, reputation management in the department, and balancing TAing with coursework and research. Many times, graduate students view their role as a TA as secondary to their coursework and research \citep{ParkChris2004Tgta, GardnerGrantE2011PPot} and are often encouraged in that view \citep{ShannonDavidM1998Tte}. One possible way to motivate TAs to put effort in their role is to recognize hard work and ``outstanding TAs'' during departmental award ceremonies. Monetary awards would be the most influential\footnote{Such as the one described in this article: \url{https://www.k-state.edu/today/announcement/?id=6260}}, however there would still be benefit just in recognition and appreciation, if funding is not possible. The awards should be given based on factors including positive student evaluations and interactions with supervising faculty. However, it should be noted that some groups of students, such as gender and racial minorities, typically receive harsher feedback in student evaluations. In a study conducted by \citet{EvansC.A.2024Gpie}, they found that when students evaluate women they tend to focus more on personality traits rather than their knowledge and ability to teach, while evaluations of men tend to be more focused on their understanding and delivery of the material. The overall ratings of men tended to be higher than that of women, which should be considered if recipients of these awards are decided based on surveyed feedback. Recognizing TAs for their accomplishments makes them feel valued and appreciated and motivates them to put effort into their TA role, which in turn enhances the experience of undergraduate students, encouraging them to remain in the program. 

\subsubsection{Implement TA Well-Being Check-Ins} One step to help socially support TAs is to implement well-being check-ins. This is a chance to check in about their workload, address any issues they may be facing or questions they may have, and provide any support they may need. Climate-testing questions for TAs could include: 
\begin{enumerate}
    \item Do you feel you have the authority/support to enforce academic policies?
    \item How manageable/balanced is the workload? 
    \item What are particularly difficult issues/aspects of the job?
    \item How comfortable are you reaching out to your overseeing faculty member/TA liaison?
    \item Do you feel like your work as a TA is recognized/valued?
    \item Do you feel like students understand your role and how they can best utilize you as a resource for their learning?
    \item What are some things that are working well/that you feel you are doing well?
\end{enumerate}

Depending on the size of the department, this can be done one-on-one, in surveys, or in small groups with the person in charge of the TAs. This should be to gather feedback and generate recommendations for change for future terms (or the current term if possible). Departments should reflect on TAs' responses and look to improve the experience for TAs. One such approach could be applying `Community of Practice', as outlined by \citet{CamaraoJoy2023“GoP}, which facilitates TA confidence in teaching through collaboration with peers. Checking in with TAs helps strengthen the TA-department communication, ensuring shared understanding about the workload, what improvements can be made, and what is working well. 

\subsubsection{Implement Clear Artificial Intelligence and Grading Policies} As Artificial Intelligence (AI) is becoming more and more prevalent and unavoidable in the world today, its use has become an issue of equity and fairness. Clear policies for handling the use of AI benefit professors, students, and TAs alike. An AI-use policy should include descriptions of acceptable use if any, such as having students cite how and where they have used AI, as well as how they know they can trust the answer -- such policies will likely be course-dependent. This ensures transparency and proper evaluation of the students' efforts in graded materials. For more detailed examples, the University of Kentucky has created a webpage with example AI-use policies for the cases of no AI use, conditional AI use, or unrestricted AI use\footnote{\url{https://celt.uky.edu/ai-course-policy-examples}}. The AI policy, along with the academic integrity policy, should be stated in the course syllabus. Simply stating ``do not use AI'' is no longer explicit enough, especially considering how much -- and how forcibly -- AI has become integrated into daily life. The policy should outline what students can and cannot use AI for, as well as the consequences for its use in contexts in which it is prohibited. The consequences of the violation should be proportionate to the violation and should not punish students incommensurately for small violations; this makes enforcing the policy easier for TAs. Providing clear policies ensures that TAs have a consistent response to violations and that students will have a clear understanding of what is considered academic misconduct. 

Having access to grading materials represents another equitable practice that can benefit TAs. Detailed solution manuals, answer keys, and rubrics allow for consistency and reduce bias during grading across students and across TAs \citep{PhysRevPhysEducRes.13.010120}. Providing detailed solutions allows TAs to understand the instructors' expectations and step-by-step reasoning, enabling them to provide direct guidance and feedback to students. Instructor-provided solutions help bridge this gap and ultimately enhance the experience for both students and TAs.

\subsection{Summary and Conclusions}
These best practices and potential actions are proposed as a framework for how departments, faculty members, and all astronomers and physicists can support each other through pedagogy across career stages. It is more important than ever for departments to help their students access support\footnote{We acknowledge that this is, at present, financially and logistically difficult for some institutions and regions.}. The solution to the challenges faced by underrepresented persons is not to place additional unpaid labor on faculty, administrative staff, and graduate students, but to balance the needs of the department as a whole in order to improve the experience and retention of all students -- especially those from underrepresented and minority groups. 

Ultimately, this chapter aimed to outline actions that individual students, departments, administrations, and faculty can take in order to improve their department culture and resources, and allow them to advocate for better underrepresented student support. A variety of examples and suggestions have been presented, with an emphasis placed on connecting students to ample, potentially lesser-known, existing resources. Below follows a summary of best practices and \textbf{core values} for retention of underrepresented students as were outlined in this document:
\begin{enumerate}
\item \textbf{Lower systemic barriers.}
\begin{enumerate}
\item In the classroom: Create opportunities for representation of women and all underrepresented persons. Promote advancement opportunities for all students.
\item Out of the classroom: Establish and champion professional societies and affinity groups that support mentoring and connecting underrepresented students. Provide support around graduate school applications and research opportunities. 
\item With TAs/graduate students: Provide comprehensive training and workshops for TAs to assist in effective managing of new responsibilities. Provide TAs with a clear grievance process, and transparency about the expected workload. Institute undergraduate peer tutoring. Be aware of TA demographics and monitor for potential imbalances in position assignments, and reward TA contributions. 
\end{enumerate}

\item \textbf{Provide social support.}
\begin{enumerate}
\item In the classroom: Use research-backed teaching methods to make class structures more interactive and inclusive. Provide department-wide training and resources about how to implement equitable learning practices. 
\item Out of the classroom: Encourage peer mentorship, especially among undergraduates, graduate students, and postdoctoral researchers. Connect students to external research opportunities and information, as well as maintaining connections to non-academic career and outreach opportunities. 
\item With TAs/graduate students: Acknowledge TA achievement, and provide mid-term wellness checks for TAs, as well as support personnel within departments. 
\end{enumerate}

\pagebreak
\item \textbf{Implement equitable practices.}
\begin{enumerate}
\item In the classroom: Adopt a rolling feedback model for courses, as well as universal and flexible policies for graded material. Ensure explicit learning goals and grading transparency are provided. 
\item Out of the classroom: Support mental health care, foster a sense of student belonging, encourage healthy work-life balance, affirm students' identities, and provide access to mental health resources. Have accurate information available pertaining to current research within your department. 
\item With TAs/graduate students: Outline clear expectations for grading in the era of generative AI. Provide clear grading criteria, including answer keys detailed solutions, and rubrics. 
\end{enumerate}
\end{enumerate}

It remains important to continually examine pedagogical dynamics and how they can support or hinder students as they progress in astronomy and physics. As explained by \citet{bell_hooks_teaching_1994} in \textit{Teaching to Transgress: Education as the Practice of Freedom}:

    \begin{mdframed}[
  leftline=true,
  rightline=false,
  topline=false,
  bottomline=false,
  linecolor=gray,
  linewidth=2pt,
  innerleftmargin=10pt,
  innerrightmargin=5pt,
  innertopmargin=6pt,
  innerbottommargin=6pt
]
\textit{``The unwillingness to approach teaching from a standpoint that includes awareness of race, sex, and class is often rooted in the fear that classrooms will be uncontrollable, that emotions and passions will not be contained. To some extent, we all know that whenever we address in the classroom subjects that students are passionate about there is always a possibility of confrontation, forceful expression of ideas, or even conflict. In much of my writing about pedagogy, particularly in classroom settings with great diversity, I have talked about the need to examine critically the way we as teachers conceptualize what the space for learning should be like. Many professors have conveyed to me their feeling that the classroom should be a ``safe'' place; that usually translates to mean that the professor lectures to a group of quiet students who respond only when they are called on. The experience of professors who educate for critical consciousness indicates that many students, especially students of color, may not feel at all ``safe'' in what appears to be a neutral setting. It is the absence of a feeling of safety that often promotes prolonged silence or lack of student engagement. Making the classroom a democratic setting where everyone feels a responsibility to contribute is a central goal of transformative pedagogy.''}
\end{mdframed}

In order to cultivate diversity and improve the retention of gender and racial minority students, we must engage in transformative pedagogy and reinvent the inaccessible systems that have enabled the leaky pipeline. By instituting policies that \textbf{address systemic barriers, provide strong social support, and promote equitable practices}, we can collectively create a more inclusive and effective learning environment for all students.

\subsection{Acknowledgments}
We would like to thank the organizing team of the Picture an Astronomer symposium for facilitating these discussions and the creation of this white paper, and our collaborators for their valuable feedback and comments. We also wish to thank Professor Oliver Fraser, from the Department of Astronomy at the University of Washington, for providing additional insights and comments.

\bibliography{references/pedagogy}{}
\bibliographystyle{aasjournal}
\pagebreak
\section{Summary of Recommendations} \label{sec:summary}

\vspace{-20pt}
\begin{center}
    \includegraphics[width=0.4\linewidth]{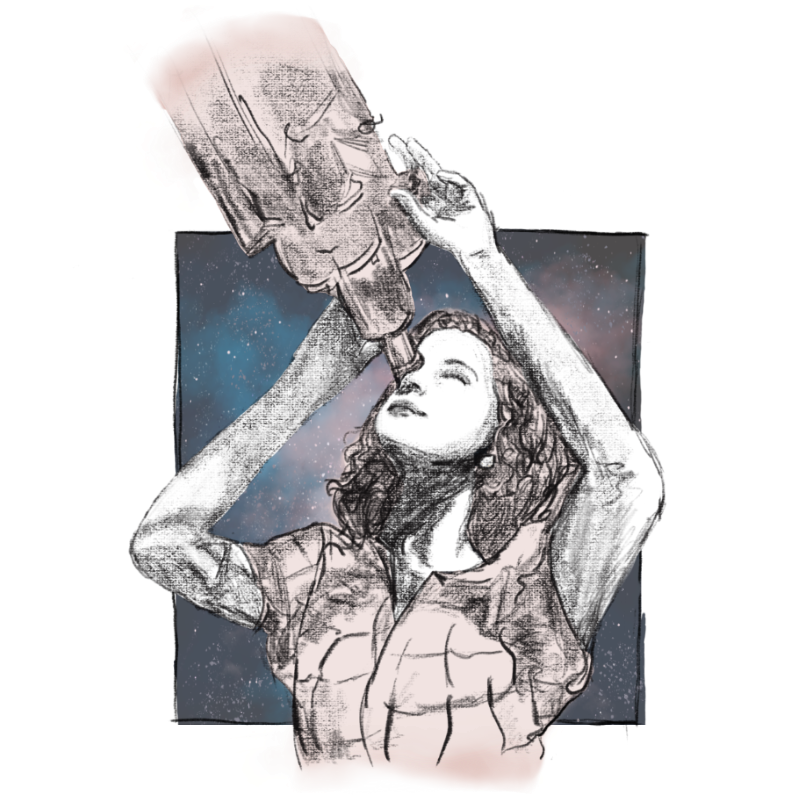}
\end{center}

Each section in this white paper is broken out based on an individual discussion focused on a typical barrier to the full and continued participation of talented women in astrophysics. However, because these challenges are often pervasive and intersecting, there is substantial overlap in recommendations that emerged from the different discussion groups. At the same time, although these recommendations were developed in the context of gendered attrition in astrophysics, they ultimately support a healthier climate 
for \textit{all} scientists. Implementing them would help retain talented researchers across genders, career stages, and disciplines,
not only women and not only astronomers. 

In what follows, we summarize the key suggestions offered in each chapter and highlight areas where proposed interventions recur across sections. We emphasize, however, that this summary is \textit{not} a substitute for the contextual discussion and supporting literature presented in the main text, and we strongly encourage readers to review the preceding chapters for a more in-depth discussion of each recommendation and its motivation.

Several themes appear repeatedly across sections. 
For example, both hiring and proposal evaluation processes should be as transparent and objective as possible, with clearly defined rubrics and dual-anonymous review to reduce biases. Because the small size of professional astronomical community and existing power structures in academia make it hard for (particularly junior) people to withstand bullying and harassment without lasting damage to their career, stronger scrutiny of letters of recommendation and meaningful, enforced codes of conduct are essential.
Women are also disproportionately asked, and often socialized, to take on  service work that is undervalued in hiring, promotion, and awards. Actively recognizing and rewarding high-quality teaching, outreach, and committee work, alongside research, can therefore promote greater equity.  

Below, we list the recommendations from the larger document (in no particular order), grouped by area of impact and accompanied by links to the sections where they are discussed in more detail. Summary ``cheat sheets'' are also available for download on the Picture an Astronomer website\footnote{\url{https://pictureanastronomer.github.io/whitepaper}}.

\subsection{On representation}

\begin{itemize}
    \item Senior women should be visible saying no, as in turning down service roles, and being outspoken, including weighing in on department culture or policies. This models what behaviors are ``allowed'', as many women fear retribution if they are not seen as agreeable and communal. (Section \ref{sec:resources})
    \item Policies that require \textit{representative} committees may overburden underrepresented community members and leave them little choice in their service obligations. This may be mitigated by allowing for more flexibility in that policy or otherwise establishing additional policies that recognize the need for representation on committees must be superseded by an equitable distribution of departmental/institutional service work. (Sections \ref{sec:bias} and \ref{sec:pedagogy})
    \item Studies have shown that when women see female scientists represented in popular media, they are more likely to view themselves as capable of becoming scientists. Similarly, who we choose to highlight in classrooms, seminars, and in other academic contexts can strongly influence whether students feel that they \textit{belong} in the field. For example, do we mention Chien-Shiung Wu alongside Tsung-Dao Lee and Chen Ning Yang when discussing the discovery of parity violation?  And when referring to the women who carried out groundbreaking work at the Harvard Observatory, do we frame their contributions appropriately rather than relying on outdated and dismissive labels like ``Pickering's harem''? 
    (Sections \ref{sec:representation}, \ref{sec:intersectional}, and \ref{sec:pedagogy}) 
    \item Professional societies and affinity groups may serve as critical support networks for scientists at all career stages. They can also serve as centralized bodies to promote change within a discipline, whether by implementing inclusive policies themselves or facilitating the broader adoption of such policies via endorsements or working group activities. (Section \ref{sec:pedagogy})
\end{itemize}

\subsection{On teaching and mentorship}

\begin{itemize}
    \item Institution- or department-level policies on mentoring, including regular check-ins or opportunities for structured feedback, can help establish mutual student-advisor expectations and address implicit bias in professional interpersonal relationships. Faculty, and others who are charged with supervising students, should attend workshops on mentoring and teaching to gain more formal insight into best practices. (Sections \ref{sec:bias}, \ref{sec:intersectional}, and \ref{sec:pedagogy})
    \item Teaching assistants are in a position both of instructing others and of being mentored (or at least supervised) by the instructor of record. First-time TAs should receive targeted training--either through a campus teaching and learning center or through discipline-specific departmental programs--along with a clear explanation of their responsibilities and the limits of their authority. They should also have regular check-ins with the instructor of record. To resolve conflict or provide a confidential sounding board for TA concerns, departments may also consider appointing a ``head TA'' or ombudsperson to serve as a peer-level resource. (Section \ref{sec:pedagogy})
    \item Students and more junior department members may benefit from ``safe'' opportunities to practice written and oral presentation of research both inside and outside of a department. Similarly, departments may wish to establish a low-stakes environment for more junior scientists to ask questions of speakers by having closed student-only question-and-answer sessions or by creating a means by which questions may be asked anonymously. (Sections \ref{sec:bias} and \ref{sec:pedagogy})
    \item Instructors should emphasize a ``growth mindset'', helping students understand that success in physics does not depend on some innate or fixed ability, 
    but rather on persistence, effort, and effective learning strategies. Similarly, instructors can also support students' self-efficacy by encouraging them to reflect on their motivations for pursuing the course or degree and to recognize their own capacity to succeed. (Section \ref{sec:representation})
    \item Incorporating active learning strategies in the classroom has been shown to improve learning outcomes for \textit{all} students, while also closing the ``achievement gap'' for underrepresented students. Active learning is a very broad designation, including everything from cold (or ``warm'') calling students to in-class group problem solving, and can be easily and organically added to most (astro)physics curricula. (Sections \ref{sec:representation}, \ref{sec:intersectional}, and \ref{sec:pedagogy})
    \item Instructors should practice transparency in connecting assessment to course content via clear learning objectives and rubrics. In addition to directly benefiting students, creating these resources also facilitates lesson planning and grading. (Sections \ref{sec:intersectional} and \ref{sec:pedagogy})
    \item Students' anonymous evaluation of instructors is heavily influenced by gender bias, even in cases of identical instruction and course policy. Teaching evaluations should accordingly be de-emphasized in merit and promotion decisions, though adopting evidence-based teaching pedagogy and evaluation processes (as provided by a formal observer from the university's center for teaching and learning) should remain a priority. (Sections \ref{sec:bias} and \ref{sec:resources}) 
    \item Everyone engaged in mentoring should undergo \textit{active} anti-bias training, which may be structured as a monthly seminar series or a collaborative learning group. This structure allows for more in-depth participation than a one-day training and greater retention of best practices among participants. (Section \ref{sec:resources})
    \item Instructors should foster inclusive classrooms, considering the impact of the nature of the environment and the means by which the content is delivered for learners from varied backgrounds and with varied needs. For instance, course materials should designed to be maximally legible for those with color blindness and dyslexia and to be readily accessible by students relying on screen readers. (Section \ref{sec:intersectional})
    \item Instructors should, on a rolling basis, act on specifically solicited mid-course feedback. Timely response to valid student concerns, which are sometimes structural, helps students currently enrolled in the course and establishes a mutual investment in classroom policies and practices. (Section \ref{sec:pedagogy}) 
    \item Departments should consider adopting homogeneous policies for addressing late or missed work or accommodating students taking medical leave. Consistency in the implementation of policies across a discipline's coursework lessens the administrative burden on both instructors and students and sets equitable expectations for performance. (Section \ref{sec:pedagogy})
    \item Instructors may wish to use ``flexible'' assessment practices, giving students the freedom to choose between comparable assignments (e.g., leaving open-ended a project resulting in either a presentation or a paper, or allowing students to choose between taking an oral or written exam). Flexible assessments more readily accommodate student preference, helping them play to their strengths while also demonstrating content mastery, and can be inclusive of different learner needs without students having to specifically seek out accommodation. (Sections \ref{sec:intersectional} and \ref{sec:pedagogy})
    \item Peer or ``near-peer'' mentorship is a viable avenue for professional and academic growth, augmenting more classical ``top-down'' mentorship as from a research advisor or course instructor. Departments may wish to establish formal programs to promote peer mentorship or otherwise encourage informal mentoring relationships, which can account for significant development outside of a research or classroom context. (Section \ref{sec:pedagogy})
    \item Institutionalized peer tutoring resources, whether through existing programs at the university- or institution-level or discipline-specific efforts facilitated by departments, provide an avenue by which students can get focused help outside of office hours and create an opportunity for more senior (or more confident) students to practice pedagogy in a small group setting. (Section \ref{sec:pedagogy})
\end{itemize}

\subsection{On hiring and awards}

\begin{itemize}
    \item Letters of recommendation should be requested at the end of the review process. While there is little value added to an application, particularly as applicant seniority increases, beside reinforcing other parts of the submitted materials, there is the risk of introducing gendered or other biased language into the assessment. (Sections \ref{sec:bias} and \ref{sec:resources})
    \item In cases where letters of recommendation are required, hiring units should establish avenues by which to exclude negative or less-than-glowing letters. This may look like providing feedback to applicants if letters are adversely impacting the overall hiring/promotion package, hiring committees looking into the background (personal and/or cultural) of individual letter writers for more complete context, and/or creating a field in job applications where applicants can indicate that they did not seek a letter from a mentor and why. (Section \ref{sec:bullying})
    \item Hiring units should be transparent about what items in a compensation package are negotiable (or have historically been changed by negotiation) and give approximate ranges for ``reasonable'' requests. (Section \ref{sec:resources})
    \item Rubrics should be created and used to ensure that the same standard is applied to all applicants. (Section \ref{sec:bias})
    \item Evaluations that include ``time to degree'' or career stage defined in years post-PhD assume a linear trajectory. Hiring (and promotions) should be flexible in accounting for gaps/delays. Similarly, department and institution leave policies should be provided to candidates regardless of their gender, age, or career stage. (Sections \ref{sec:bias} and \ref{sec:family})
    \item Hiring committee members--and anyone else interacting with candidates--should undergo anti-bias training. To ensure that the training is effective, it should be conducted in person with limited opportunity for distraction. (Sections \ref{sec:bias}, \ref{sec:resources}, and \ref{sec:pedagogy})
    \item Anyone engaging with job candidates should also be trained in what questions are permissible and what questions are illegal in hiring to ensure equal treatment of all candidates. (Section \ref{sec:family})
    \item Awards and solicitations for nominations should be advertised at the department level so that everyone is aware of opportunities without relying on informal information networks that may exclude some scientists. Similarly, advertisements of job opportunities, including for the most junior department members, should be circulated widely. (Sections \ref{sec:resources} and \ref{sec:pedagogy})
    \item Employers should, where legal, disclose anonymized information about their employee demographics, including pay, promotion, and retention information. Tracking (and sharing) salient hiring and retention information ensures that concerning patterns can be remedied, while simultaneously offering an avenue to quantify progress. Departments should similarly consider tracking the demographics of students funded through teaching vs. research assistantships or independent fellowships. (Sections \ref{sec:intersectional} and \ref{sec:pedagogy})
    \item Hiring and promotion committees should consider positively a demonstrated commitment to accessibility and equity, such as developing inclusive research tools or effectively teaching and mentoring students with diverse needs.  (Sections \ref{sec:intersectional} and \ref{sec:pedagogy})
    \item When hiring or recruiting for undergraduate research positions, clearly describing the prerequisite skills and/or coursework and setting explicit expectations for the project can make the process more equitable. Departments may also wish to ensure that internally offered research opportunities span a range of required experience levels, so that there are always low barrier, entry-level projects available for students who are just beginning to develop their research skills. (Section \ref{sec:pedagogy}) 
\end{itemize}

\subsection{On benefits and conditions}

\begin{itemize}
    \item Departments should, where possible, reserve a number of guaranteed spots in university-affiliated childcare centers for their members, including postdocs and graduate students. For these groups, childcare often represents a disproportionately large financial burden relative to their salaries, and guaranteed access can significantly reduce stress and barriers to participation. (Section \ref{sec:family})
    \item Institutions, departments, groups, and collaborations should clearly define the time commitment, scope, and assignment of service roles. Departments should also communicate the typical time required for advising, service, and teaching at different levels of engagement (minimum, average, and heavy workloads). Doing so helps faculty benchmark their own commitments and identify when expectations may be exceeding reasonable norms. (Sections \ref{sec:bias}, \ref{sec:resources}, and \ref{sec:pedagogy})
    \item For groups undergoing unionization, it is critical to advocate for family-friendly policies for those in the collective bargaining unit. (Section \ref{sec:family})
    \item Where possible, universities should offer postdocs, including independent fellows, and graduate students the opportunity to be hired as employees so that they may take advantage of employee protections and leave policies, as well as other benefits. (Section \ref{sec:family})
    \item Institutions should provide students and postdocs who take parental leave an additional year on their employment to both diminish the deleterious effect of time taken off of research while at a precarious career stage, and to ensure that the end of their appointment is aligned with the timing of the academic job cycle. In the same vein, institutions should also extend tenure clocks accordingly. (Section \ref{sec:family})
    \item Departments (and larger institutions) should make clear how all members can access institutional accommodations or support, including disability services and healthcare. (Sections \ref{sec:intersectional} and \ref{sec:pedagogy})
\end{itemize}

\subsection{On gatherings and travel}

\begin{itemize}
    \item Robust hybrid components to meetings allow individuals with family commitments and other unyielding responsibilities, health concerns, and visa/travel constraints to better participate in the global scientific community. (Sections \ref{sec:family}, \ref{sec:intersectional}, and \ref{sec:global})
    \item Conference organizers should take into account state/province and national policies when selecting a conference location to be maximally inclusive of potential attendees. For instance, women who are, or may become, pregnant may avoid conferences in places that have curtailed reproductive rights for their safety. Similarly, LGBTQ+ scientists may not be able to travel to places where their existence is not accepted. (Section \ref{sec:intersectional})
    \item Meeting should, when possible, be scheduled during the traditional work day (and ideally during regular school hours) to accommodate people's commitments outside of work. This can be especially challenging when working in international collaborations when telecons/calls have to work for potentially disparate time zones; in those cases, efforts should be made to minimize requirements for synchronous attendance, potentially recording or transcribing meetings for people who need to catch up asynchronously outside of the set meeting time. (Section~\ref{sec:family})
    \item Employees should feel empowered to request accommodations related to field deployments and professional travel. Groups should also consider having a contingency plan in place, including scientific understudies, for extended travel. (Section~\ref{sec:family})
    \item Conference organizers should provide flexible childcare funding, which can be used either for on-site care at the conference location or for care at the parent's preferred location. Priority should be given to early-career scientists and those for whom childcare poses a greater financial burden. (Section \ref{sec:family})
    \item At conferences, session chairs/moderators should be trained in addressing bullying or bias in question-and-answer sessions, where they are expected to guide the discussion and navigate potentially uncomfortable power dynamics. (Section \ref{sec:bullying})
    \item Venues and workspaces--including offices--should be selected and set up with inclusivity and accessibility in mind. Both professional and public outreach events should offer reasonable accommodations to ensure that those with different accessibility needs can fully participate. (Section \ref{sec:intersectional})
    \item Institutions should offer specific help to international scholars navigating visa processes, including providing funding for visa fees (and costs associated with travel for visa appointments), endorsement of longer term and/or expedited visas, and logistical support with the administrative aspects of the application. Conference organizers should be conscious of the uncertainty of visa applications when offering early invitation letters and flexible registration refunds. (Section \ref{sec:global})
\end{itemize}

\subsection{On norms in the field}

\begin{itemize}
    \item Senior department/institute members must lead by example, upholding and enforcing high standards (and, at minimum, codes of conduct) in interpersonal and professional behavior. Bias, micro- and macro-aggressions, bullying, and harassment cannot be tolerated, and intervention--from the least vulnerable people present--must be immediate in the moment the transgression occurs. (Sections \ref{sec:bias}, \ref{sec:bullying}, and \ref{sec:intersectional})
    \item Groups and collaborations should, in addition to adopting clear and enforceable public codes of conduct, have in place obvious support structures for victims and maintain databases to track repeated incidents or retaliation for reports. (Sections~\ref{sec:bullying} and \ref{sec:intersectional})
    \item Students--and more senior department members--should be given both the opportunity and incentive to engage in pedagogy, outreach, and service. Service roles should be assigned on a first come, first serve basis, with preferred tasks going to those who volunteer for them, thereby cutting down on assignments made to more agreeable individuals because of others' weaponized incompetence. Awards that highlight and acknowledge achievement in teaching and community engagement can help recognize effort and demonstrate the importance of non-research skills. (Sections \ref{sec:bias}, \ref{sec:intersectional}, and \ref{sec:pedagogy})
    \item Effective presentation skills should be valued by departments and the larger community, and effort should be made to recognize and communicate that discussing science without jargon is indicative of a deep understanding rather than disinterest or lack of expertise. (Section \ref{sec:bias})
    \item Department members should receive training in teaching/mentorship, bystander intervention, and in recognizing and confronting bullying and implicit/explicit bias. All trainings should be conducted in person to minimize distractions, and, where relevant, should be heavily incentivized or made mandatory. Trainings regarding bullying and bias could be augmented by a short video series based on realistic interactions within astronomy to move beyond treating the workshop as a purely intellectual exercise. (Sections~\ref{sec:bias}, \ref{sec:resources}, and \ref{sec:intersectional})
    \item Departments should regularly offer (or point members to university-wide) training focused on self-advocacy and self-promotion. This may include opportunities to learn how to engage with the university's media office; gain peer feedback or advice on grant writing, CVs, or research/teaching statements; or discussion of effective negotiation strategies. (Section \ref{sec:resources}) 
    \item Institutions should cultivate a vocally supportive culture where it is clear that people can ask for accommodations, including, but not limited to, family life or caregiving, without repercussions. (Section \ref{sec:family}) 
    \item Professional societies like the American Astronomical Society or the International Astronomical Union should maintain lists of ongoing equity initiatives, so that members may be made aware of existing resources within the community. At an institutional level, this goal can be supported by implementing standardized onboarding procedures at all career stages that includes both logistical information and clear guidance on codes of conduct and grievance policies. (Section~\ref{sec:bullying})
    \item The larger astronomical community may want to develop an independent resource, external to any individual department or institution, to help guide people impacted by bullying and harassment through a potential grievance process without necessarily engaging with a mandated reporter at their home institution. Such a resource would also be able to offer advice on situations that do not rise to the level of requiring a formal report, while being sensitive to the politics and power dynamics of academic astronomy. Such an effort could potentially be housed within a professional society. (Section \ref{sec:bullying})
    \item Individuals and institutions should facilitate recognition of name changes, whether formal or informal, without fanfare, as they may be due to family, personal, or social circumstances. For instance, most journals now allow retroactive name changes on published articles for continuity. In the classroom and at conferences, it is simple to implement a ``preferred'' name policy. (Section \ref{sec:intersectional})
    \item Departments should use regular climate/community surveys as a means of tracking internal perception of institutional culture. Such surveys also serve as a starting point for planning department-level interventions. (Section \ref{sec:intersectional})
    \item Conferences, workshops, and other professional opportunities, including grants/fellowships and facilities' calls for proposal, should be shared widely within departments and institutions. Leaving people to find opportunities themselves can exacerbate gaps in information networks and lead to substantial inequity. (Section \ref{sec:pedagogy})
    \item Websites--both for departments and for individuals--should be kept up-to-date, highlighting recent work. This ensures that students and junior researchers are able to familiarize themselves with potential opportunities for employment or collaboration. Such a practice may have the added benefit that external parties are then aware of ongoing projects for potential collaborations or awards. (Section \ref{sec:pedagogy}) 
    \item Universities do not exist outside of the context of their local communities. Departments should actively cultivate relationships with cultural institutions, local businesses, and community organizations to both facilitate outreach efforts and engage more broadly across the town-gown\footnote{The term ``town–gown'' refers to the collaborative and sometimes tense relationship between a university (where the term ``gown'' is based on the traditional academic dress) and the surrounding non-university community.} divide. (Section \ref{sec:pedagogy})
    \item Departments should encourage their members to share preferred name pronunciation and pronouns to promote a culture of cross-cultural respect. Instructors can encourage that this information also be shared in learning management system or in class at the start of a new term. (Sections \ref{sec:intersectional} and \ref{sec:pedagogy})
    \item Introductory coursework and/or some department-level events should include the basic details of an academic career, such as typical time spent at each career stage; expectations for graduate, postdoctoral, and pre-tenure work; norms around funding structures, such as guaranteed compensation vs. resources allocated from earning grants, fellowships, and awards; and proposal for, and use of, major facilities. It is best if such discussion also includes practical elements like working through the steps of writing a strong proposal. Standardizing such instruction across a department improves professional outcomes and helps to make explicit some of the ``hidden curriculum''. (Sections \ref{sec:global} and \ref{sec:pedagogy})
    \item Individual collaborators, research groups, and professional journals can improve access to academic publishing for non-native English speakers by offering help with proofreading and editing. Supporting non-native speakers in preparing for presentations by listening to, and giving feedback on, practice talks, can help allay concerns about speaking publicly in a second language. (Section \ref{sec:global})
    \item Creating astrophysics glossaries, translations of paper summaries, and public engagement talks in multiple languages--whether on an institutional- or country-level--helps scholars who are trying to do native language outreach by alleviating some of the potentially duplicated administrative work that comes with such efforts. (Section \ref{sec:global})
\end{itemize}

\pagebreak
\section{Further Reading and Resources} \label{sec:reading}

\vspace{-30pt}
\begin{center}
    \includegraphics[width=0.4\linewidth]{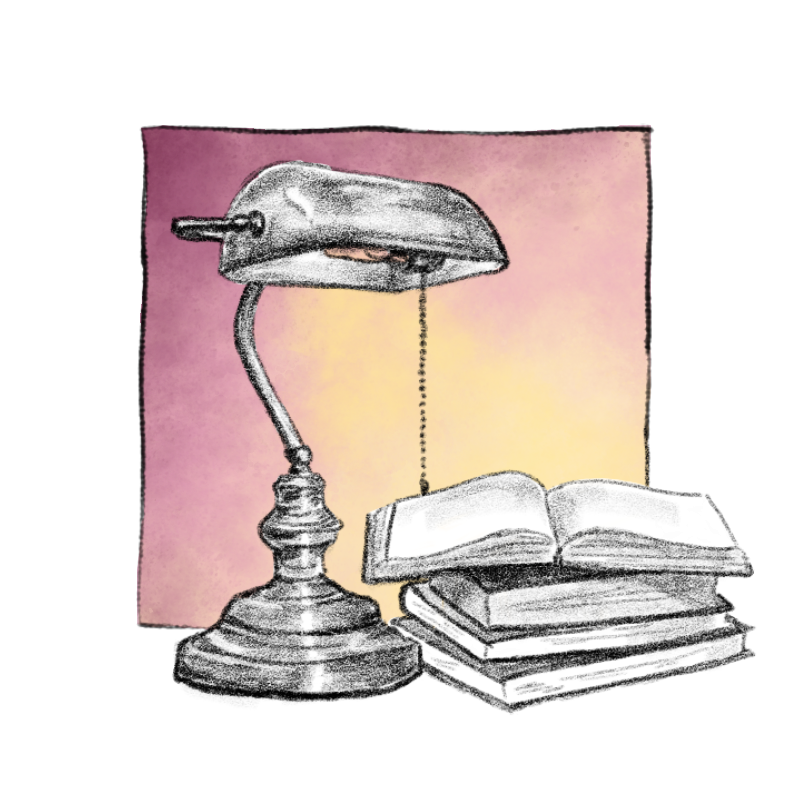}
\end{center}
\vspace{-30pt}

Though this white paper is an effort to collect in one place a number of actionable community-level interventions aimed at improving the retention of women in astrophysics, it is very challenging to be truly comprehensive in approaching a challenge as large as the chronic underrepresentation of 50\% of the population. To facilitate future discussions, in addition to the literature presented within each section, we list here some resources (many of which were also referenced within individual sections) that offer additional perspectives or information on these same issues.

\begin{itemize}
    \item ``Report to the council of the AAS from the working group on the status of women in astronomy'' \citep{Cowley.etal.1974}
    \item Proceedings from \textit{Women at Work: A Meeting on the Status of Women in Astronomy} \citep{Urry.etal.1992}
    \item The Baltimore Charter for Women in Astronomy\footnote{\url{https://www.stsci.edu/files/live/sites/www/files/home/events/event-assets/1992/_documents/1992-workshop-women-at-work-charter.pdf}}
    \item \textit{Gender Differences at Critical Transitions in the Careers of Science, Engineering, and Mathematics Faculty} \citep{NAP12062}
    \item \href{https://aas.org/comms/cswa/news/pasadenarecs}{Equity Now: The Pasadena Recommendations for Gender Equality in Astronomy}
    \item \textit{Blazing the Trail: Essays by Leading Women in Science} \citep{Ideal.Meharchand.2013}
    \item ``Science and gender: scientists must work harder on equality'' \citep{Urry.2015}
    \item ``Gender bias in academia: A lifetime problem that needs solutions'' \citep{Llorens2021}
    \item  \textit{The Sky is for Everyone: Women Astronomers in Their Own Words} \citep{Trimble.Weintraub.2022} and the many excellent references therein
    \item ``Closing the gender gap in the Australian astronomy workforce'' \citep{Kewley.2021} and ``The achievement of gender parity in a large astrophysics research centre'' \citep{Kewley.etal.2023}
    \item \textit{Astronomy as a Field: A Guide for Aspiring Astrophysicists} \citep{Polzin.etal.2023}
    \item Recommendations to the American Astronomical Society by the AAS Committee on the Status of Women in Astronomy \citep{Wexler.etal.2025a, Wexler.etal.2025b}
\end{itemize}

\bibliography{references/furtherreading}{}
\bibliographystyle{aasjournal}

\end{document}